\documentstyle[12pt,epsfig,psfig]{mnthesis}
\def\lord{$ \raisebox{-.3ex}{$\stackrel{<}{_{\sim}}$} $}

\input{epsf.sty}

\newcommand{\beeq}{\begin{equation}}
\newcommand{\eneq}{\end{equation}}
\newcommand{\beqn}{\begin{eqnarray}}
\newcommand{\eeqn}{\end{eqnarray}}

\def\mybig{\displaystyle \strut }

\def\dd{\partial}
\def\la{\raise.16ex\hbox{$\langle$}\lower.16ex\hbox{}  }
\def\ra{\, \raise.16ex\hbox{$\rangle$}\lower.16ex\hbox{} }
\def\go{\rightarrow}

\def\next{{~,~~~}}
\def\onehalf{ \hbox{${1\over 2}$} }

\def\eff{{\rm eff}}
\def\P{{\rm P}}

\def\vphi{\varphi}
\def\ep{\epsilon}

\def\myfrac#1#2{{\mybig #1\over \mybig #2}}

\author{Ramin G. Daghigh}
\title{Microscopic Black Holes and Cosmic Shells}
\phd
\abstract{\vspace{-0.8cm}

In the first part of this thesis the relativistic viscous fluid equations describing the outflow of high 
temperature matter created via Hawking radiation from microscopic black holes 
are solved numerically for a realistic equation of state.  We focus on black 
holes with initial temperatures greater than 100 GeV and lifetimes less than 6 
days.  The spectra of direct photons and photons from $\pi^0$ decay are 
calculated for energies greater than 1 GeV.  We calculate the diffuse gamma 
ray spectrum from black holes distributed in our galactic halo.  However, the 
most promising route for their observation is to search for point sources 
emitting gamma rays of ever-increasing energy.  We also calculate the spectra of all three flavors of neutrinos arising from 
direct emission from the fluid at the neutrino-sphere and from the decay of 
pions and muons from their decoupling at much larger radii and smaller 
temperatures for neutrino energies between 
1 GeV and the Planck energy.  The results for neutrino spectra may be applicable for the last few 
hours and minutes of the lifetime of a microscopic black hole.  In the second part of this thesis the combined field equations of gravity and a scalar field are studied.  When a potential for a scalar field has two local minima there arise
spherical shell-type solutions of the classical field equations
due to gravitational attraction. We establish such solutions numerically
in a space which is asymptotically de Sitter.  It generically
arises when the energy scale characterizing the scalar field potential
is much less than the Planck scale. It is shown that the mirror image of
the shell appears in the other half of the Penrose diagram.  The configuration 
is smooth everywhere with no physical singularity.}
\words{282}
\program{Physics}
\director{J. I. Kapusta}
\copyrightpage

\dedication{To Paula, Nader, Eghbaleh and Hamoon.}
\acknowledgements{I would like to express my deep appreciation to Joseph Kapusta for his guidance and insight throughout this research.  With his kindness and patience, he surpassed the role of adviser.  He is and will be a role model and a mentor throughout my life.  I would also like to express my deep appreciation to Yutaka Hosotani for his contributions in this thesis.  Finally, I would like to thank my wife, Paula, for her patience during the years of graduate school.  }

\begin{document}

\beforepreface

\afterpreface

\chapter{Introduction}

\hspace{0.2in}  Black holes can be created in the collapse of a
star which was originally 10 to 20 times the mass of the sun.  There is strong
evidence that there are huge black holes of about 1 million solar
masses near the center of galaxies.  Another possibility
are primordial black holes of almost arbitrarily small mass which were created in the early universe.  Depending on the time of the formation of primordial black holes in the early universe, they can have a wide range of masses.  The black holes which were formed at the the Planck time ($10^{-43}$ s) would have the Planck mass ($10^{-5}$ g), and those were formed at $1$ s would be as massive as $10^5$ solar masses.  It was shown by Hawking in 1974 that a black hole radiates thermally with a temperature that is inversely proportional to its mass.  This phenomenon, which is called Hawking radiation, becomes important for black holes with small enough mass.  In fact, primordial black holes with a mass less than about $10^{15}$ g have already evaporated within the lifetime of the universe, which is $10^{10}$ yrs.  The size of such black holes is less than $10^{-16}$ m, making them microscopic.  Primordial black holes with an initial mass a little more than $10^{15}$ g are evaporating at the present time at a rate which makes it possible to detect them.  In the first part of this thesis we study the particle spectra emitted from primordial microscopic black holes in the last 6 days of their existence.  We assume that at very high black hole temperatures the particles emitted from black hole via Hawking radiation will scatter from each other enough to create a shell of thermalized matter surrounding these black holes.  This shell of hot matter will change the emergent spectrum from microscopic black holes.  More particles with lower average energy will be emitted out of the thermalized matter compared to fewer particles with higher average energy emitted directly via Hawking radiation.  This phenomenon is similar to the photosphere of stars.  Because of the photosphere the temperature on the outer surface of a star is much lower than the temperature at the core of the star.  In this thesis we show that the most promising route for the observation 
of microscopic black holes is to search for point sources emitting gamma rays 
of ever-increasing energy. 
A black hole 
with a temperature above $100$ GeV and a Schwarzschild radius less than $10^{-4}$ fm will get brighter on a time scale of 6 days 
and then disappear.  Such an observation would be remarkable, possibly unique, because astrophysical sources of gamma rays normally cool at late times.  
This would 
directly reflect the increasing Hawking temperature as the black hole explodes 
and disappears.  Since the highest temperatures in the present universe exist in the vicinity of microscopic black holes, by studying the emergent spectrum from these black holes we can learn about the physics at very high temperatures which is not achievable in terrestrial experiments.  Observation of microscopic black holes will shed light on the biggest problems in physics today.  By studying and observing microscopic black holes we will be able to test the physics above the electroweak scale and beyond the four dimensional Standard Model. 

It is possible that the source of the dark energy in our universe is a scalar field.  Therefore, it is important to study the solutions to the combined field equations of gravity and a scalar field.  In the second part of this thesis we report the discovery of static shell-like solutions to the combined field equations of gravity and a scalar field with a potential which has two non-degenerate minima.  The absolute minimum of the potential is the true vacuum and the other minimum is the false vacuum.  Both inside and outside of the shells are de Sitter space and their radii are comparable to the size of the horizon of the universe.  The shells are spherical and very thin compared to their radii.    The energy density in the shells is much higher than the energy density of the de Sitter space inside and outside of the shells.  If anything 
like these structures exist in nature they most likely would have been created 
in the early universe and are therefore cosmological.  We know of no other way 
to produce them.

In chapter 2 of this thesis we present the work done in collaboration with J. Kapusta which can be found in \cite{Ramin1} and \cite{Ramin2}.  In chapter 3 we present the work done in collaboration with Y. Hosotani, J. Kapusta and T. Nakajima which can be found in \cite{Ramin3}.

\section{Microscopic Black Holes}

\hspace{0.2in}  Hawking radiation from black holes \cite{Hawk} is of fundamental interest
because it is a phenomenon arising out of the union of general relativity and quantum mechanics, a situation that could 
potentially be observed.  It is also of great interest because of the 
temperatures involved.  A black hole with mass $M$ radiates thermally with a 
Hawking temperature $T_H = m_{\rm P}^2/8\pi M$
where $m_{\rm P} = G^{-1/2} = 1.22\times 10^{19}$ GeV is the Planck mass.
(Units are $\hbar = c = k_{\rm B} = 1$.)  In order for the black hole
to evaporate rather than accrete it must have a temperature greater than that
of 
the present-day blackbody radiation of the universe of 2.7 K = 2.3$\times
10^{-
4}$ eV.  This implies that $M$ must be less than $1\%$ of the mass of the
Earth.

Such small black holes most likely would have been formed primordially; there
is 
no other mechanism known to form them.  As the black hole radiates, its mass 
decreases and its temperature increases until $T_H$ becomes comparable to the
Planck mass, at which point the semiclassical calculation breaks down and the 
regime of full quantum gravity is entered.  Only in two other situations are 
such enormous temperatures achievable: in the early universe and in central 
collisions of heavy nuclei like gold or lead.  Even then only about $T = 500$ 
MeV is reached at the RHIC (Relativistic Heavy Ion Collider) just completed at
Brookhaven National Laboratory and $T = 1$ GeV is expected at the LHC (Large
Hadron Collider) at CERN to be completed in 2006.  Supernovae and newly formed
neutron stars only reach temperatures of a few tens of MeV.
To set the scale from fundamental physics, we note that the spontaneously
broken
chiral symmetry of QCD gets restored in a phase transition or rapid crossover at a
temperature around 170 MeV, while the spontaneously broken gauge symmetry in
the
electroweak sector of the standard model gets restored in a phase
transition or rapid crossover at a temperature around 100 GeV.  The fact that
temperatures of the latter order of magnitude will never be achieved in a
terrestrial experiment should motivate us to study the fate of microscopic
black
holes during the final days, hours and minutes of their lives when their 
temperatures have risen to 100 GeV and above.  In this thesis we shall focus on 
Hawking temperatures greater than 100 GeV.  The fact that microscopic black 
holes have not yet been observed \cite{review} should not be viewed as a 
deterrent, but rather as a challenge for the new millennium!

There is some uncertainty over whether the particles scatter from each
other after being emitted, perhaps even enough to allow a fluid description of
the wind coming from the black hole.  Let us examine what might happen as the
black hole mass decreases and the associated Hawking temperature increases.

When $T_H \ll m_{\rm e}$ (electron mass) only photons, gravitons,
and neutrinos will be created with any significant probability.  These
particles will not significantly interact with each other, and will
freely propagate away from the black hole with energies of order $T_H$.
Even when $T_H \approx m_{\rm e}$ the Thomson cross section is too
small to allow the photons to scatter very frequently in the rarified
electron-positron plasma around the black hole.  This may change when
$T_H \approx 100$ MeV when muons and charged pions are created in
abundance.  At somewhat higher temperatures hadrons are copiously produced and
local thermal equilibrium may be achieved, although exactly how is an unsettled
issue.  Are hadrons emitted directly by the black hole?  If so, they will be
quite abundant at temperatures of order 150 MeV because their mass spectrum
rises exponentially (Hagedorn growth as seen in the Particle Data Group tables
\cite{PDG}).  Because they are so massive they move nonrelativistically and may
form a very dense equilibrated gas around the black hole.  But hadrons are
composites of quarks and gluons, so perhaps quark and gluon jets are emitted
instead?  These jets must decay into the observable hadrons on a typical
proper length scale of 1 fm and a typical proper time scale of 1 fm/c.
This was first studied by
MacGibbon and Webber \cite{MW} and MacGibbon and Carr \cite{MC}.  Subsequently
Heckler \cite{Ha} argued that since the emitted quarks and gluons are so
densely
packed outside the event horizon they are not actually fragmenting into
hadrons in vacuum but in something more like a quark-gluon plasma, so perhaps
they thermalize.  He also argued that QED bremsstrahlung and pair production
were sufficient to lead to a thermalized QED plasma when $T_H$ exceeded 45 GeV
\cite{Hb}.  These results are somewhat controversial and need to be confirmed.
The issue really is how to describe the emission of wave packets via the Hawking
mechanism when the emitted particles are (potentially) close enough to be
mutually interacting.  A more quantitative treatment of the particle
interactions on a semiclassical level was carried out by Cline, Mostoslavsky
and
Servant \cite{cline}.  They solved the relativistic Boltzmann equation with QCD
and QED interactions in the relaxation-time approximation. It was found that
significant particle scattering would lead to a photosphere though not perfect
fluid flow.

Rather than pursuing the Boltzmann transport equation Kapusta applied 
relativistic viscous fluid equations to the problem assuming sufficient
particle 
interaction \cite{me}. It was found that a self-consistent description emerges 
of a fluid just marginally kept in local thermal equilibrium, and that
viscosity 
is a crucial element of the dynamics.  The fluid description has been used to calculate the particle spectra emitted in heavy ion collisions  and has been very successful in this application \cite{Cser}.  The purpose of this thesis is a more 
extensive analysis of these equations and their observational consequences. 

The 
plan of the second chapter of this thesis is as follows.  In Sec. 2.1 we give a brief review of Hawking radiation 
sufficient for our uses.  In Sec. 2.2 we give the set of relativistic viscous 
fluid equations necessary for this problem along with the assumptions that go 
into them.  In Sec. 2.3 we suggest a relatively simple parametrization of the 
equation of state for temperatures ranging from several MeV to well over 100 
GeV.  We also suggest a corresponding parametrization of the bulk and shear 
viscosites.  In Sec. 2.4 we solve the equations numerically, study the scaling behavior of the solutions, and check their physical self-consistency.
In Sec. 2.5 we estimate where the transition from viscous fluid flow to
free-streaming takes place.  In Sec. 2.6 we calculate the instantaneous and 
time-integrated spectra of high energy photons from the two dominant sources: 
direct and neutral pion decay.  In Sec. 2.7 we study the diffuse gamma ray 
spectrum from microscopic black holes distributed in our galactic halo.  We
also 
study the systematics of gamma rays from an individual black hole, should we be so fortunate to observe one.  In Sec. 2.8 we calculate the instantaneous and 
time-integrated spectra of high energy neutrinos from the three dominant 
sources: direct, pion decay and muon decay.  In Sec. 2.9 we compare the spectra from all neutrino sources graphically.  We also compare our results with the spectra of neutrinos emitted directly as Hawking 
radiation without any subsequent interactions.  In Sec. 2.10 we
study the possibility of observing neutrinos from a microscopic black 
hole directly.

\section{Cosmic Shells}

\hspace{0.2in} Gravitational interactions, which are inherently attractive for ordinary
matter, can produce soliton-like objects even when they
are strictly forbidden in flat space.  They are possible as
a consequence of the balance between repulsive and attractive forces.
One such example is a monopole or dyon solution in the pure
Einstein-Yang-Mills theory in the asymptotically anti-de Sitter
space \cite{Hosotani1,Radu,Galtsov0}.  In the pure Yang-Mills theory in
flat space  there can be no static solution at all \cite{Deser} but once
gravitational interaction is included there arise particle like solutions
\cite{Bartnik}. Whereas all solutions are unstable in the asymptotically
flat or de Sitter space, there appear a continuum of stable monopole and
dyon solutions in the asymptotically anti-de Sitter space.  The stable
solutions are cosmological in nature; their size is typically of order
$|\Lambda|^{-1/2}$ where $\Lambda$ is the cosmological constant.

The possibility of  false vacuum black holes has also been
explored. Suppose that the potential in a scalar field
theory has two minima, one corresponding to the  true vacuum and the other
to the false vacuum.   If the universe is in the false vacuum, a bubble
of the true vacuum is created by quantum tunneling which expands with
accelerated velocity.  The configuration is called a bounce \cite{Coleman}.  
Now flip the configuration \cite{Kapusta}.  The universe is in the
true vacuum with potential $V=0$ and the inside of a sphere is excited to the 
false vacuum with $V>0$. Is such a de Sitter lump in Minkowski space possible?  If the lump is too small it would be totally unstable.  The energy localized
inside the lump can dissipate to spatial infinity.  If the lump is big
enough the Schwarzschild radius becomes larger than the lump radius so that
the lump is inside a black hole.  The energy cannot escape to infinity.
It looks like a soliton in Minkowski space.  However, as a black hole it is a 
dynamical object.  The configuration is essentially time dependent.
This false vacuum black hole configuration, however, does not solve
the static field equations at the horizon.  It has been recently proven that there
can be no such everywhere-regular black hole solution \cite{Galtsov,Bronnikov}.
Rather, false vacuum lumps in flat space evolve dynamically 
\cite{Maeda,Sato,Guth}.

The purpose of this part of thesis is to report new solutions to the coupled
equations of  gravity and scalar field theory which display a spherically
symmetric shell structure \cite{Hosotani2}.  We demonstrate that such
a structure appears when the potential for a scalar field has two local minima
and the space is asymptotically de Sitter.  In the examples we present, both
the inside and outside of the shells are de Sitter space with the same
cosmological constant.   The structure becomes possible only when the energy
scale of the scalar field potential becomes small compared with the Planck 
scale.  While such shell structures might not be easy to create in the present 
universe,  it is quite plausible that they could have been created during a
phase  transition early in the universe \cite{Maeda}.
A similar configuration has been investigated in Ref.\ \cite{Dymnikova}.

The 
plan of the third chapter of this thesis is as follows.
In Sec. 3.1 we precisely state the 
problem, solve the field equations in those regions of space-time where they can 
be linearized, and sketch the solution in the nonlinear shell region in static 
coordinates.  In Sec. 3.2 we solve the nonlinear equations in the shell region 
and display the dependence on the parameters of the theory.  In Sec. 3.3 we 
extend the solution from static coordinates, which have a coordinate 
singularity, to other coordinate systems that do not, thereby displaying the 
existence and character of the solution throughout the full space-time manifold.  
In Sec. 3.4 we study the stability of the classical solution to quantum 
fluctuations.  Finally in chapter 4 we will conclude and summarize both parts of this thesis.

\chapter{High Temperature Matter and Particle Spectra from Microscopic Black Holes}

\section{Hawking Radiation}

\hspace{0.2in} There are at least two intuitive ways to think about Hawking radiation from
black holes.  One way is vacuum polarization.  Particle-antiparticle pairs are
continually popping in and out of the vacuum, usually with no observable
effect.
In the presence of matter, however, their effects can be observed.  This is the
origin of the Lamb effect first measured in atomic hydrogen in 1947.  When
pairs
pop out of the vacuum near the event horizon of a black hole one of them may be
captured by the black hole and the other by necessity of conservation laws will
escape to infinity with positive energy.  The black hole therefore has lost
energy - it radiates.  Due to the general principles of thermodynamics applied
to black holes it is quite natural that it should radiate thermally.  An
intuitive argument that is more quantitative is based on the uncertainty
principle.  Suppose that we wish to confine a massless particle to the vicinity
of a black hole.  Given that the average momentum of a massless particle at
temperature $T$ is approximately $\pi T$, the uncertainty principle requires
that confinement to a region the size of the Schwarzschild diameter places a
restriction on the minimum value of the temperature
\begin{equation}
\pi T \cdot 2 r_S > 1/2.
\end{equation}
The minimum is actually attained for the Hawking temperature.  The various
physical quantities are related as $r_S = 2M/m_{\rm P}^2 =
1/4\pi T_H$.

The number of particles of spin $s$ emitted with energy $E$ per unit time is
given by the formula
\begin{equation}
\frac{dN_s}{dEdt} = \frac{\Gamma_s}{2\pi} \,
\frac{1}{\exp(E/T_H)-(-1)^{2s}} \, .
\end{equation}
All the computational effort really goes into calculating the absorption
coefficient $\Gamma_s$ from a relativistic wave equation in the presence of a
black hole.  Integrating over all particle species yields the luminosity
\begin{equation}
L = -\frac{dM}{dt} = \alpha(M) \frac{m_{\rm P}^4}{M^2} =
64 \pi^2 \alpha(T_H) T_H^2 \, .
\end{equation}
Here $\alpha(M)$ is a function reflecting the species of particles available
for creation in the gravitational field of the black hole.  It is generally
sufficient to consider only those particles with mass less than $T_H$;
more massive particles are exponentially suppressed by the Boltzmann factor.
Then
\begin{equation}
\alpha = 2.011\times 10^{-8} \left[ 3700 N_0 + 2035 N_{1/2} + 835 N_1 + 95 N_2
\right]
\end{equation}
where $N_s$ is the net number of polarization degrees of freedom for all
particles with spin $s$ and with mass less than $T_H$.  The coefficients for
spin 1/2, 1 and 2 were computed by Page \cite{Page} and for spin 0 by Sanchez
\cite{Sanchez}\footnote{In \cite{Ramin1} the coefficient for spin 0 particles was taken as 4200.  In Eq. (2.4) we have reduced this value to 3700 which results in a small correction of 0.7$\%$ in the numerical value of $\alpha(T_H > 100 \,{\rm GeV})$.  We thank Harald Anlauf for providing us with this more accurate coefficient}.  In the standard model $N_0 = 4$ (Higgs boson), $N_{1/2} = 90$ (three
generations of quarks and leptons), $N_1 = 24$ [SU(3)$\times$SU(2)$\times$U(1)
gauge theory], and $N_2 = 2$ (gravitons).  This assumes $T_H$ is greater than
the temperature for the electroweak gauge symmetry restoration.
Numerically $\alpha(T_H > 100 \,{\rm GeV}) = 4.4\times 10^{-3}$.  Starting
with
a black hole of temperature $T_H$, the time it takes to evaporate or explode is
\begin{equation}
\Delta t = \frac{m_{\rm P}^2}{3 \alpha(T_H) (8\pi T_H)^3} \, .
\end{equation}
This is also the characteristic time scale for the rate of change of the
luminosity of a black hole with temperature $T_H$.

At present a black hole will explode if $T_H > 2.7$ K and correspondingly
$M < 4.6\times 10^{25}$ g which is approximately 1\% of the mass of the Earth.
More massive black holes are cooler and therefore will absorb more matter and
radiation than they radiate, hence grow with time.  Taking into account
emission
of gravitons, photons, and neutrinos a critical mass black hole today has a
Schwarszchild radius of 68 microns and a lifetime of $2\times10^{43}$ years.

\section{Relativistic Viscous Fluid Equations}

\hspace{0.2in} The relativistic
imperfect fluid equations describing a steady-state, spherically symmetric flow
with no net baryon number or electric charge and neglecting gravity
(see below) are $T^{\mu\nu}_{\;\;\;\;;\nu} =$ {\em black hole source}.  The
nonvanishing components of the energy-momentum tensor in radial coordinates are
\cite{MTW}
\begin{eqnarray}
T^{00}&=& \gamma^2 (P+\epsilon) -P + v^2 \Delta T_{\rm diss} \nonumber \\
T^{0r}&=& v\gamma^2 (P+\epsilon) + v \Delta T_{\rm diss} \nonumber \\
T^{rr}&=& v^2\gamma^2 (P+\epsilon) +P + \Delta T_{\rm diss}
\end{eqnarray}
representing energy density, radial energy flux, and radial momentum flux,
respectively, in the rest frame of the black hole.  Here
$v$ is the radial velocity with $\gamma$ the corresponding Lorentz factor, $u =
v\gamma$, $\epsilon$ and $P$ are the local energy density and pressure, and
\begin{equation}
\Delta T_{\rm diss} = -\frac{4}{3}\eta \gamma^2 \left( \frac{du}{dr}
-\frac{u}{r}\right) - \zeta \gamma^2 \left( \frac{du}{dr}
+\frac{2u}{r}\right) \, ,
\end{equation}
where $\eta$ is the shear viscosity and $\zeta$ is the bulk viscosity.  A
thermodynamic identity gives $Ts = P + \epsilon$ for zero chemical potentials,
where $T$ is temperature and $s$ is entropy density.  There are two independent
differential equations of motion to solve for the functions $T(r)$ and $v(r)$.
These may succinctly written as
\begin{eqnarray}
\frac{d}{dr} \left( r^2 T^{0r} \right) &=& 0 \nonumber \\
\frac{d}{dr} \left( r^2 T^{rr} \right) &=& 0 \, .
\end{eqnarray}

An integral form of these equations is sometimes more useful since it can 
readily incorporate the input luminosity from the black hole.  The first
represents the equality of the energy flux passing through a sphere of radius r
with the luminosity of the black hole:
\begin{equation}
4\pi r^2 T^{0r} = L.
\end{equation}
The second follows from integrating a linear combination of the differential
equations.  It represents the combined effects of the entropy from the black
hole together with the increase of entropy due to viscosity:
\begin{equation}
4\pi r^2 u s  = 4\pi \int_{r_0}^r dr' \, r'^2 \frac{1}{T}\left[
\frac{8}{9} \eta \left( \frac{du}{dr'} - \frac{u}{r'} \right)^2
+ \zeta \left( \frac{du}{dr'} + \frac{2u}{r'} \right)^2 \right]
+ \frac{L}{T_H}.
\end{equation}
The term $L/T_H$ arises from equating the entropy per unit time lost by the
black hole $-d S_{\rm bh}/dt$ with that flowing into the matter.  Using the
area
formula for the entropy of a black hole, $S_{\rm bh} = m_{\rm P}^2 \pi r_S^2 =
4\pi M^2/m_{\rm P}^2$, and identifying $-dM/dt$ with the luminosity, the
entropy
input from the black hole is obtained.

The above pair of equations are to be applied beginning at some radius $r_0$
greater than the Schwarzschild radius $r_S$, that is, outside the quantum
particle production region of the black hole.
The radius $r_0$ at which the imperfect fluid equations are first applied
should
be chosen to be greater than the Schwarzschild radius, otherwise the
computation
of particle creation by the black hole would be invalid.  It should not be too
much greater, otherwise particle collisions would create more entropy than is
accounted for by the equation above.  The energy and entropy flux into the
fluid 
come from quantum particle creation by the black hole at temperature $T_H$. 
Gravitational effects are of order $r_S/r$, hence
negligible for $r > (5-10)r_S$.

\section{Equation of State and Transport Coefficients}

\hspace{0.2in} Determination of the equation of state as well as the two viscosities for
temperatures ranging from MeV to TeV and more is a formidable task.  Here we
shall present some relatively simple parametrizations that seem to contain the 
essential physics.  Improvements to these can certainly be made, but probably 
will not change the viscous fluid flow or the observational consequences very
much. 

\begin{figure}[htb]
\centerline{\epsfig{figure=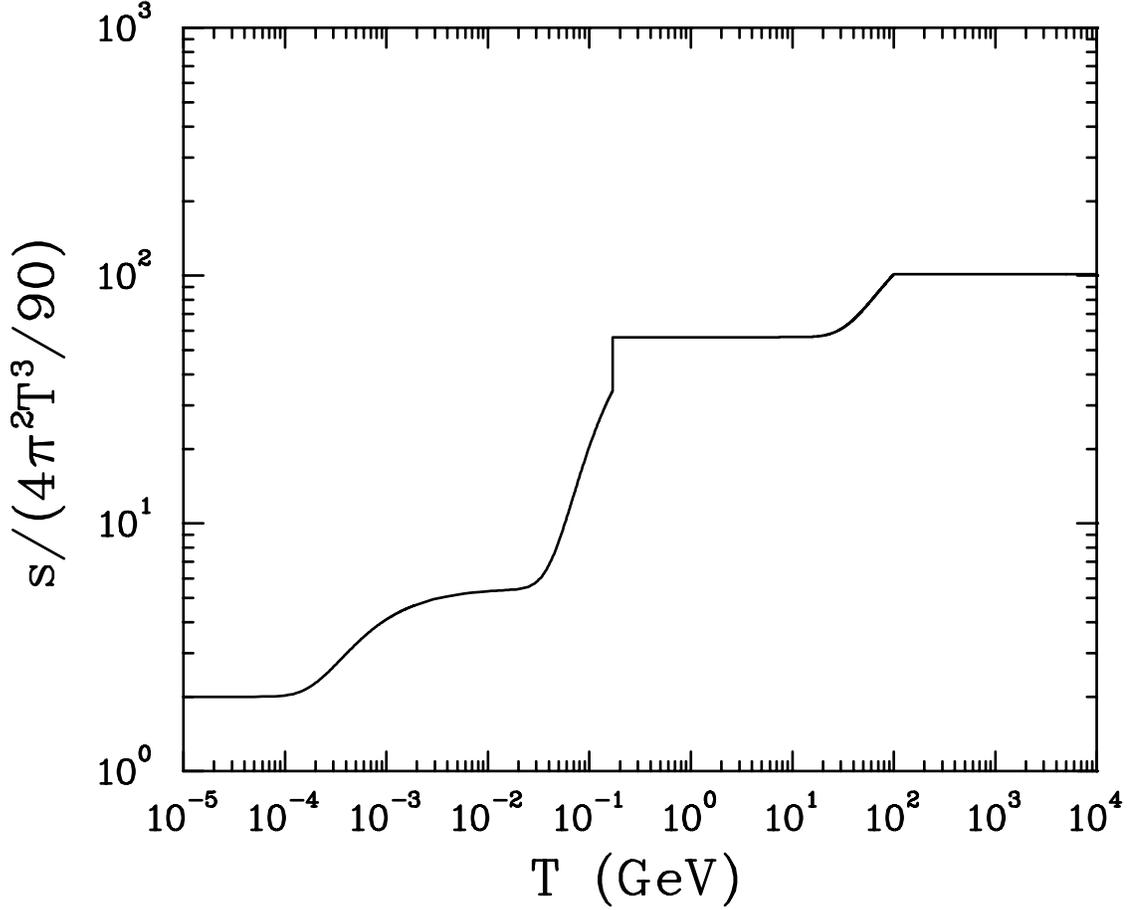,width=12.0cm,angle=90}}
\caption{Entropy density as a function of temperature, excluding neutrinos
and gravitons.  It is assumed that the QCD phase transition is first order
and the EW phase transition is second order.}
\label{bhgamma-f1}
\end{figure}

The hot shell of matter surrounding a primordial black hole provides a
theoretical testing ground rivaled only by the big bang itself. 
To illustrate this we have plotted a semi-realistic parametrization of the 
equation of state in figure \ref{bhgamma-f1}.  Gravitons and neutrinos are not included.  We 
assume a second order electroweak phase transition at a temperature of 
$T_{EW}$ = 100 GeV.  Above that temperature the standard model has 101.5 
effective massless bosonic degrees of freedom (as usual fermions count as 7/8
of 
a boson).  We assume a first order QCD phase transition at a 
temperature of $T_{QCD}$ = 170 MeV.  The number 
of effective massless bosonic degrees of freedom changes from 47.5 just above 
this critical temperature (u, d, s quarks and gluons) to 7.5 just below it 
(representing the effects of all the massive hadrons in the Particle Data Group
tables) \cite{OK}.  Below 30 MeV only electrons, positrons, and photons remain, and finally below a few hundred keV only photons survive in any appreciable 
number.  The explicit parametrization shown in figure \ref{bhgamma-f1} is as follows:

\begin{eqnarray}
s(T) &=& \frac{4\pi^2}{90}T^3 \,
\left\{ \begin{array}{ll}
101.5, \hspace{2.0in} T_{EW} \leq T, \\
56.5 + 45 \,{\rm e}^{-(T_{EW}-T)/T},  \hspace{0.5in} T_{QCD} \leq T < T_{EW}, \\
2 + 3.5 \,{\rm e}^{-m_e/T} + 27.25 \,{\rm e}^{-(T_{QCD}-T)/T},
  \hspace{0.3in} T < T_{QCD}.
\end{array} 
\right.
\end{eqnarray}

It may very well be that there are no true thermodynamic phase transitions in 
the standard model but only rapid crossovers from one phase to the other.  None of our calculations or results depend on such details.
A word about neutrinos:  It is quite possible that they should be considered in approximate equilibrium at temperatures above 100 GeV where the electroweak 
symmetry is restored.  Still there is some uncertainty about this.  Since they 
provide only a few effective degrees of freedom out of more than 100 their 
neglect should cause negligible error.

Now we turn to the viscosities.  The shear viscosity was calculated in 
\cite{tau} for the full standard model in the symmetry restored phase, meaning 
temperatures above 100 GeV or so, using the relaxation time approximation.  The result is
\begin{equation}
\eta(T > 100 \, {\rm GeV}) = 82.5 \, T^3
\end{equation}
when numerical values for coupling constants etc. are put in.  The shear 
viscosity for QCD degrees of freedom only was calculated to leading order in
the 
QCD coupling $\alpha_s$ in \cite{baym} to be
\begin{equation}
\eta({\rm QCD}) = \frac{0.342 (1+1.7N_f)}{(1+N_f/6)\alpha_s^2
\ln(\alpha_s^{-1})} \, T^3
\end{equation}
where $N_f$ is the number of quark flavors whose mass is less than $T$. An 
improved calculation for gauge theories was given in \cite{arnold}; for
QCD there is very little difference with \cite{baym}.  We observe 
that the ratio of the shear viscosity to the entropy density, as appropriate
for 
the above two cases, is dimensionless and has about the same numerical value in both.  Therefore, as a practical matter we assume that the shear viscosity 
always scales with the entropy density for all temperatures of interest.  We 
take the constant of proportionality from the full standard model cited above:
\begin{equation}
\eta = \frac{82.5}{101.5} \left( \frac{s}{4\pi^2/90} \right).
\end{equation}
There is even less known about the bulk viscosity at the temperatures of 
interest to us. The bulk viscosity is zero for point particles with no internal degrees of freedom and with local interactions among them.  In renormalizable 
quantum field theories the interactions are not strictly local.  In particular, the coupling constants acquire temperature dependence according to the 
renormalization group.  For example, to one loop order the QCD coupling has the functional dependence $\alpha_s \sim 1/\ln(T/\Lambda)$ where $\Lambda$ is the 
QCD scale.  On account of this dependence the bulk viscosity is nonzero.  We 
estimate that
\begin{equation}
\zeta \approx 10^{-4}\, \eta
\end{equation}
and this is what we shall use in the numerics.      

Overall we have a modestly realistic description of the equation of state and 
the viscosities that are still a matter of theoretical uncertainty.
One needs $s(T), \eta(T), \zeta(T)$ over a huge range of $T$.  Of course, these are some of the quantities one hopes to obtain experimental information on from observations of exploding black holes.

\section{Numerical Solution and Scaling}

\hspace{0.2in} Several limiting cases of the relativistic viscous fluid equations were
studied in \cite{me}.  The most realistic situation used the equation of
state $\epsilon = aT^4$, $s = (4/3)aT^3$ and viscosities
$\eta = b_ST^3$, $\zeta = b_BT^3$ with the coefficients $a$, $b_S$, $b_B$
all constant.  A scaling solution, valid at large radii when $\gamma \gg 1$,
was found to be $T(r) = T_0 (r_0/r)^{2/3}$ and $\gamma(r) = \gamma_0 
(r/r_0)^{1/3}$.  The constants must be related by $36aT_0r_0 = 
(32b_S + 441b_B)\gamma_0$.  This $r$-dependence of $T$ and $\gamma$ is exactly 
what was conjectured by Heckler \cite{Hb}.

It was shown in \cite{me} that if approximate local thermal equilibrium is 
achieved it can be maintained, at least for the semi-realistic situation 
described above.  The requirement is that the inverse of the local volume 
expansion rate $\theta = u^{\mu}_{\;\; ;\mu}$ be
comparable to or greater than the relaxation time for thermal equilibrium
\cite{MTW}.  Expressed in terms of a local volume element $V$ and proper time
$\tau$ it is $\theta = (1/V)dV/d\tau$, whereas in the rest frame of the black
hole the same quantity can be expressed as $(1/r^2)d(r^2 u)/dr$.  Explicitly
\begin{equation}
\theta = \frac{7\gamma_0}{3r_0}\left(\frac{r_0}{r}\right)^{2/3}
= \frac{7\gamma_0}{3r_0T_0} T \, .
\end{equation}
Of prime importance in achieving and maintaining local thermal equilibrium in a
relativistic plasma are multi-body processes such as $2 \rightarrow 3$ and
$3 \rightarrow 2$, etc.  This has been well-known when calculating quark-gluon
plasma formation and evolution in high energy heavy ion collisions
\cite{klaus,S}
and has been emphasized in Refs. \cite{Ha,Hb} in the context of black hole
evaporation.  This is a formidable task in the standard model with its 16
species of particles.  Instead three estimates for the requirement that
local thermal equilibrium be maintained were made. The first and simplest 
estimate is to require that the thermal de Broglie wavelength of a massless 
particle, $1/3T$, be less than $1/\theta$.   The second estimate is to require 
that the Debye screening length for each of the gauge groups in the standard 
model be less than $1/\theta$. The Debye screening length is the inverse of the Debye screening mass $m^{\rm D}_n$ where $n =1, 2, 3$ for the gauge groups
U(1),
SU(2), SU(3).  Generically $m^{\rm D}_n \propto g_nT$ where $g_n$ is the gauge
coupling constant and the coefficient of proportionality is essentially the
square root of the number of charge carriers \cite{kapbook}.  For example, for
color SU(3) $m^{\rm D}_3 = g_3 \sqrt{1+N_{\rm f}/6}\,T$
where $N_{\rm f}$ is the number of light quark flavors at the temperature $T$.
The numerical values of the gauge couplings are: $g_1 = 0.344$, $g_2 = 0.637$,
and $g_3 = 1.18$ (evaluated at the scale $m_Z$) \cite{PDG}.  So within a factor
of about 2 we have $m^{\rm D} \approx T$. The third and most relevant estimate
is the mean time between two-body collisions in the standard model for
temperatures greater than the electroweak symmetry restoration temperature.
This mean time was calculated in \cite{tau} in the process of
calculating the viscosity in the relaxation time approximation.  Averaged over
all particle species in the standard model one may infer from that paper an
average time of $3.7/T$.  Taking into account multi-body reactions would
decrease that by about a factor of two to four.  All three of these estimates
are consistent within a factor of 2 or 3.  The conclusion to be drawn is that
local thermal equilibrium should be achieved when
$\theta \lord T$. Once thermal equilibrium is achieved it is not lost because 
$\theta / T$ is independent of $r$.  The picture that emerges is that of an 
imperfect fluid just marginally kept in local equilibrium by viscous forces.

The results quoted above are only valid at large $r$ and for the equation of 
state $s \propto T^3$.  To know the behavior of the solution at non-asymptotic 
$r$ and for the more sophisticated equation of state and viscosities described 
in Sec. 2.3 requires a numerical analysis.  We have found that the most 
convenient form of the viscous fluid equations for numerical evaluation are
\begin{equation}
4\pi r^2 \left[ \gamma u T s -\frac{4}{3} \eta \gamma u
\left( \frac{du}{dr} - \frac{u}{r} \right) -
\zeta \gamma u \left( \frac{du}{dr} + \frac{2u}{r} \right)
\right] = L
\end{equation}
for energy conservation [from Eq. (2.9)] and
\begin{equation}
\frac{d}{dr} \left( 4\pi r^2 u s \right) = \frac{4\pi r^2}{T}
\left[ \frac{8}{9} \eta \left( \frac{du}{dr} - \frac{u}{r} \right)^2
+ \zeta \left( \frac{du}{dr} + \frac{2u}{r} \right)^2 \right]
\end{equation}
for entropy flow [from Eq. (2.10)].  Obviously the entropy flux is a
monotonically 
increasing function of $r$ because of dissipation.

Mathematically the above pair of equations apply for all $r > 0$, although 
physically we should only apply them beyond the Schwarzschild radius $r_S$.
Let us study them first in the limit $r \rightarrow 0$, which really means the 
assumption that $v \ll 1$.  Then $u \approx v$ and $\gamma \approx 1$.  We also consider black hole temperatures greater than $T_{EW}$ so that the equation of 
state and the viscosities no longer change their functional forms.  It is 
straightforward to check that a power solution satisfies the equations, with
\begin{eqnarray}
u(r) &=& u_i (r/r_i)^{2/5} \nonumber \\
T(r) &=& T_i (r_i/r)^{3/5}
\end{eqnarray}
\begin{figure}[htb]
\centerline{\epsfig{figure=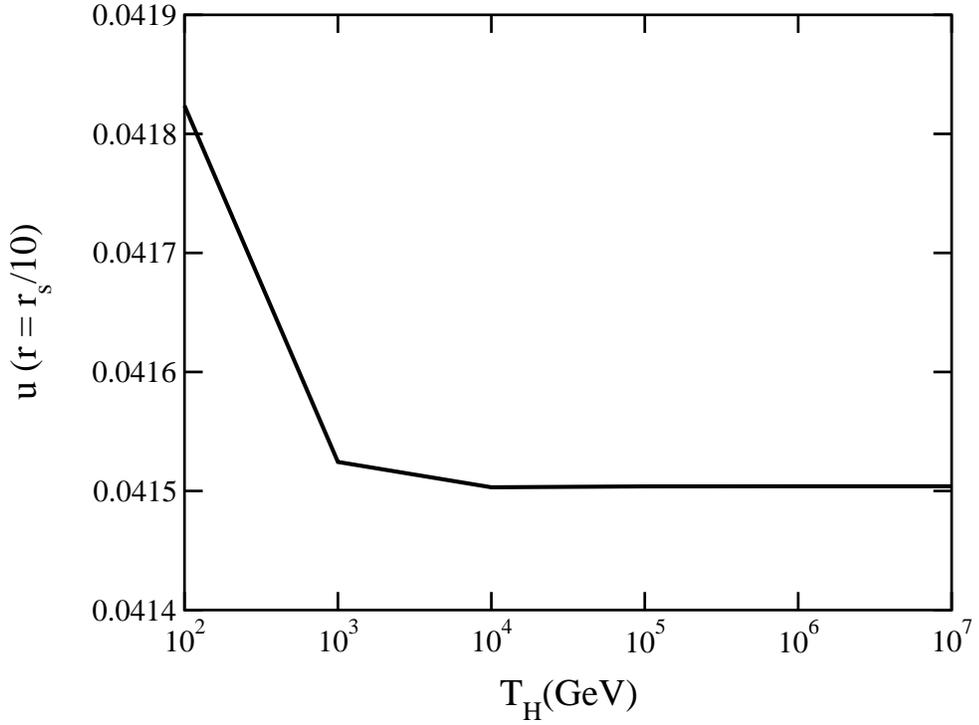,width=12.0cm,angle=270}}
\caption{The value of $u = v\gamma$ at one-tenth the Schwarzschild radius
as determined by numerical solution.  The physical applicability of the
numerical solution begins at radii greater than $r_S$.}
\label{bhgamma-f2}
\end{figure}
where $r_i$ is some reference radius.  If the luminosity and the reference 
radius are given then $u_i$ and $T_i$ are determined by the fluid equations.

The numerical solution for all radii needs some initial conditions.  Typically 
we begin the solution at one-tenth the Schwarzschild radius.  At this radius
the 
$u_i$, as determined above, is small enough to serve as a good first estimate.  However, it needs to be fine-tuned to give an acceptable solution at large $r$. For example, at large $r$ there is an approximate but false solution:
$T=$ const with $u \sim r$.  The problem is that we need a solution valid 
over many orders of magnitude of $r$.  If Eq. (2.17) is divided by
$r^2$ and if the right hand side is neglected in the limit
$r \rightarrow \infty$ then the left hand side is forced to be zero.
We have used both 
Mathematica and a fourth order Runge-Kutta method with adaptive step-size to 
solve the equations.  They give consistent results.  For more details on numerical calculations, see appendix A.

\begin{figure}[htb]
\centerline{\epsfig{figure=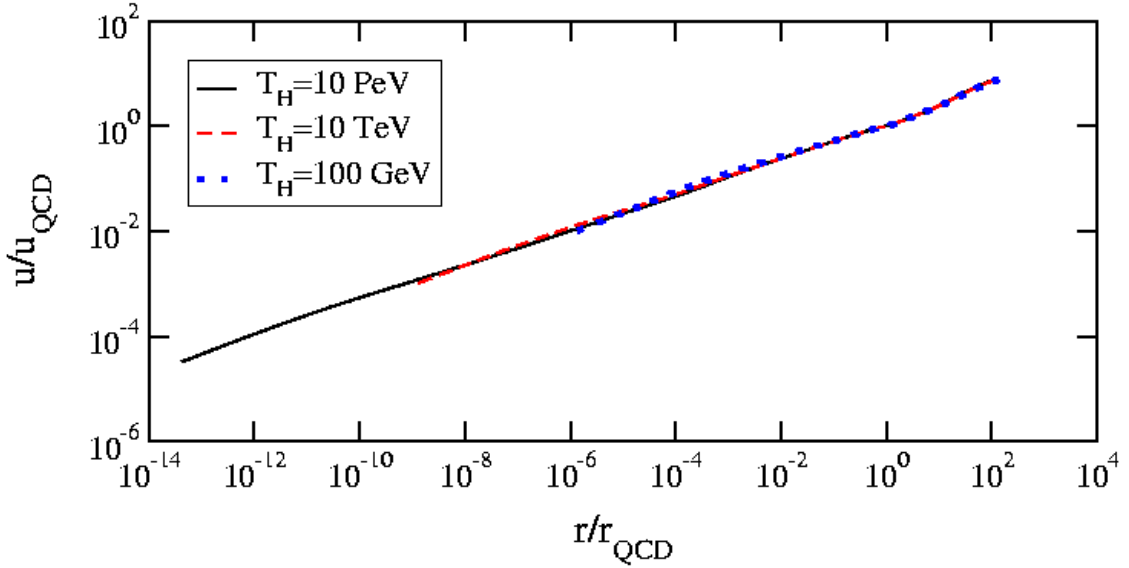,width=13.0cm,angle=270}}
\caption{The radial dependence of $u$ for three different Hawking temperatures.
The curves begin at $r_S/10$ and terminate when the local temperature reaches
10 MeV.}
\label{bhgamma-f3}
\end{figure}
\begin{figure}[htb]
\centerline{\epsfig{figure=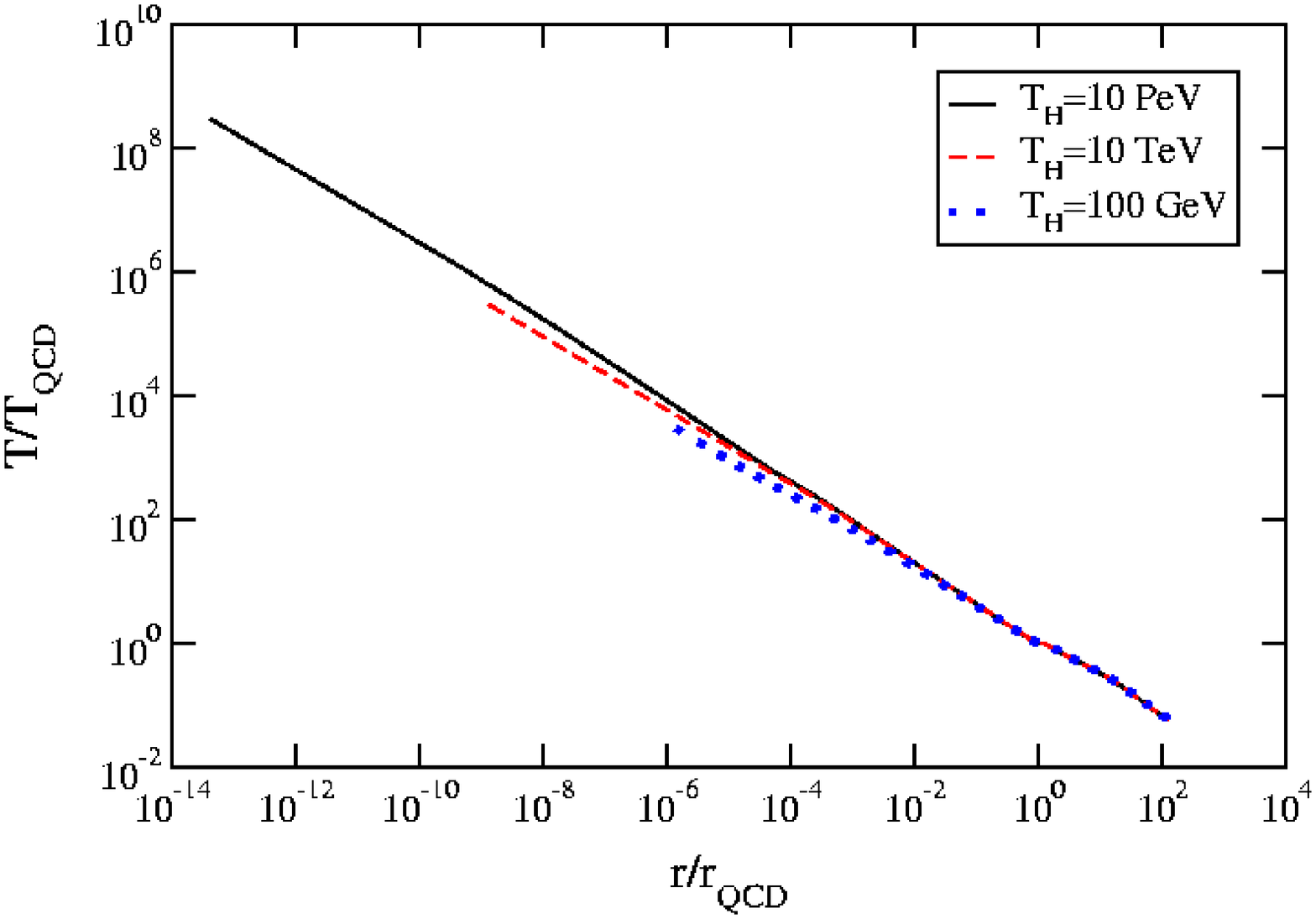,width=13.0cm,angle=270}}
\caption{ The radial dependence of $T$ for three different Hawking
temperatures.
The curves begin at $r_S/10$ and terminate when the local temperature reaches
10 MeV.}
\label{bhgamma-f4}
\end{figure}

\begin{figure}[htb]
\centerline{\epsfig{figure=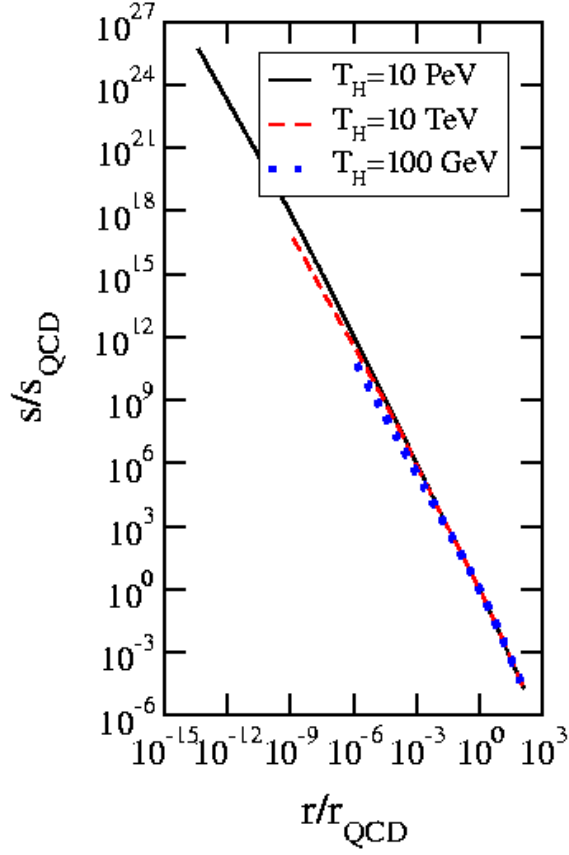,width=14.0cm,angle=270}}
\caption{The radial dependence of $s$ for three different Hawking temperatures.
The curves begin at $r_S/10$ and terminate when the local temperature reaches
10 MeV.}
\label{bhgamma-f5}
\end{figure}

In figure \ref{bhgamma-f2} we plot $u(r_S/10)$ versus $T_H$.  It is essentially constant for 
all $T_H > T_{EW}$ with the value of 0.0415.  In figure \ref{bhgamma-f3} we plot the function
$u(r)$ versus $r$ for three different black hole temperatures.  The radial 
variable $r$ is expressed in units of its value when the temperature first 
reaches $T_{QCD}$, and $u$ is expressed in units of its value at that same 
radius.  This allows us to compare different black hole temperatures.  To
rather 
good accuracy these curves seem to be universal as they essentially lie on top 
of one another.  The curves are terminated when the temperature reaches 10 MeV. The function $u(r)$ behaves like $r^{1/3}$ until temperatures of order 100 MeV 
are reached.  The simple parametrization
\begin{equation}
u(r) = u_S \left( r/r_S \right)^{1/3}
\end{equation}
with $u_S = 0.10$
will be very useful when studying radiation from the surface of the fluid.

In figure \ref{bhgamma-f4} we plot the temperature in units of $T_{QCD}$ versus the radius in 
units of $r_{QCD}$ for the same three black hole temperatures as in figure \ref{bhgamma-f3}.  
Again the curves are terminated when the temperature drops to 10 MeV.  The 
curves almost fall on top of one another but not perfectly.  The temperature 
falls slightly slower than the power-law behavior $r^{-2/3}$ expected on the 
basis of the equation of state $s = (4/3)aT^3$.  The reason is that the 
effective number of degrees of freedom is falling with the temperature.  The 
entropy density is shown in figure \ref{bhgamma-f5}.  It also exhibits an imperfect degree of 
scaling similar to the temperature.

\begin{figure}[htb]
\centerline{\epsfig{figure=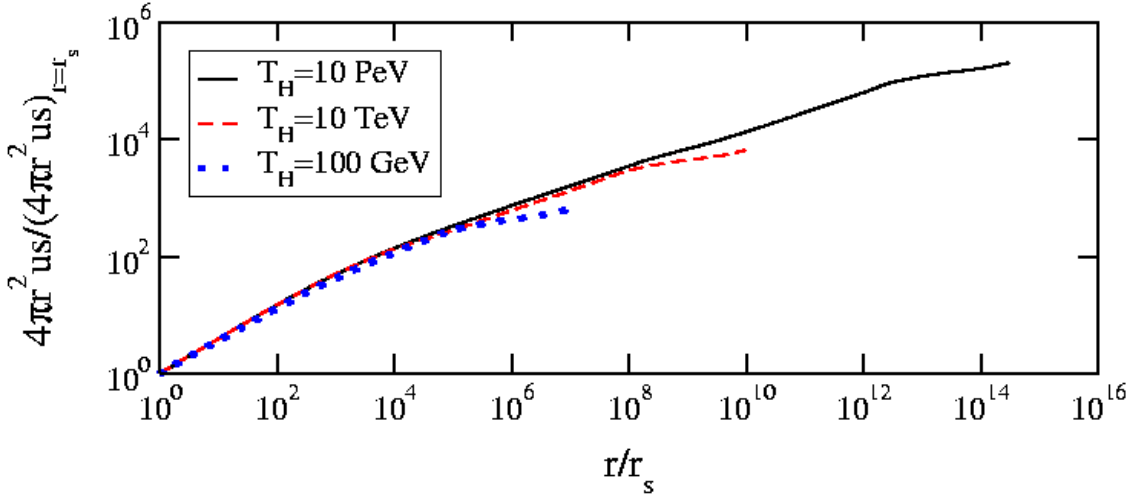,width=13.0cm,angle=270}}
\caption{The radial dependence of the entropy flow for three different Hawking 
temperatures.  The curves begin at $r_S/10$ and terminate when the local 
temperature reaches 10 MeV.}
\label{bhgamma-f6}
\end{figure}

Since viscosity plays such an important role in the outgoing fluid we should 
expect significant entropy production.  In figure \ref{bhgamma-f6} we plot the entropy flow 
$4\pi r^2 u s$ as a function of radius for the same three black hole 
temperatures as in figures \ref{bhgamma-f3}-\ref{bhgamma-f5}.  It increases by several orders of magnitude.  
The fluid flow is far from isentropic.

\section{Onset of Free-Streaming}

\begin{figure}[htb]
\centerline{\epsfig{figure=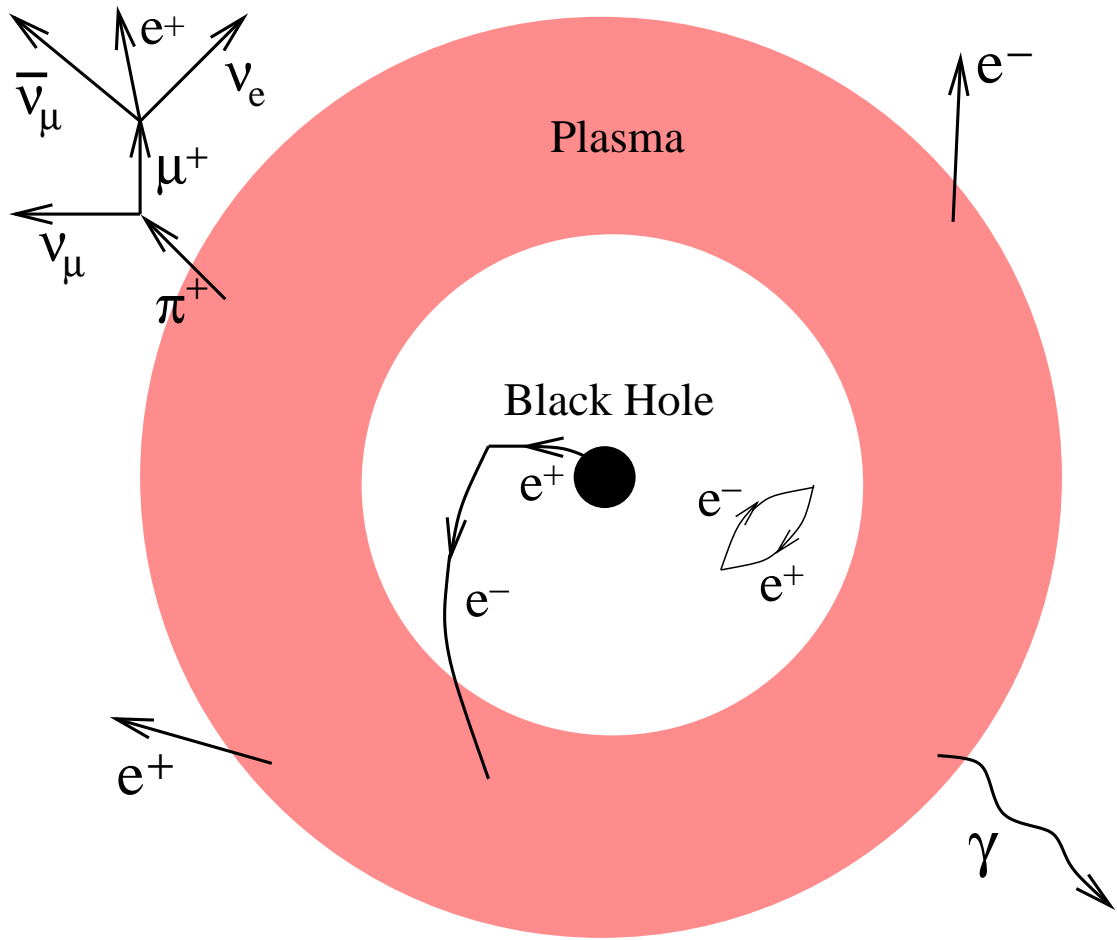,width=14.0cm,angle=270}}
\caption{Free-streaming of particles at freeze-out surface.}
\label{poster1}
\end{figure}

\hspace{0.2in} Eventually the fluid expands so rapidly that the particles composing the fluid 
lose thermal contact with each other and begin free-streaming as is shown in figure \ref{poster1}.  In heavy ion 
physics this is referred to as thermal freeze-out, and in astronomy it is 
usually associated with the photosphere of a star.  
In 
the sections above we argued that thermal contact should occur for all 
particles, with the exception of gravitons and neutrinos, down to temperatures 
on the order of $T_{QCD}$.  Below that temperature the arguments given no
longer 
apply directly; for example, the relevant interactions are not those of 
perturbative QCD.

\begin{figure}[htb]
\centerline{\epsfig{figure=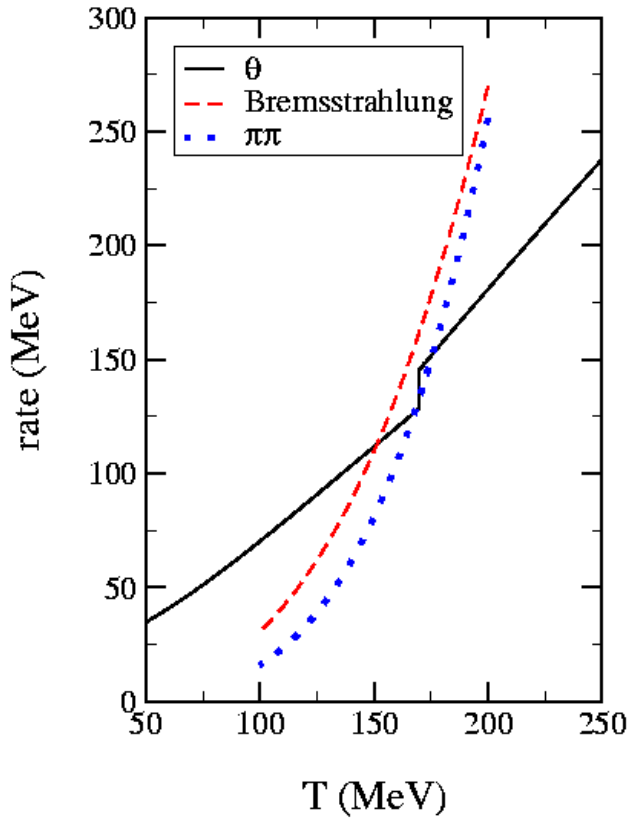,width=13.0cm,angle=270}}
\caption{The rate for $\pi\pi$ scattering and for the bremsstrahlung reaction
$ee \rightarrow ee\gamma$ are compared to the local volume expansion rate.
The Hawking temperature is 10 TeV.}
\label{bhgamma-f7}
\end{figure}

Extensive studies have been made of the interactions among hadrons at finite 
temperature.  Prakash {\it et al.} \cite{Prakash} used experimental information to construct scattering amplitudes for pions, kaons and nucleons and from them 
computed thermal relaxation rates.  The relaxation time for $\pi - \pi$ 
scattering can be read off from their figures and simply parametrized as
\begin{equation}
\tau^{-1}_{\pi \pi} \approx 16 \left( \frac{T}{100 \, {\rm MeV}} \right)^4
\,\, {\rm MeV}
\end{equation}
which is valid for $100 < T < 200$ MeV.  This rate is compared to the volume
expansion rate $\theta$ (see Sec. 2.4) in figure \ref{bhgamma-f7}.  From the figure it is 
clear that pions cannot maintain thermal equilibrium much below 160 MeV or so.  Since pions are the lightest hadrons and therefore the most abundant at low 
temperatures, it seems unlikely that other hadrons could maintain thermal 
equilibrium either.

Heckler has argued vigorously that electrons and photons should continue to 
interact down to temperatures on the order of the electron mass \cite{Ha,Hb}.  
Multi-particle reactions are crucial to this analysis.  Let us see how it 
applies to the present situation.  Consider, for example, the cross section for $ee \rightarrow ee\gamma$.  The energy-averaged cross section is \cite{Hb}
\begin{equation}
\overline{\sigma}_{\rm brem} = 8 \alpha_{EM} r_0^2 \ln(2E/m_e)
\end{equation}
where $m_e$ is the electron mass, $r_0 = \alpha_{EM}/m_e$ is the classical 
electron radius, and $E \gg m_e$ is the energy of the incoming electrons in the
center-of-momentum frame.  (If one computes the rate for a photon produced with the specific energies $0.1E$, $0.25E$, or $0.5E$ the cross section would be 
larger by a factor 4.73, 2.63, or 1.27, respectively.)  The rate using the 
energy-averaged cross section is
\begin{equation}
\tau^{-1}_{\rm brem} \approx \left[ \frac{3}{\pi^2} \xi(3) T^3 \right]
\left[\frac{8\alpha^3_{EM}}{m_e^2} \ln (6T/m_e) \right]
\end{equation}
where we have used the average energy $\langle E \rangle \approx 3T$ for 
electrons with $m_e \ll T$.  This rate is also plotted in figure \ref{bhgamma-f7}.  It is
large 
enough to maintain local thermal equilibrium down to temperatures on the order 
of 140 MeV.  Of course, there are other electromagnetic many-particle reactions which would increase the overall rate.  On the other hand, as pointed out by 
Heckler \cite{Hb}, these reactions are occurring in a high density plasma with 
the consequence that dispersion relations and interactions are renormalized by 
the medium.  If one takes into account only renormalization of the electron 
mass, such that $m_{eff}^2 \approx m_e^2 + e^2 T^2/3$ when $m_e \ll T$, then
the 
rate would be greatly reduced.

Does this mean that photons and electrons are not in thermal equilibrium at the temperatures we have been discussing?  Consider bremsstrahlung reactions in the QCD plasma.  There are many $2 \rightarrow 3$ reactions, such as: $q_1 q_2
\rightarrow q_1 q_2 \gamma$, $q_1 \overline{q}_2 \rightarrow q_1 \overline{q}_2
\gamma$, $gg \rightarrow q \overline{q} \gamma$, and so on.  Here the
subscripts 
label the quark flavor, which may or may not be the same.  The rate for these 
can be estimated using known QED and QCD cross sections 
\cite{Jauch,Haug,Cutler}.  Using an effective quark mass given by $gT$ we find 
that the rate is $\alpha_s T$ with a coefficient of order or larger than unity. Since $\alpha_s$ becomes of order unity near $T_{QCD}$ we conclude that photons are in equilibrium down to temperatures of that order at least.  To make the 
matter even more complicated we must remember that the expansion rate $\theta$ 
is based on a numerical solution of the viscous fluid equations which assume a 
constant proportionality between the shear and bulk viscosities and the entropy density.  Although these proportionalities may be reasonable in QCD and 
electroweak plasmas at high temperatures they may fail at temperatures below 
$T_{QCD}$.  The viscosities should be computed using the relaxation times for 
self-consistency of the transition from viscous fluid flow to free-streaming, 
which we have not done.  For example, the first estimate for the shear
viscosity 
for massless particles with short range interactions is $T^4 \tau$ where $\tau$ is the relaxation time.  For pions we would get $\eta \sim$ const, not
$\eta \sim T^3$.  As another example, we must realize that the bulk viscosity 
can become significant when the particles can be excited internally.  This is, 
in fact, the case for hadrons.  Pions, kaons and nucleons are all the lowest 
mass hadrons each of which sits at the base of a tower of resonances
\cite{PDG}.
See, for example, \cite{Weinberg} and references therein.

\begin{figure}[htb]
\centerline{\epsfig{figure=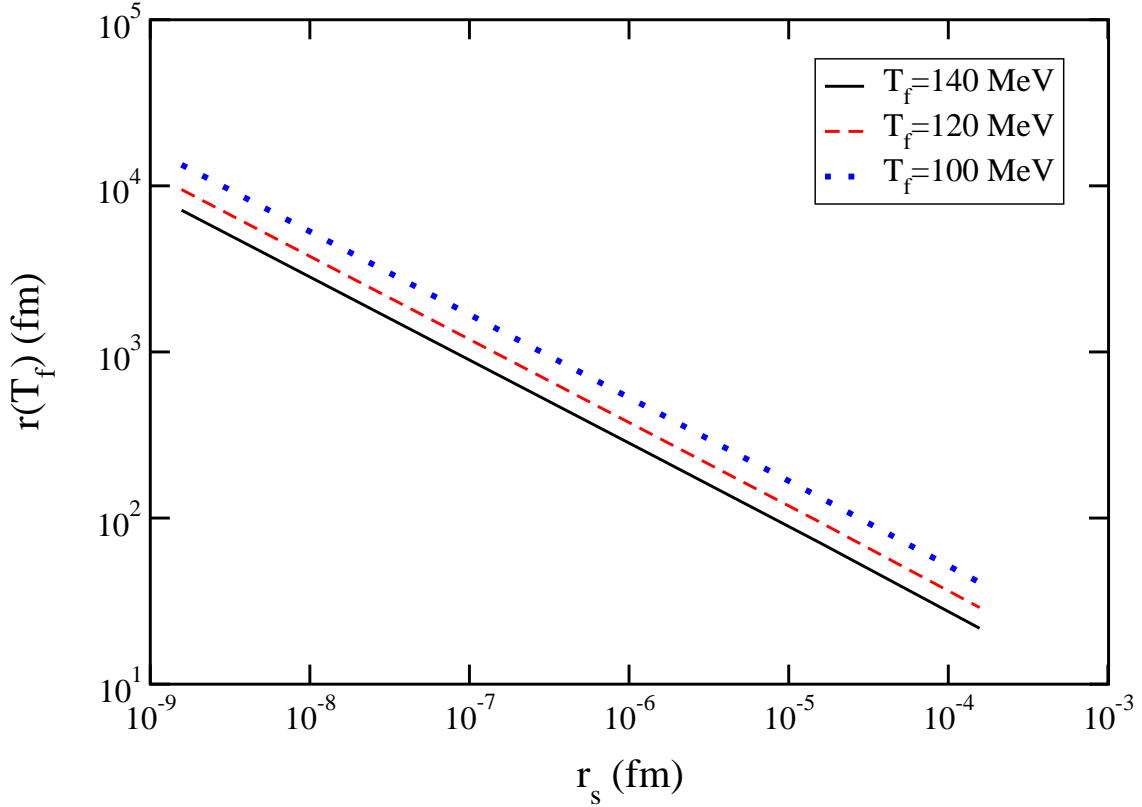,width=13.0cm,angle=270}}
\caption{The freeze-out or free-streaming radius as a function of the
Schwarzschild radius for three different freeze-out temperatures.}
\label{bhgamma-f8}
\end{figure}

In order to do gamma ray phenomenology we need a practical criterion for the 
onset of free-streaming.  We shall assume that this happens suddenly at a 
temperature $T_f$ in the range 100 to 140 MeV.  We shall assume that particles 
whose mass is significantly greater than $T_f$ have all annihilated, leaving 
only photons, electrons, muons and pions.  In figure \ref{bhgamma-f8} we plot the freeze-out 
radius $r_f = r(T_f)$ for $T_f =$ 100, 120 and 140 MeV versus the Schwarzschild radius.  The fact that $r_f$ increases as $T_f$ decreases is an obvious 
consequence of energy conservation.  More interesting is the power-law scaling:
$r_f \sim r_S^{-1/2} \sim T_H^{1/2}$.  This scaling can be understood as 
follows.

The luminosity from the decoupling or freeze-out surface is
\begin{equation}
L_f = 4\pi r_f^2 \left( \frac{2\pi^2}{45} \gamma_f^2 T_f^4 \right) d_f
\end{equation}
where the quantity in parentheses is the surface flux for one massless bosonic 
degree of freedom and $d_f$ is the total number of effective massless bosonic 
degrees of freedom.  For the particles listed above we have $d_f = 12$.  By 
energy conservation this is to be equated with the Hawking formula for the
black 
hole luminosity,
\begin{equation}
L_h = 64\pi^2 \alpha_h^{eff} T_H^2 \, ,
\end{equation}
where $\alpha_h^{eff}$ does not include the contribution from gravitons and 
neutrinos.  Together with the scaling function for the flow velocity, Eq. (2.20), we can solve for the radius
\begin{equation}
r_f = \frac{2}{\pi} \left( \frac{45 \pi \alpha_h^{eff}}{2 u_S^2 d_f} 
\right)^{3/8} \sqrt{\frac{T_H}{T_f^3}}
\end{equation}
and for the boost
\begin{equation}
\gamma_f T_f = 2u_S \left( \frac{45 \pi \alpha_h^{eff}}{2 u_S^2 d_f} 
\right)^{1/8} \sqrt{T_f T_H} \approx 0.22 \sqrt{T_f T_H} \, .
\end{equation}
From these we see that the final radius does indeed scale like the inverse of 
the square-root of the Schwarzschild radius or like the square-root of the
black 
hole temperature, and that the average particle energy (proportional to
$\gamma_f T_f$) scales like the square-root of the black hole temperature.  One important observational effect is that the average energy of the outgoing 
particles is reduced but their number is increased compared to direct Hawking 
emission into vacuum \cite{Ha,Hb}.

\section{Photon Emission}

\hspace{0.2in} Photons observed far away from the black hole primarily come from 
one of two sources.  Either they are emitted directly in the form of a boosted 
black-body spectrum, or they arise from neutral pion decay.  We will consider 
each of these in turn.

\subsection{Direct photons}

\hspace{0.2in} Photons emitted directly have a Planck distribution in the local rest frame
of the fluid.  The phase space density is
\begin{equation}
f(E') = \frac{1}{{\rm e}^{E'/T_f} -1} \, .
\end{equation}
The energy appearing here is related to the energy as measured in the rest
frame 
of the black hole and to the angle of emission relative to the radial vector by
\begin{equation}
E' = \gamma_f (1-v_f \cos\theta) E \, .
\end{equation}
No photons will emerge if the angle is greater than $\pi/2$.  Therefore the 
instantaneous distribution is
\begin{eqnarray}
\frac{d^2N_{\gamma}^{\rm dir}}{dE dt} &=& 
4\pi r_f^2 \left(\frac{E^2}{2\pi^2}\right)
\int_0^1 d(\cos\theta) \cos\theta f(E,\cos\theta) \nonumber \\ & \approx &
- \frac{2 r_f^2 T_f E}{\pi \gamma_f} \ln \left(1 - 
{\rm e}^{-E/2 \gamma_f T_f} \right)
\end{eqnarray}
where the second equality holds in the limit $\gamma_f \gg 1$.  This limit is 
actually well satisfied for us and is used henceforth.

The instantaneous spectrum can be integrated over the remaining lifetime of the black hole straightforwardly.  The radius and boost are both known in terms of 
the Hawking temperature $T_H$, and the time evolution of the latter is simply 
obtained from solving Eq. (2.3).  For a black hole that disappears at time $t = 0$ we have
\begin{equation}
T_H(t) = - \frac{1}{8\pi} \left( \frac{m_{\rm P}^2}{3 \alpha_h t} \right)^{1/3} \, .
\end{equation}
Here $\alpha_h$ is approximately 0.0044 for 
$T_H > 100$ GeV which includes the contribution from gravitons and neutrinos.
Starting with a black hole whose temperature is $T_0$ we obtain the spectrum
\begin{equation}
\frac{dN_{\gamma}^{\rm dir}}{dE} = \frac{360 u_S^2}{\pi^5 d_f}
\left( \frac{45 \pi \alpha_h^{eff}}{2 u_S^2 d_f} \right)^{1/4}
\frac{m_{\rm P}^2 T_f}{E^4}
\sum_{n=1}^{\infty} \frac{1}{n} \int_0^{E/2 \gamma_f(T_0) T_f}
dx \, x^4 \, {\rm e}^{-nx} \, .
\end{equation}
Here we have ignored the small numerical difference between $\alpha_h^{eff}$
and $\alpha_h$. In the high energy limit, namely, when $E \gg  2\gamma_f(T_0) 
T_f$, the summation yields the pure number $4(2\pi^6/315)$.  Note the power-law behavior $E^{-4}$.  This has important observational consequences.

\subsection{$\pi^0$ decay photons}

\hspace{0.2in} The neutral pion decays almost entirely into two photons: $\pi^0
\rightarrow \gamma \gamma$.  In the rest frame of the pion the single photon 
Lorentz invariant distribution is
\begin{equation}
E \frac{d^3N_{\gamma}}{d^3p} = \frac{\delta(E-m_{\pi}/2)}{\pi m_{\pi}}
\end{equation}
which is normalized to 2.  This must be folded with the distribution of $\pi^0$ to obtain the total invariant photon distribution
\begin{equation}
E \frac{d^4N_{\gamma}}{d^3p dt} = \int_{m_{\pi}}^{\infty} dE_{\pi}
\frac{d^2N_{\pi^0}}{dE_{\pi}dt} \frac{1}{\pi m_{\pi}}
\delta \left( \frac{ EE_{\pi} - {\bf p} \cdot {\bf p}_{\pi} }
{m_{\pi}} - \frac{m_{\pi}}{2} \right).
\end{equation}
After integrating over angles we get
\begin{equation}
\frac{d^2N_{\gamma}^{\pi^0}}{dE dt} = 2 \int_{E_{\rm min}}^{\infty}
\frac{dE_{\pi}}{p_{\pi}} \frac{d^2N_{\pi^0}}{dE_{\pi}dt}
\end{equation}
where $E_{\rm min} = (E^2 + m_{\pi}^2/4)/E$.  In the limit $E \gg m_{\pi}$ we 
can approximate $E_{\rm min} = E$ and evaluate $d^2N_{\pi^0}/dE_{\pi}dt$ in 
the same way as photons.  This leads to the relatively simple expression
\begin{equation}
\frac{d^2N_{\gamma}^{\pi^0}}{dE dt} = \frac{4 r_f^2 T_f^2}{\pi} 
\sum_{n=1}^{\infty} \frac{1}{n^2} {\rm e}^{-nE/2\gamma_f T_f} \, .
\end{equation}
The time-integrated spectrum is computed in the same way as for direct photons
\begin{equation}
\frac{dN_{\gamma}^{\pi^0}}{dE} = \frac{360 u_S^2}{\pi^5 d_f}
\left( \frac{45 \pi \alpha_h^{eff}}{2 u_S^2 d_f} \right)^{1/4}
\frac{m_{\rm P}^2 T_f}{E^4}
\sum_{n=1}^{\infty} \frac{1}{n^2} \int_0^{E/2 \gamma_f(T_0) T_f}
dx \, x^3 \, {\rm e}^{-nx}.
\end{equation}
In the high energy limit, namely, when $E \gg  2\gamma_f(T_0) T_f$, the 
summation yields the pure number $2\pi^6/315$. 
\begin{figure}[htb]
\centerline{\epsfig{figure=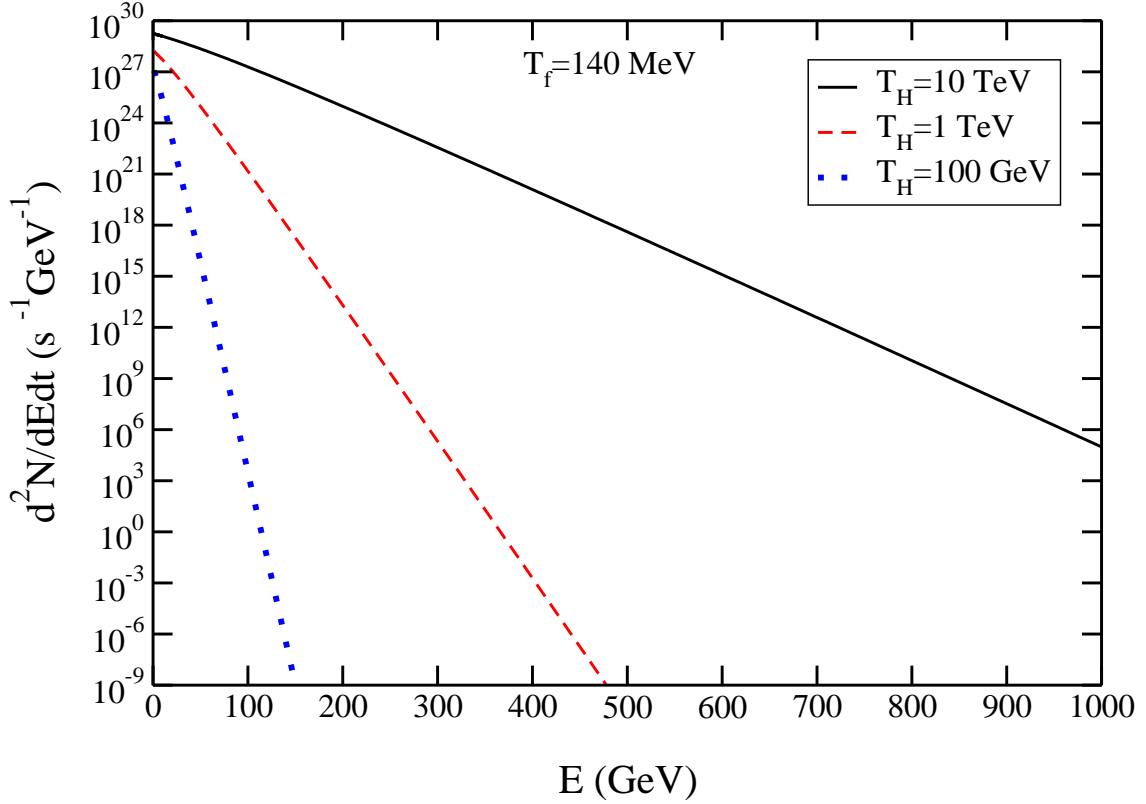,width=13.0cm,angle=270}}
\caption{The instantaneous gamma ray spectrum for three different
Hawking temperatures assuming $T_f = 140$ MeV.}
\label{bhgamma-f9}
\end{figure}

\begin{figure}[htb]
\centerline{\epsfig{figure=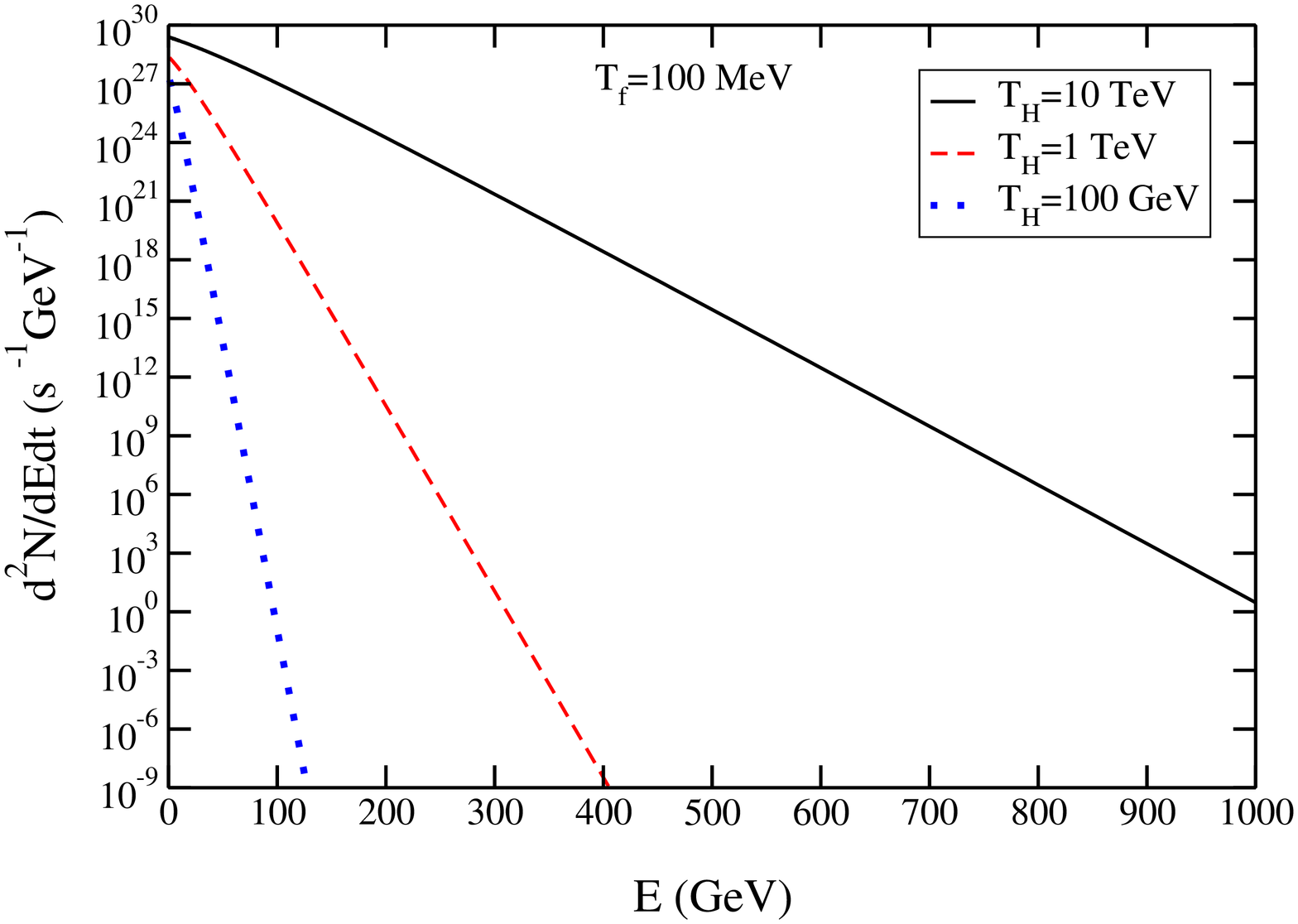,width=13.0cm,angle=270}}
\caption{Same as figure 9 but with $T_f = 100$ MeV.}
\label{bhgamma-f10}
\end{figure}  

\subsection{Instantaneous and integrated photon spectra}

\hspace{0.2in} The instantaneous spectra of high energy gamma rays, arising from both direct 
emission and from $\pi^0$ decay, are plotted in figures \ref{bhgamma-f9} (for $T_f = 140$ MeV) and  \ref{bhgamma-f10}(for $T_f = 100$ MeV).  In each figure there are three curves 
corresponding to Hawking temperatures of 100 GeV, 1 TeV and 10 TeV.  
The photon spectra are essentially exponential above a few GeV with inverse 
slope $2 \gamma_f(T_H) T_f \propto \sqrt{T_f T_H}$.  If these instantaneous 
spectra could be measured they would tell us a lot about the equation of state, the viscosities, and how energy is processed from first Hawking radiation to 
final observed gamma rays.  Even the time evolution of the black hole
luminosity 
and temperature could be inferred.

\begin{figure}[htb]
\centerline{\epsfig{figure=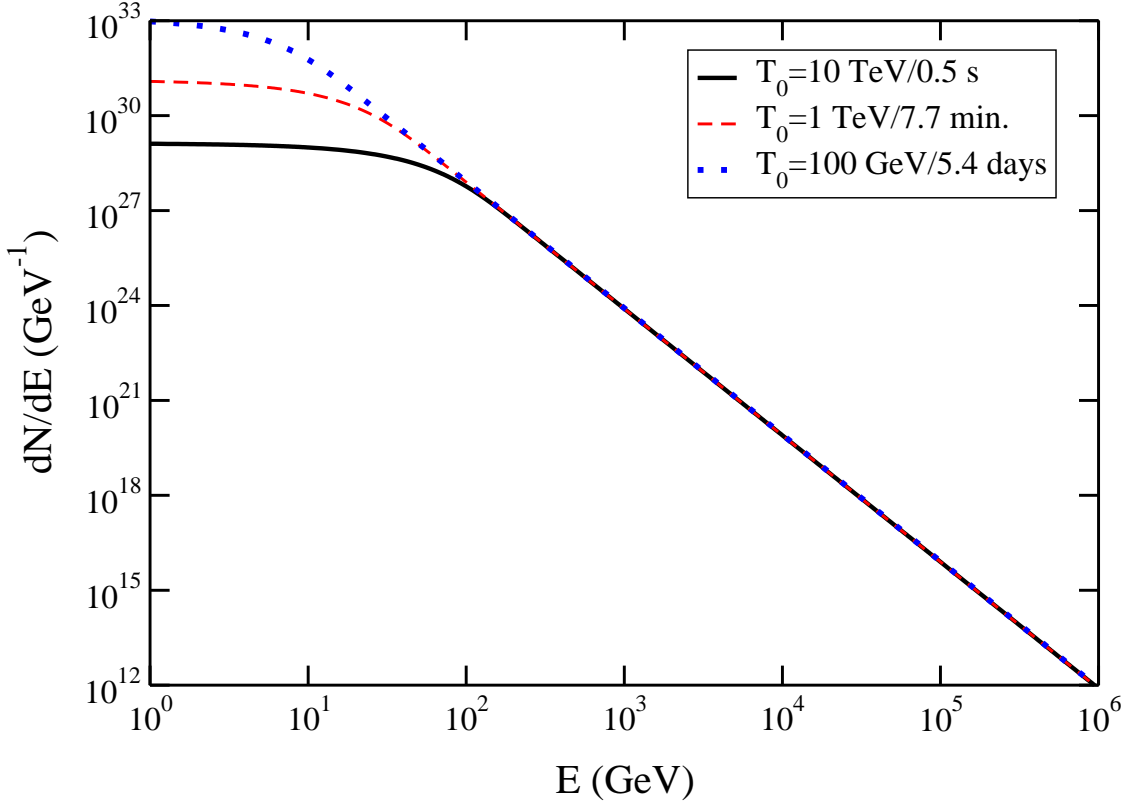,width=13.0cm,angle=270}}
\caption{The time-integrated gamma ray spectrum starting from the
indicated Hawking temperature.  Here $T_f = 140$ MeV.}
\label{bhgamma-f11}
\end{figure}

The time integrated spectra for $T_f = 140$ MeV are plotted in figure \ref{bhgamma-f11} for
three initial temperatures $T_0$.  A black hole with a Hawking temperature of 
100 GeV has 5.4 days to live, a black hole with a Hawking temperature of 1 TeV 
has 7.7 minutes to live, and a black hole with a Hawking temperature of 10 TeV 
has only 1/2 second to live.  The high energy gamma ray spectra are represented by
\begin{equation}
\frac{dN}{dE} = \frac{m_{\rm P}^2 T_f}{26 E^4} \, .
\end{equation}
It is interesting that the contribution from $\pi^0$ decay comprises 20\% of
the 
total while direct photons contribute the remaining 80\%.  The $E^{-4}$
fall-off 
is the same as that obtained by Heckler \cite{Ha}, whereas Halzen {\it et al.} 
\cite{Hal} and MacGibbon and Carr \cite{MC} obtained an $E^{-3}$ fall-off on
the 
basis of direct fragmentation of quarks and gluons with no fluid flow and no 
photosphere.

\section{Observability of Gamma Rays}

\hspace{0.2in} The most obvious way to observe the explosion of a microscopic black hole is by high energy gamma rays.  We consider their contribution to the diffuse gamma
ray 
spectrum in Sec. 2.7.1, and in Sec. 2.7.2 we study the systematics of a single identifiable explosion.

\subsection{Diffuse spectra from the galactic halo}

\hspace{0.2in} Suppose that microscopic black holes were distributed about our galaxy in some 
fashion.  Unless we were fortunate enough to be close to one so that we could 
observe its demise, we would have to rely on their contribution to the diffuse 
background spectrum of high energy gamma rays.

The flux of photons with energy greater than 1 GeV at Earth can be computed
from 
the results of Sec. 2.6 together with the knowledge of the rate density 
$\dot{\rho}({\bf x})$ of exploding black holes.  It is
\begin{equation}
\frac{d^3N_{\rm Earth}}{dE dA dt} = \frac{m_{\rm P}^2 T_f}{26 E^4}
\int d^3x \frac{\dot{\rho}({\bf x})}{4\pi d^2({\bf x})}
{\rm e}^{-d({\bf x})/\lambda_{\gamma \gamma}(E)}
\end{equation}
where $d({\bf x})$ is the distance from the black hole to the Earth.  The 
exponential decay is due to absorption of the gamma ray by the black-body 
radiation \cite{Gould}.  The mean free path $\lambda_{\gamma \gamma}(E)$ is 
highly energy dependent.  It has a minimum of about 1 kpc around 1 PeV, and is 
greater than $10^5$ kpc for energies less than 100 TeV.

We need a model for the rate density of exploding black holes.  We shall assume they are distributed in the same way as the matter comprising the halo of 
our galaxy.  Thus we take
\begin{equation}
\dot{\rho}({\bf x}) = \dot{\rho}_0 \,
\frac{R_c^2}{x^2+y^2 +q^2 z^2 + R_c^2}
\end{equation}
where the galactic plane is the $x-y$ plane, $R_c$ is the core radius, and $q$ 
is a flattening parameter.  For numerical calculations we shall take the core 
radius to be 10 kpc.  The Earth is located a distance $R_E = 8.5$ kpc from the 
center of the galaxy and lies in the galactic plane.  Therefore
$d^2 = (x-R_E)^2 + y^2 + z^2$.

\begin{figure}[htb]
\centerline{\epsfig{figure=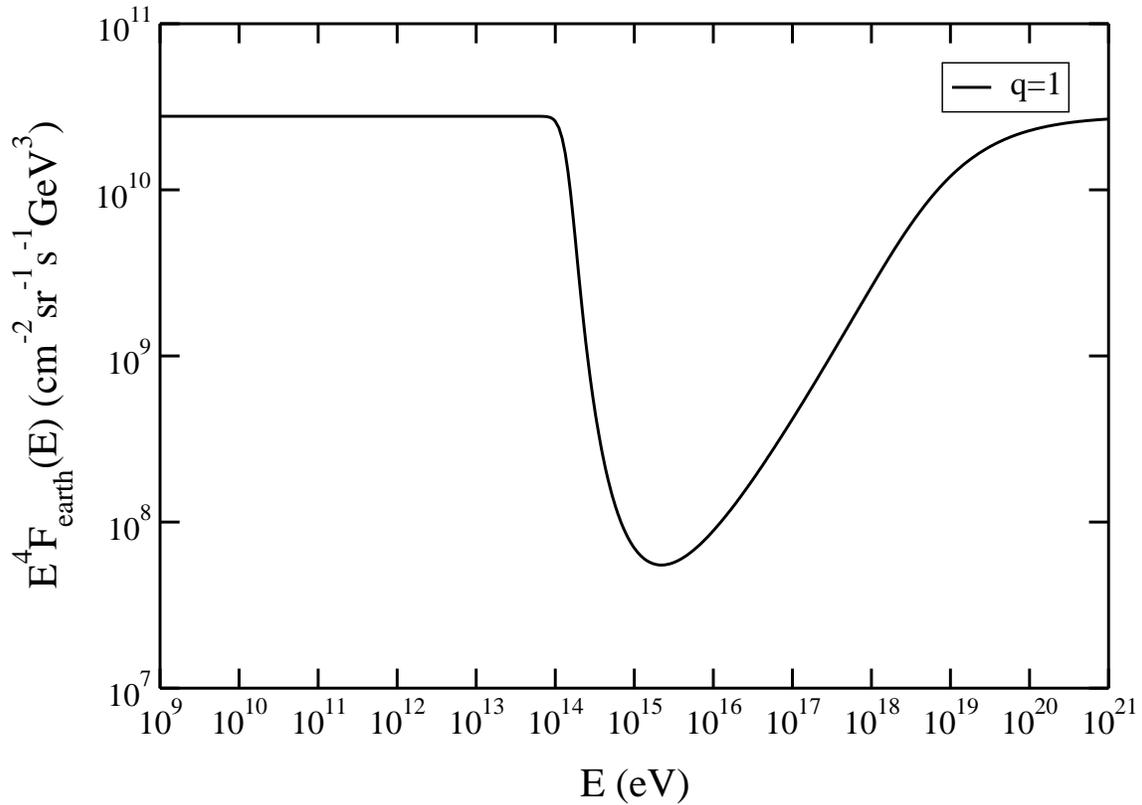,width=13.0cm,angle=270}}
\caption{The flux of diffuse gamma rays coming from our galactic halo.
The normalization is $\dot{\rho}_0 = 1$ pc$^{-3}$ yr$^{-1}$.
The halo is assumed to be spherically symmetric, $q = 1$; the results
for a flattened halo with $q = 2$ are very similar.}
\label{bhgamma-f12}
\end{figure}

The last remaining quantity is the normalization of the rate density 
$\dot{\rho}_0$.  This is, of course, unknown since no one has ever knowingly 
observed a black hole explosion.  The first observational limit was determined 
by Page and Hawking \cite{PH}.  They found that the local rate density
$\dot{\rho}_{\rm local}$ is less than 1 to 10 per cubic parsec per year on the 
basis of diffuse gamma rays with energies on the order of 100 MeV.  This limit 
has not been lowered very much during the intervening twenty-five years.  For 
example, Wright \cite{W} used EGRET data to search for an anisotropic
high-lattitude component of diffuse gamma rays in the energy range from 30 MeV 
to 100 GeV as a signal for steady emission of microscopic black holes.  He 
concluded that $\dot{\rho}_{\rm local}$ is less than about 0.4 per cubic parsec per year.  (For an alternative point of view on the data see \cite{DBCline}.)
In our numerical calculations we shall assume a value
$\dot{\rho}_0 = 1$ pc$^{-3}$ yr$^{-1}$ corresponding to
$\dot{\rho}_{\rm local} \approx 0.58$ pc$^{-3}$ yr$^{-1}$.  This makes for easy scaling.  Estimating the quantity of dark matter in our galaxy as
$M_{\rm halo}/\rho_{0, \, {\rm halo}} = 4.7\times 10^4$ kpc$^3$ means that we 
could have up to $47\times 10^{12}$ microscopic black hole explosions per year 
in our galaxy.

Figure \ref{bhgamma-f12} shows the calculated flux at Earth, multiplied by $E^4$.  Of course 
this curve would be flat if it were not for absorption on the microwave 
background radiation.  There is a relative suppression of three orders of 
magnitude centered between $10^{15}$ and $10^{16}$ eV.  This means that it is 
unlikely to observe exploding black holes in the gamma ray spectrum above 
$10^{14}$ eV.  Even below that energy it is unlikely because they have not been observed at energies on the order of 100 MeV, and the spectrum falls 
faster than the primary cosmic ray spectrum $\propto E^{2.7}$.  The curve 
displayed in figure \ref{bhgamma-f12} assumes a spherical halo, $q = 1$, but there is hardly 
any difference when the halo is flattened to $q = 2$.

\subsection{Point source systematics}

\hspace{0.2in} Given the unfavorable situation for observing the effects of exploding 
microscopic black holes on the diffuse gamma ray spectrum, we now turn to the 
consequences for observing one directly. How far away could one be seen?  Let
us 
call that distance $d_{\rm max}$.  We assume that
$d_{\rm max} < \lambda_{\gamma\gamma}$ for simplicity, although that assumption can be relaxed if necessary.  Let $A_{\rm det}$ denote the effective area of
the 
detector that can measure gamma rays with energies equal to or greater than 
$E_{\rm min}$.  The average number of gamma rays detected from a single 
explosion a distance $d_{\rm max}$ away is
\begin{equation}
\langle N_{\gamma}(E > E_{\rm min}) \rangle =
\frac{ A_{\rm det} }{4\pi d_{\rm max}^2} \int_{E_{\rm min}}^{\infty}
\frac{d N_{\gamma}}{dE} dE = \frac{ A_{\rm det} }{4\pi d_{\rm max}^2}
\frac{m_{\rm P}^2 T_f}{78 E_{\rm min}^3} \, .
\end{equation}
Obviously we should have $E_{\rm min}$ as small as possible to get the largest
number, but it cannot be so small that the simple $E^{-4}$ behavior of the
emission spectrum is invalid.  See figure \ref{bhgamma-f11}.

A rough approximation to the number distribution of detected gamma rays is a 
Poisson distribution.
\begin{equation}
P(N_{\gamma}) = \frac{\langle N_{\gamma} \rangle ^{N_{\gamma}}}
{N_{\gamma} !} {\rm e}^{-\langle N_{\gamma} \rangle }
\end{equation}
The exact form of the number distribution is not so important.  What is 
important is that when $\langle N_{\gamma}(E > E_{\rm min}) \rangle > 1$ we 
should expect to see multiple gamma rays coming from the same point in 
the sky.  Labeling these gamma rays according to the order in which they
arrive, 
1, 2, 3, etc. we would expect their energies to increase with time:
$E_1 < E_2 < E_3 < \cdot \cdot \cdot$ .  Such an observation would be remarkable, possibly 
unique, because astrophysical sources normally cool at late times.  This would 
directly reflect the increasing Hawking temperature as the black hole explodes 
and disappears.

\begin{figure}[htb]
\centerline{\epsfig{figure=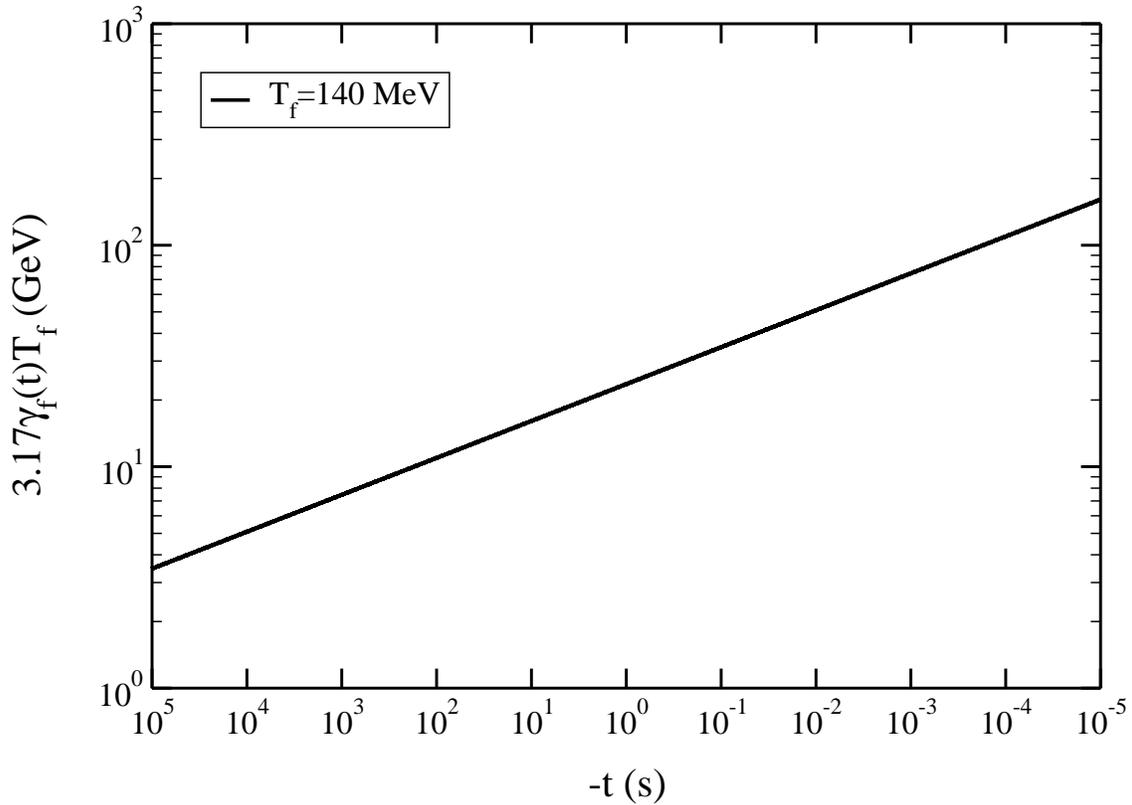,width=13.0cm,angle=270}}
\caption{The average gamma ray energy as a function of the remaining
lifetime of the black hole.  The times spanned correspond to approximately
400 GeV $< T_H <$ 200 TeV.}
\label{bhgamma-f13}
\end{figure}

It is interesting to know how the average gamma ray energy increases with time.
Using Eqs. (2.30) and (2.36) we compute the average energy of direct photons to be
$4 \gamma_f T_f \zeta(4)/\zeta(3)$ and the average energy of $\pi^0$ decay 
photons to be one-half that.  The ratio of direct to decay photons turns out
to be $\pi$.  Therefore the average gamma ray energy is $3.17 \gamma_f(t) T_f$.
This average is plotted in figure \ref{bhgamma-f13} for $10^5 > t > 10^{-5}$ seconds.  The 
average gamma ray energy ranges from about 4 to 160 GeV.

The maximum distance can now be computed.  Using some characteristic numbers we find
\begin{equation}
d_{\rm max} \approx 150 \sqrt{\frac{A_{\rm det}}{1 \, {\rm km}^2}}
\left( \frac{10 \, {\rm GeV}}{E_{\rm min}}\right)^{3/2} \, \, {\rm pc} \, .
\end{equation}
If we take the local rate density of explosions to be 0.4 pc$^{-3}$ yr$^{-1}$
then within 150 pc of Earth there would be $5\times 10^6$ explosions per year. 
These would be distributed isotropically in the sky.  Still, it suggests that 
the direct observation of exploding black holes is feasible if they are near to the inferred upper limit to their abundance in our neighborhood.  We should 
point out that a search for 1 s bursts of ultrahigh energy gamma rays from
point 
sources by CYGNUS has placed an upper limit of $8.5 \times 10^5$ pc$^{-3}$
yr$^{-1}$ \cite{CYGNUS}.  However, as we have seen in figure \ref{bhgamma-f13} and
elsewhere, this is not what should
be expected if our calculations bear any resemblance to reality.
Rather than a burst, the luminosity and average gamma ray energy increase
monotonically over a long period of time.

\section{Neutrino Emission}

\hspace{0.2in} In this section we focus on high energy 
neutrino emission from black holes with Hawking temperatures greater than 100 
GeV and corresponding masses less than 10$^8$ kg.  It is at these and higher 
temperatures that new physics will arise.  Such a study is especially important 
in the context of high energy neutrino detectors under construction or planned 
for the future.  Previous notable studies in this area have been carried out by 
MacGibbon and Webber \cite{MW,nu4} and by Halzen, Keszthelyi and Zas \cite{nu5}, 
who calculated the instantaneous and time-integrated spectra of neutrinos 
arising from the decay of quark and gluon jets. 

The source of neutrinos in the viscous fluid picture is quite varied.  Neutrinos 
should stay in thermal equilibrium, along with all other elementary particles, 
when the local temperature is above 100 GeV.  The reason is that 
at energies above the electroweak scale of 100 GeV neutrinos should have 
interaction cross sections similar to those of all other particles.  Thus the 
neutrino-sphere, where the neutrinos decouple, ought to exist where the local 
temperature falls below 100 GeV.  
The spectra of these direct neutrinos are calculated in Sec. 2.8.1.

Neutrinos also come from decays
involving pions and muons.  The relevant processes are (i) a thermal pion decays 
into a muon and muon-neutrino, followed by the muon decay $\mu \rightarrow e 
\nu_e \nu_{\mu}$, and (ii) a thermal muon decays in the same way. 
The spectra of neutrinos arising from pion decay are calculated in Sec. 2.8.2 while those arising from direct or indirect muon decay are calculated in Sec. 2.8.3. 

The spectra from all of these sources are compared graphically in Sec. 2.9.  
We also compare with the spectra of neutrinos emitted directly as Hawking 
radiation without any subsequent interactions.  The main result is that the 
time-integrated direct Hawking spectrum falls at high energy as $E^{-3}$ whereas 
the time-integrated neutrino spectrum coming from a fluid or from pion and muon 
decays all fall as $E^{-4}$.  Thus the fluid picture predicts more neutrinos at 
lower energies than the direct Hawking emission picture.  If a microscopic black 
hole is near enough the instantaneous spectrum could be measured, and its shape 
and magnitude would provide information on the number of degrees of freedom in 
the nature on mass scales exceeding 100 GeV.

\subsection{Directly Emitted Neutrinos}

In this section we first review the emission of neutrinos by the Hawking 
mechanism unmodified by any rescattering.  Then we estimate the spectra of 
neutrinos which rescatter in the hot matter.  The last scattering surface should 
be represented approximately by that radius where the temperature has dropped to 
100 GeV.  The reason is that neutrinos with energies much higher than that have 
elastic and inelastic cross sections that are comparable to the cross sections 
of quarks, gluons, electrons, muons, and tau leptons.  Much below that energy 
the relevant cross sections are greatly suppressed by the mass of the exchanged 
vector bosons, the W and Z.   Furthermore, the electroweak symmetry is broken 
below temperatures of this order, making it natural to place the last scattering 
surface there.  A much better treatment would require the solution of transport 
equations for the neutrinos, an effort that is perhaps not yet justified.  All 
the formulas in this section refer to one flavor of neutrino.  Our current 
understanding is that there are three flavors, each available as a particle or 
antiparticle.  The sum total of all neutrinos would then be a factor of 6 larger 
than the formulas presented here.  

\subsubsection{Direct neutrinos}

The emission of neutrinos by the Hawking mechanism is usually calculated on the 
basis of detailed balance.  It involves a thermal flux of neutrinos incident on 
a black hole.  The Dirac equation is solved and the absorption coefficient is 
computed.  This involves numerical calculations \cite{Page, Sanchez, nu2}.
The number emitted per unit time per unit energy is given by
\begin{eqnarray}
\frac{d^2N_{\nu}^{\rm dir}}{dEdt} = \frac{\Gamma_{\nu}}{2\pi} \,
\frac{1}{\exp(E/T_H)+1} \,,
\end{eqnarray}
where $\Gamma_{\nu}$ is an energy-dependent absorption coefficient.  There is no 
simple analytic formula for it. For our purposes it is sufficient to parametrize 
the numerical results.  A fair representation is given by
\begin{eqnarray}
\Gamma_{\nu} = \frac{27}{64\pi^2} \frac {E^2}{T_H^2} \left( 0.075+\frac{0.925}
{\exp(5-aE/T_H)+1} \right),
\end{eqnarray}
and $a \approx 1.607$.  The exact expression has very small amplitude 
oscillations arising from the essentially black disk character of the black 
hole.  The parametrized form does not have these oscillations, but otherwise is 
accurate to within about 5\%.

If there are no new degrees of freedom present in nature, other than those 
already known, then the time dependence of the black hole mass and temperature 
are easily found.  The relationship between the time and the temperature [Eq. (2.31)] allows 
us to compute the time-integrated spectrum, starting from the moment when $T_H = 
T_0$.  There is a one-dimensional integral to be done numerically.
\begin{equation}
\frac{dN_{\nu}^{\rm dir}}{dE}=\frac{27m_{\rm P}^2}{16(4\pi)^6\alpha_h E^3}
\int_{0}^{E/T_0}dx\frac{x^4}{\exp(x)+1}\left( 0.075+\frac{0.925}
{\exp(5-ax)+1} \right)
\end{equation}
In the high energy limit, meaning $E \gg T_0$, the upper limit can be taken to 
infinity with the result
\begin{equation}
\frac{dN_{\nu}^{\rm dir}}{dE} \rightarrow \frac{29.9m_{\rm P}^2}{(4\pi)^6\alpha_h E^3} 
\approx \frac{m_{\rm P}^2}{565 E^3}\, .
\end{equation}

\subsubsection{Direct neutrinos from an expanding fluid}

Neutrinos emitted from the decoupling surface have a Fermi distribution in the 
local rest frame of the fluid.  The phase space density is
\begin{equation}
f(E') = \frac{1}{{\rm e}^{E'/T_{\nu}} +1} \, .
\end{equation}
The decoupling temperature of neutrinos is denoted by $T_{\nu}$.  The energy 
appearing here is related to the energy as measured in the rest frame 
of the black hole and to the angle of emission relative to the radial vector by
\begin{equation}
E' = \gamma_{\nu} (1-v_{\nu} \cos\theta) E \, .
\end{equation}
No neutrinos will emerge if the angle is greater than $\pi/2$.  Therefore the 
instantaneous distribution is
\begin{displaymath}
\frac{d^2N_{\nu}^{\rm fluid}}{dE dt} = 
4\pi r_{\nu}^2 \left(\frac{E^2}{2\pi^2}\right)
\int_0^1 d(\cos\theta) \cos\theta f(E,\cos\theta) =
\frac{r^2_{\nu}T_{\nu}}{\pi u_{\nu}} E 
\sum_{n=1}^{\infty}\frac{(-1)^{n+1}}{n}
\end{displaymath}
\begin{equation} 
\left\{ \left( 1 - \frac{T_{\nu}}{n u_{\nu}E} \right)
\exp[-nE (\gamma_{\nu}-u_{\nu})/T_{\nu}]
+ \frac{T_{\nu}}{n u_{\nu}E} \exp[-nE\gamma_{\nu}/T_{\nu}]
\right\} \, ,
\end{equation}
where $r_{\nu}$ is the radius of the decoupling suface and $u = v \gamma$.
Integration over the energy gives the luminosity (per neutrino).

We need to know how the radius and radial flow velocity at neutrino decoupling 
depend on the Hawking temperature or, equivalently, the black hole mass; we have 
already argued that the neutrino temperature at decoupling is about $T_{\nu} = 
100$ GeV.  We also know the relation between $r_{\nu}$ and $u_{\nu}$ from Eq. (2.20):
\begin{eqnarray}
r_{\nu}= \frac{1}{4\pi T_H} \left( \frac{u_{\nu}}{u_S} \right)^3 \, .
\end{eqnarray}
The final piece of information is to recognize that each type of neutrino will 
contribute 7/8 effective bosonic degree of freedom to the total number of 
effective bosonic degrees of freedom in the energy density of the fluid.  After 
integrating the instantaneous neutrino energy spectrum to obtain its luminosity 
we equate it with the appropriate fraction of the total luminosity. 
\begin{equation}
L_{\nu}^{\rm fluid}=\frac{7}{8} 
\frac{\pi^3 r_{\nu}^2 T_{\nu}^4}{90} 
\frac{3-v_{\nu}}{(1-v_{\nu})^3 \gamma_{\nu}^4}
= \frac{7/8}{106.75} L^{\rm Hawking}
\end{equation}
where
\begin{equation}
L^{\rm Hawking}= 64\pi^2\alpha_h^{eff} T_H^2 \, .
\end{equation}
The number 106.75 counts all effective bosonic degrees of freedom in the 
standard model excepting gravitons.  Here $\alpha_h^{eff}$ does not include the contribution from gravitons.  This results in an equation which determines 
$v$, equivalently $u$, in terms of the Hawking temperature:
\begin{equation}
\frac{(3-v_{\nu})v_{\nu}^4u_{\nu}^2}{(1-v_{\nu})^3} =
2^{13} \frac{45 \pi}{427} \alpha_h^{eff}
u_S^6 \left( \frac{T_H}{T_{\nu}}\right)^4 =
C \left( \frac{T_H}{T_{\nu}}\right)^4.
\end{equation}
Numerically the constant $C = 1.14\times 10^{-5}$.

We are interested in black holes with $T_H > T_{\nu} = 100$ GeV.  The 
corresponding range of neutrino-sphere flow velocities corresponds to
$0.2 < u_{\nu} < \infty$.  There is no simple analytical expression for the 
solution to Eq. (2.54) for this wide range of the variable.  At asymptotically 
high temperatures the left side has the limit $16 u_{\nu}^8$.  For intermediate 
values the left side may be approximated by $22 u_{\nu}^7$.  We approximate the 
left side by the former for $u_{\nu} > 11/8$ and by the latter for $0.2 < 
u_{\nu} < 11/8$.  Thus       
\begin{eqnarray}
u_{\nu}&=&\left( \frac {CT_H^4}{22T_{\nu}^4}\right)^{1/7} \;\;\; {\rm for}
\;\; u_{\nu} < 11/8 \\
u_{\nu}&=&\left( \frac {CT_H^4}{16T_{\nu}^4}\right)^{1/8} \;\;\; {\rm for}
\;\; u_{\nu} > 11/8 \, .
\end{eqnarray}
This approximation is valid to better than 20\% within the range mentioned.

The time-integrated spectrum can be calculated on the basis of Eqs. (2.50), (2.31), 
(2.51) and either (2.54), which is exact, or with the approximation of (2.55-56) which results in 
\begin{eqnarray}
\lefteqn{\frac{dN_{\nu}^{\rm fluid}}{dE} \;\;=\;\; 
\frac{m_{\rm P}^2T_{\nu}E}{2^{13}\pi^6\alpha_hu_s^6}
\sum_{n=1}^{\infty}(-1)^{n+1}} \nonumber\\ 
&&\times  \Biggl\{  
\left(\frac{C}{22T_{\nu}^4}\right)^{5/7}
\int_{T_0}^{T_{\star}}\frac{dT_H}{T_H^{22/7}}\frac{\exp[-n\frac{E}{T_{\nu}}
(\gamma_{1}-u_{1})]}{n} \nonumber\\
&&+\frac{T_{\nu}}{E}\left(\frac{C}{22T_{\nu}^4}\right)^{4/7}\int_{T_0}^{T_{\star}}\frac{dT_H}{T_H^{26/7}}\left(
\frac{\exp[-n\frac{E}{T_{\nu}}\gamma_{1}]}{n^2}
- \frac{\exp[-n\frac{E}{T_{\nu}}(\gamma_{1}-u_{1})]}{n^2}  \right)    
  \nonumber\\ 
&& +\left(\frac{C}{16T_{\nu}^4}\right)^{5/8}
\int_{T_{\star}}^{\infty}\frac{dT_H}{T_H^{7/2}}\frac{\exp[-n\frac{E}{T_{\nu}}
(\gamma_{2}-u_{2})]}{n} \nonumber\\
&&+\frac{T_{\nu}}{E}\left(\frac{C}{16T_{\nu}^4}\right)^{4/8}\int_{T_{\star}}^{\infty}\frac{dT_H}{T_H^{4}}\left(
\frac{\exp[-n\frac{E}{T_{\nu}}\gamma_{2}]}{n^2}
- \frac{\exp[-n\frac{E}{T_{\nu}}(\gamma_{2}-u_{2})]}{n^2}  \right)
\Biggl \},
\end{eqnarray}
where $T_{\star}=2(\frac{11}{8})^2T_{\nu}/C^{1/4}$.  
What is interesting here is the high energy limit, $E \gg 10T_{\nu} = 1$ TeV.  
This is determined wholly by the first of the two exponentials in Eq. (2.50) with 
the coefficient of 1.  Using also Eqs. (2.31), (2.51) and (2.55) we find the limit
\begin{equation}
\frac{dN_{\nu}^{\rm fluid}}{dE} \rightarrow
\frac{4185 \zeta(6) C^{1/4}}{854 \pi^5} \frac{m_{\rm P}^2 T_{\nu}}{E^4}
\approx \frac{m_{\rm P}^2 T_{\nu}}{1040 E^4}.
\end{equation}
Here we have ignored the small numerical differences between $\alpha_h^{eff}$ and $\alpha_h$. This $E^{-4}$ spectrum is characteristic of all neutrino sources in the viscous 
fluid description of the microscopic black hole wind.

\subsection{Neutrinos from Pion Decay}

\hspace{0.1in} Muon-type neutrinos will come 
from the decays of charged pions, namely $\pi^+ \rightarrow \mu^+ \nu_{\mu}$ and 
$\pi^- \rightarrow \mu^- \bar{\nu}_{\mu}$.  In what follows we calculate the 
spectrum of muon neutrinos; the spectrum for muon anti-neutrinos is of course 
identical.

In the rest frame of the decaying pion the muon and the neutrino have momentum q 
determined by energy and momentum conservation.
\begin{equation}
m_{\pi} = q + \sqrt{m_{\mu}^2+q^2}
\end{equation}
Numerically $q = 0.2134 m_{\pi}$.
In the pion rest frame the spectrum for the neutrino, normalized to one, is
\begin{equation}
E^{\prime} \frac{d^3N_{\nu}^{\pi}}{d^3p^{\prime}} =
\frac{\delta(E^{\prime}-q)}{4\pi q} \, .
\end{equation}
The Lorentz invariant rate of emission is obtained by folding together the 
spectrum with the rate of emission of $\pi^+$.
\begin{equation}
E \frac{d^4N_{\nu}^{\pi}}{d^3p dt} = \int_{m_{\pi}}^{\infty} dE_{\pi}
\left( \frac{d^2N_{\pi}}{dE_{\pi}dt} \right) \frac{1}{4\pi q}
\delta \left( \frac{ EE_{\pi} - {\bf p} \cdot {\bf p}_{\pi} }
{m_{\pi}} - q \right)
\end{equation}
The spectrum of pions is computed in the same way as the spectrum of direct 
neutrinos from the expanding fluid in Sec. 2.8.1 but with one difference and one 
simplification.  The difference is that the pion has a Bose distribution as 
opposed to the Fermi distribution of the neutrino.  The simplification is that 
the fluid at pion decoupling has a highly relativistic flow velocity with $u_f 
\approx \gamma_f \gg 1$.  The analog to Eq. (2.50) is
\begin{equation}
\frac{d^2N_{\pi}}{dE_{\pi} dt} = 
- \frac{r_f^2 T_f p_{\pi}}{\pi \gamma_f} \ln \left(1 - 
{\rm e}^{-E_{\pi}/2 \gamma_f T_f} \right) \, .
\end{equation}
The instantaneous energy spectrum of the neutrino is reduced to a single 
integral.
\begin{equation}
\frac{d^2N_{\nu}^{\pi}}{dE dt} = - \frac{m_{\pi} T_f r_f^2}{2\pi q \gamma_f}
\int_{E_{\rm min}}^{\infty} dE_{\pi} 
\ln\left( 1 - {\rm e}^{-E_{\pi}/2\gamma_f T_f} \right)
\end{equation}
Here $E_{\rm min} = m_{\pi}(E^2+q^2)/2Eq$ is the minimum pion energy that will 
produce a neutrino with energy $E$.  We are interested only in high energy 
neutrinos with $E \gg m_{\pi}$, in which case $E_{\rm min} = m_{\pi}E/2q$ is an 
excellent approximation.  The integral can also be expressed as an infinite 
summation
\begin{equation}
\frac{d^2N_{\nu}^{\pi}}{dEdt}=\frac{m_{\pi}r_f^2T_f^2}{\pi q}
\sum_{n=1}^{\infty}\frac{1}{n^2}
\exp\left(-\frac{nm_{\pi}E}{4\gamma_fT_fq}\right)
\end{equation}
When $E \gg \gamma_f T_f$ only the first term in the summation in Eq. (2.64) 
is important.

The instantaneous spectrum is easily integrated over time because the flow 
velocity at pion decoupling is highly relativistic.  The spectrum arising from 
the last moments when the Hawking temperature exceeds $T_0$ is
\begin{equation}
\frac{dN_{\nu}^{\pi}}{dE}=\frac{64u_s^4q^3}{\pi^6\alpha_hm_{\pi}^3}
\left( \frac{45 \pi \alpha_h^{eff}}{2u_s^2d_f} \right)^{5/4}
\frac{m_{\rm P}^2T_f}{E^4}
\sum_{n=1}^{\infty}\frac{1}{n^2}
\int_0^{m_{\pi}E/4\gamma_f(T_0)T_fq}dx x^3 \, 
{\rm e}^{-nx}.
\end{equation}
In the limit that $E \gg \gamma_f(T_0) T_f$ the upper limit on the integral may be taken to infinity.  After ignoring the small numerical difference between $\alpha_h^{eff}$ and $\alpha_h$, we can express the result in terms of the constant $C$ as:
\begin{equation}
\frac{dN_{\nu}^{\pi}}{dE} \rightarrow \frac{464\pi}{427}
\left(\frac{q}{m_{\pi}}\right)^3
\left(\frac{203C}{2d_f}\right)^{1/4}
\frac{m_{\rm P}^2T_f}{E^4}
\approx \frac{m_{\rm P}^2T_f}{300E^4}.
\end{equation}
This is much smaller than the direct neutrino emission from the fluid because 
the neutrino decoupling temperature $T_{\nu}$ is much greater than the pion 
decoupling temperature $T_f$.  Here $\alpha_h^{eff}$ does not include gravitons and neutrinos.

\subsection{Neutrinos from Muon Decay}

Muons can be emitted directly or indirectly by the weak decay of pions:
$\pi^- \rightarrow \mu^- \nu_e \bar{\nu}_{\mu}$
plus the charge conjugated decay.  Both sources 
contribute to the neutrino spectrum.  The invariant distribution of the electron 
neutrino (to be specific) in the rest frame of the muon is
\begin{equation}
E^{\prime} \frac{d^3N_{\nu_e}}{d^3p^{\prime}} =
\frac{4}{\pi m_{\mu}^4} (3m_{\mu}-4E^{\prime})E^{\prime}
\end{equation}
where the electron mass has been neglected in comparison to the muon mass.  This 
distribution is used in place of the delta-function distribution of Eq. (2.60) for 
both direct and indirect muons, as evaluated in the following two subsections.

\subsubsection{Neutrinos from direct muons}

The instantaneous spectrum of $\nu_e$, $\bar{\nu}_e$, $\nu_{\mu}$ or 
$\bar{\nu}_{\mu}$ arising from muons in thermal equilibrium until the decoupling 
temperature of $T_f$ can be computed by folding together the spectrum of muons 
together with the decay spectrum of neutrinos in the same way as Eq. (2.61) was 
obtained.  Using Eqs. (2.62) and (2.67) results in
\begin{eqnarray}
\frac{d^2N_{\nu}^{{\rm dir} \; \mu}}{dEdt}&=&
\frac{2r_f^2 T_fE}{3\pi\gamma_f}
\sum_{n=1}^{\infty}\Biggl \{ -{\rm Ei}
 \left( -\frac{nE}{2\gamma_fT_f}\right)\left[9\frac{nE}{2\gamma_fT_f}
+2\left(\frac{nE}{2\gamma_fT_f}\right)^2\right] \nonumber\\
&&+\exp\left(-\frac{nE}{2\gamma_fT_f}\right)\left[\frac{10\gamma_fT_f}{nE}-7-
2\frac{nE}{\gamma_fT_f}\right]
\Biggl \},
\end{eqnarray}
where ${\rm Ei}$ is the exponential-integral function.  In the high energy 
limit, defined here by $E \gg 2\gamma_fT_f$, the spectrum simplifies to
\begin{equation}
\frac{d^2N_{\nu}^{{\rm dir} \; \mu}}{dEdt}=\frac{2r_f^2T_fE}{3\pi\gamma_f}
\exp{\left(-\frac{E}{2\gamma_fT_f}\right)} \, .
\end{equation}

The time-integrated spectrum can be calculated in a fashion analogous to that 
followed in Sec. 2.8.2.  Thus
\begin{eqnarray}
\lefteqn{\frac{dN_{\nu}^{{\rm dir} \; \mu}}{dE}= 
\frac{16u_s^4}{3 \pi^6 \alpha_h}
\left( \frac{45 \pi \alpha_h^{eff}}{2u_s^2d_f} \right)^{5/4}
\frac{m_{\rm P}^2T_f}{E^4}}\nonumber\\ 
&&\times \sum_{n=1}^{\infty} \int_0^{E/2\gamma_f(T_0)T_f} \Biggl\{
-{\rm Ei} \left( -\frac{nE}{2\gamma_fT_f}\right)\left[9\frac{nE}{2\gamma_fT_f}
+2\left(\frac{nE}{2\gamma_fT_f}\right)^2\right] \nonumber\\
&&+\exp\left(-\frac{nE}{2\gamma_fT_f}\right)\left[\frac{10\gamma_fT_f}{nE}-7-
2\frac{nE}{\gamma_fT_f}\right]
\Biggl\} \, . 
\end{eqnarray}
In the high energy limit this simplifies to
\begin{eqnarray}
\frac{dN_{\nu}^{{\rm dir} \; \mu}}{dE} \rightarrow 
\frac{11745}{488\pi^5 d_f} \left[ 225 \zeta(7)-217 \zeta(6) \right]
\left(\frac{203C}{2d_f}\right)^{1/4}
\frac{m_{\rm P}^2T_f}{E^4} \approx \frac{m_{\rm P}^2T_f}{250 E^4} \, .
\end{eqnarray}

\subsubsection{Neutrinos from indirect muons}

The spectrum of neutrinos coming from the decay of muons which themselves came 
from the decay of pions proceeds exactly as in the previous subsection, but with 
the replacement of the direct muon energy spectrum by the indirect muon energy 
spectrum.  In the rest frame of the pion the muon distribution is
\begin{equation}
E^{\prime} \frac{d^3N_{\mu}^{\pi}}{d^3p^{\prime}} =
\frac{\sqrt{m_{\mu}^2+q^2}}{4\pi q^2} \delta(p^{\prime}-q) \, ,
\end{equation}
which is the finite mass version of Eq. (2.60).  Folding together this spectrum 
with the spectrum of pions (2.62) yields the spectrum of indirect muons.
\begin{equation}
\frac{d^2N_{\mu}^{\pi}}{dEdt}=\frac{m_{\pi}r_f^2T_f^2}{\pi q}
\sum_{n=1}^{\infty}\frac{1}{n^2}
\exp\left(-\frac{nm_{\pi}E}{4\gamma_fT_f\sqrt{m_{\mu}^2+q^2}}\right)
\end{equation}
This spectrum is now folded with the decay distribution of neutrinos to obtain
\begin{eqnarray}
\frac{d^2N_{\nu}^{{\rm indir} \; \mu}}{dEdt} &=&
\frac{m_{\pi}r_f^2 T_f^2}{3\pi q}
\sum_{n=1}^{\infty} \frac{1}{n^2} \Biggl\{
{\rm Ei}(-nx)
\left(5-\frac{27}{6}n^2x^2-\frac{2}{3}n^3x^3\right)
\nonumber\\
&-&\left(-\frac{19}{6}+\frac{23}{6}nx+\frac{2}{3}n^2x^2\right)
{\rm e}^{-nx}
\Biggl\}_{x_-}^{x_+}
\end{eqnarray}
where
\begin{eqnarray}
x_{\pm}=\frac{m_{\pi}E}{2\gamma_fT_f} \left[ 
\frac{(m_{\mu}^2+q^2)^{1/2}\pm q}{m_{\mu}^2}
\right] \, .
\end{eqnarray}
The high energy limit $E \gg \gamma_fT_f$ is
\begin{eqnarray}
\frac{d^2N_{\nu}^{{\rm indir} \; \mu}}{dEdt} \rightarrow
\frac{m_{\pi}r_f^2T_f^2}{3\pi q}
\left[\frac{{\rm e}^{-x_-}}{x_-^2}-\frac{{\rm e}^{-x_+}}{x_+^2}\right] \, .
\end{eqnarray}

The integral over time can be done in the usual way.  The resulting 
expression is

\begin{eqnarray}
\lefteqn{\frac{dN_{\nu}^{{\rm indir} \; \mu}}{dE} \;\; = \;\;
\frac{4u_s^4}{3\pi^6\alpha_hm_{\pi}^3q}
\left( \frac{45 \pi \alpha_h^{eff}}{2u_s^2d_f} \right)^{5/4}
\frac{m_{\rm P}^2T_f}{E^4}}
\nonumber\\
&&\times \sum_{n=1}^{\infty} \frac{1}{n^2} \Biggl\{
\left( \frac{m_{\mu}^2}{(m_{\mu}^2+q^2)^{1/2}-q} \right)^{4}
\int_0^{x_{0-}}dx_-x_-^3  \nonumber\\
&&\Biggl[  \exp(-nx_-)(-\frac{19}{6}+\frac{23}{6}nx_-+\frac{4}{6}n^2x_-^2)
-Ei(-nx_-)
(5-\frac{27}{6}n^2x_-^2-\frac{4}{6}n^3x_-^3) \Biggl]\nonumber\\
&&-\left( \frac{m_{\mu}^2}{(m_{\mu}^2+q^2)^{1/2}+q} \right)^{4}
\int_0^{x_{0+}}dx_+x_+^3  
\nonumber\\
&&\Biggl[  
\exp(-nx_+)(-\frac{19}{6}+\frac{23}{6}nx_++\frac{4}{6}n^2x_+^2)
-Ei(-nx_+)
(5-\frac{27}{6}n^2x_+^2-\frac{4}{6}n^3x_+^3)\Biggl] 
\Biggl\},\nonumber\\
\end{eqnarray}
where $x_{0\pm}=x_{\pm}(T_0)$.
The high energy limit is simple and of the familiar form 
$1/E^4$.  
\begin{equation}
\frac{dN_{\nu}^{{\rm indir} \; \mu}}{dE} \rightarrow 
\frac{174 \pi}{2989 d_f} \left(\frac{203C}{2d_f}\right)^{1/4}
\frac{\sqrt{m_{\mu}^2+q^2} \, (m_{\mu}^2+2q^2)}{m_{\pi}^3}
\frac{m_{\rm P}^2T_f}{E^4}\nonumber\\
\approx \frac{m_{\rm P}^2T_f}{1250 E^4}
\end{equation}

\section{Comparison of Neutrino Sources}

In this section we compare the different sources of neutrinos that were computed 
in the previous sections.  All the figures presented display one type of 
neutrino or anti-neutrino.  That type should be clear from the context.  For 
example, equal numbers of $\nu_e$, $\bar{\nu}_e$, $\nu_{\mu}$, 
$\bar{\nu}_{\mu}$, $\nu_{\tau}$, and $\bar{\nu}_{\tau}$ are produced as Hawking 
radiation and by direct emission by the fluid at the neutrino decoupling 
temperature $T_{\nu}$.  Only electron and muon type neutrinos are produced by 
muon decay, and only muon type neutrinos by pion decay.  These differences could 
help to distinguish the decay of a microscopic black hole from other neutrino 
sources if one happened to be within a detectable distance.

\begin{figure}[htb]
\centerline{\epsfig{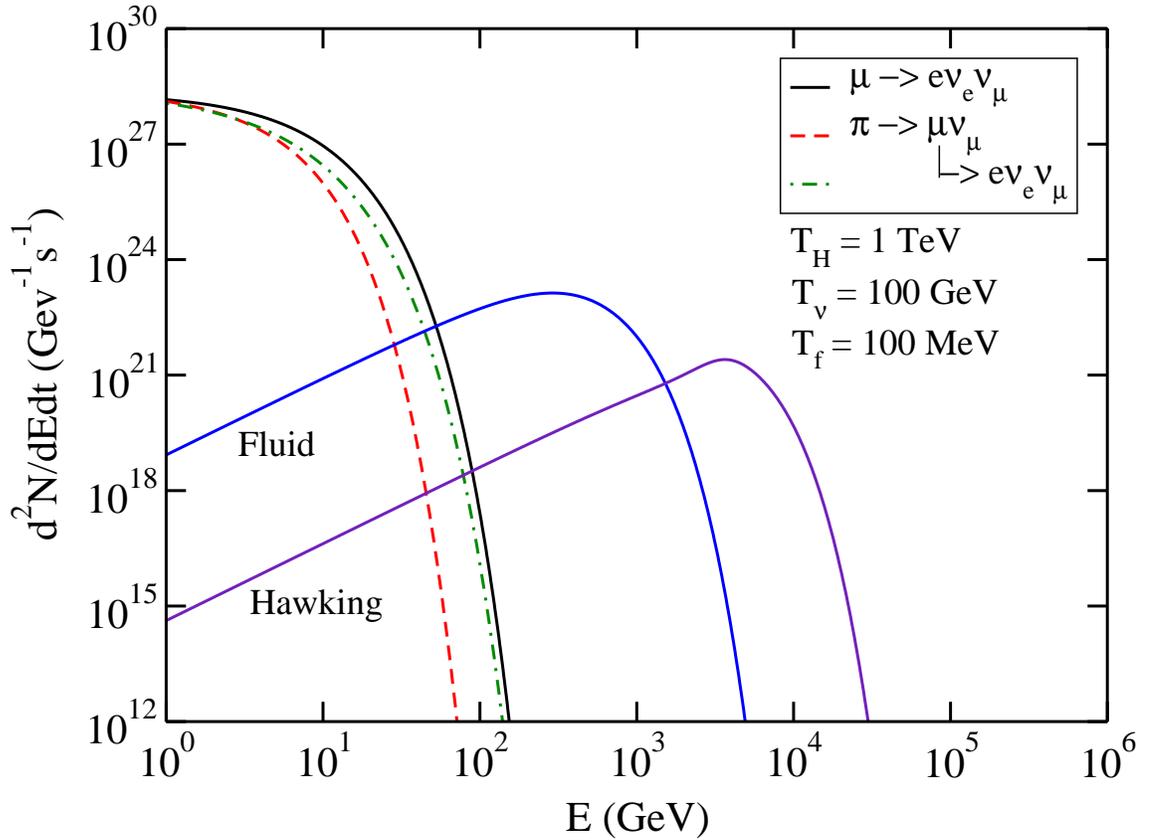}}
\caption{The instantaneous direct neutrino spectra emerging from the  
fluid with neutrino-sphere located at $T_{\nu}=100$ GeV compared to the direct 
Hawking radiation.  Also shown are neutrinos arising from  
pion, direct muon and indirect muon decays  at a decoupling temperature of 
$T_f=100$ MeV.  Here the black hole temperature is $T_H=1$ TeV.
All curves are for one flavor of neutrino or anti-neutrino.}
\label{bhnu-f1}
\end{figure}

\begin{figure}[htb]
\centerline{\epsfig{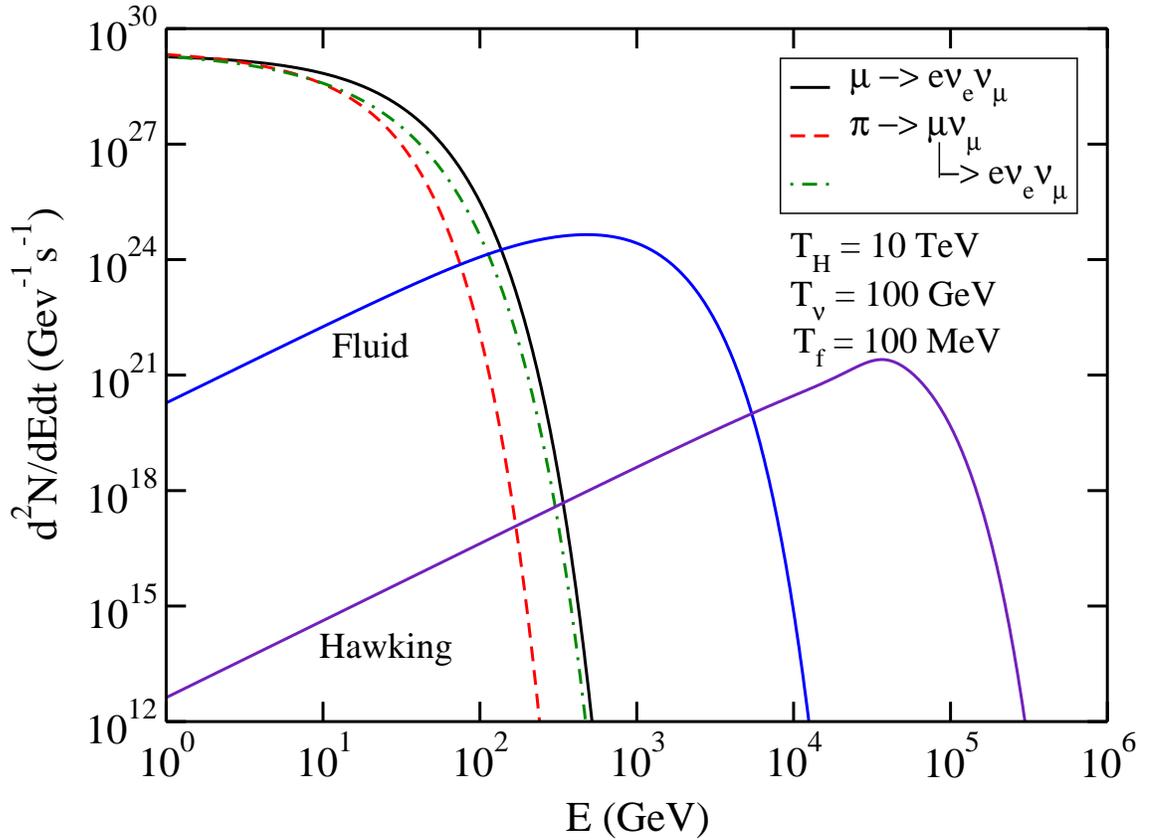}}
\caption{Same as figure \ref{bhnu-f1} but with $T_H=1$ TeV.}
\label{bhnu-f2}
\end{figure}

\begin{figure}[htb]
\centerline{\epsfig{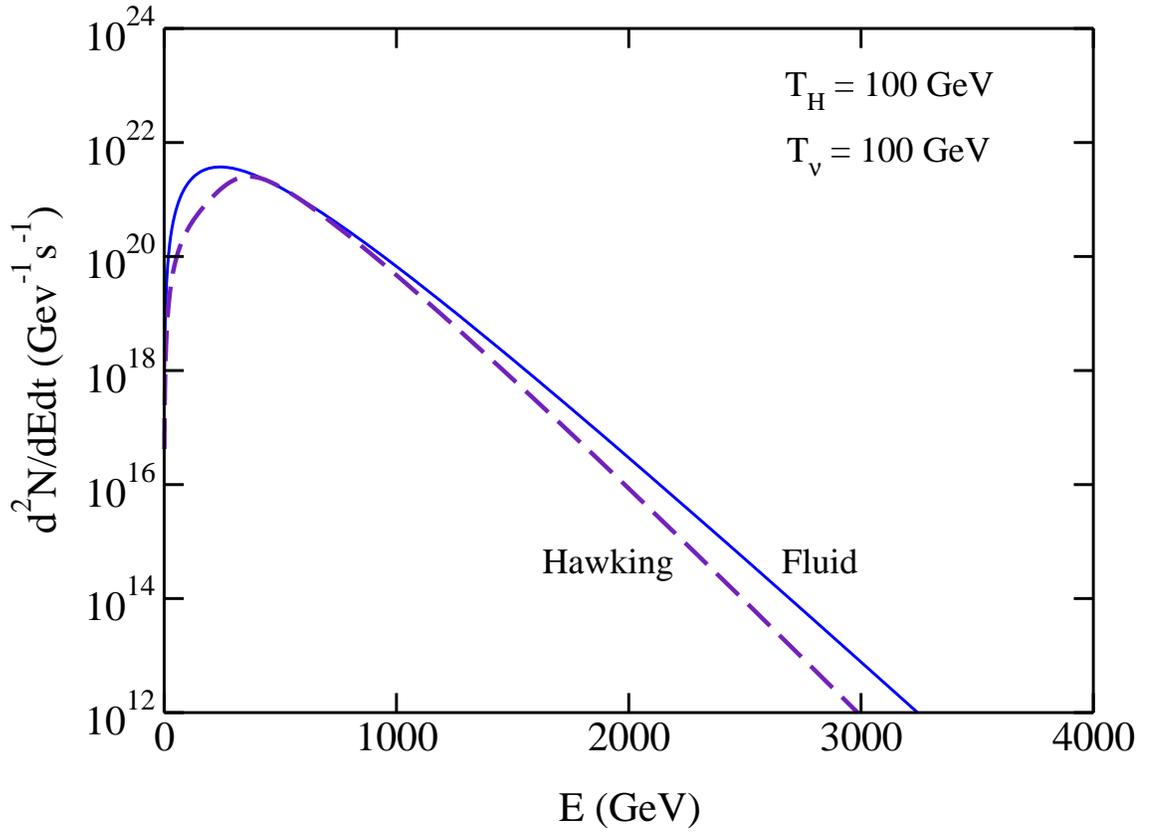}}
\caption{The instantaneous direct neutrino spectra emerging from the  
fluid with neutrino-sphere located at $T_{\nu}=100$ GeV compared to the direct 
Hawking radiation.  Both curves are for one flavor of neutrino or anti-neutrino.}
\label{2-nu-mu}
\end{figure}
\begin{figure}[htb]
\centerline{\epsfig{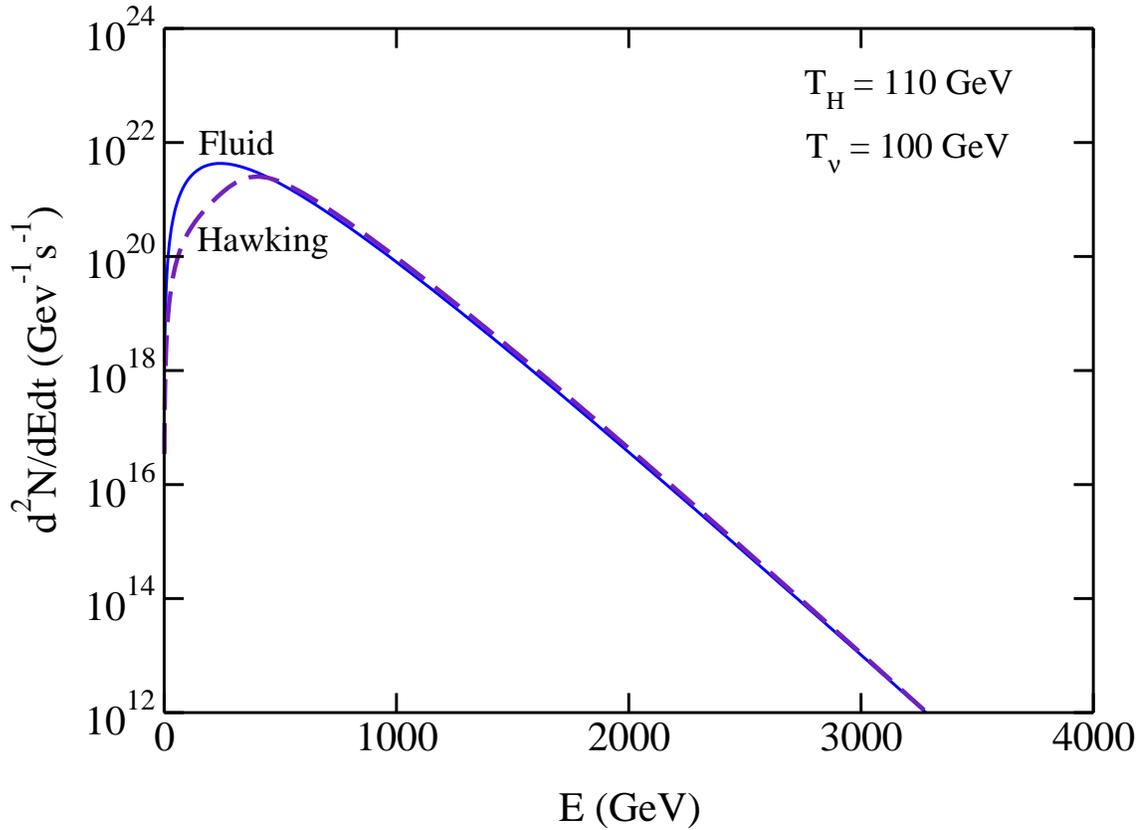}}
\caption{Same as figure  \ref{2-nu-mu} but with $T_H=110$ GeV.}
\label{110-nu-mu}
\end{figure}
\begin{figure}[htb]
\centerline{\epsfig{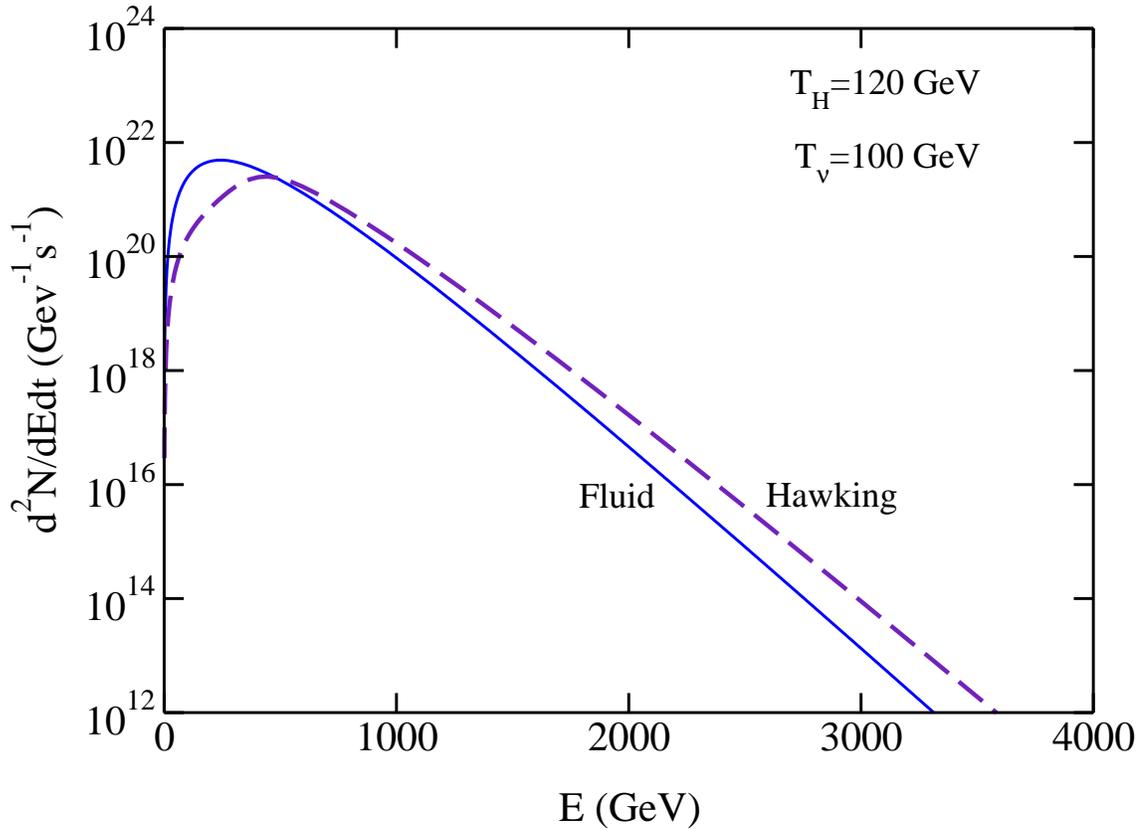}}
\caption{Same as figure \ref{2-nu-mu} but with $T_H=120$ GeV.}
\label{120-nu-mu}
\end{figure}

The instantaneous spectra are displayed in figure \ref{bhnu-f1} for a Hawking temperature of 1 
TeV corresponding to a black hole mass of $10^7$ kg and a lifetime of 7.7 
minutes. The instantaneous spectra for a Hawking temperature of 10 TeV 
corresponding to a black hole mass of $10^6$ kg and a lifetime of 0.5 seconds 
are displayed in figure \ref{bhnu-f2}.  There are several important features of these spectra.  
One feature is that the spectrum of direct neutrinos emitted by the fluid, at 
the decoupling temperature of $T_{\nu} = 100$ GeV, peaks at a lower energy than 
the spectrum of neutrinos that would be emitted directly as Hawking radiation.  
The peaks are located approximately at $\sqrt{T_{\nu}T_H}$ for the fluid and at 
$5T_H$ for the Hawking neutrinos.  The reason is that the viscous flow degrades 
the average energy of particles composing the fluid, but the number of particles 
is greater as a consequence of energy conservation.  In the viscous fluid 
picture of the black hole explosion direct neutrinos are emitted as Hawking 
radiation without any rescattering when $T_H < T_{\nu}$, whereas when $T_H > 
T_{\nu}$ they are assumed to rescatter and then be emitted from the 
neutrino-sphere located at $T_{\nu}$.  It is incorrect to add the two curves 
shown in these figures.  In reality, of course, it would be better to use 
neutrino transport equations to describe what happens when $T_H \approx 
T_{\nu}$.  That is beyond the scope of this thesis, and perhaps worth doing only if and when there is observational evidence for microscopic black 
holes.  

Here we would like to investigate the discontinuity in the transition from Hawking radiation to neutrino emission from fluid.  In order to see the discontinuity, we plot the instananeous direct neutrino spectra emitted by the fluid and Hawking radiation in figures \ref{2-nu-mu}, \ref{110-nu-mu} and \ref{120-nu-mu} for black hole temperatures of 100, 110 and 120 GeV.  As one can see, the difference in the slope of both process at high energy reaches a minimum when the black hole temperature is around $T_H=110$ GeV (figure \ref{110-nu-mu}).  We also calculate the total number of neutrinos per unit time for these three graphs numerically.  For $T_H=100$ GeV  
\begin{eqnarray}
\frac{dN^{\rm dir}_{\nu}}{dt}\approx4.13\times10^{29}, \nonumber\\
\frac{dN^{\rm fluid}_{\nu}}{dt}\approx5.45\times10^{29}.
\end{eqnarray}
For $T_H=110$ GeV  
\begin{eqnarray}
\frac{dN^{\rm dir}_{\nu}}{dt}\approx1.84\times10^{29}, \nonumber\\
\frac{dN^{\rm fluid}_{\nu}}{dt}\approx2.34\times10^{29}.
\end{eqnarray}
For $T_H=120$ GeV  
\begin{eqnarray}
\frac{dN^{\rm dir}_{\nu}}{dt}\approx2.18\times10^{29}, \nonumber\\
\frac{dN^{\rm fluid}_{\nu}}{dt}\approx2.69\times10^{29}.
\end{eqnarray}
In this energy range the total number of emitted neutrinos from both process is roughly the same.  Therefore if the transition from Hawking radiation to fluid emission happens when $T_H$ is around 100 GeV the discontinuity will be small, otherwise the discontinuity is big.

Another feature of figures \ref{bhnu-f1} and \ref{bhnu-f2} to note is that the average energy of neutrinos 
arising from pion and muon decay is much less than that of directly emitted 
neutrinos.  Again, the culprit is viscous fluid flow degrading the average 
energy of particles, here the pion and muon, until the time of their 
decoupling at $T_f \approx 100$ MeV.  On the other hand their number is 
greatly increased on account of energy conservation.  The average energies 
of the neutrinos are 
somewhat less than their parent pions and muons because energy must be shared 
among the decay products.  The spectrum of muon-neutrinos coming from the decay
$\pi \rightarrow \mu \nu$ is the softest because the pion and muon masses
are very close, leaving very little energy for the neutrino.

\begin{figure}[htb]
\centerline{\epsfig{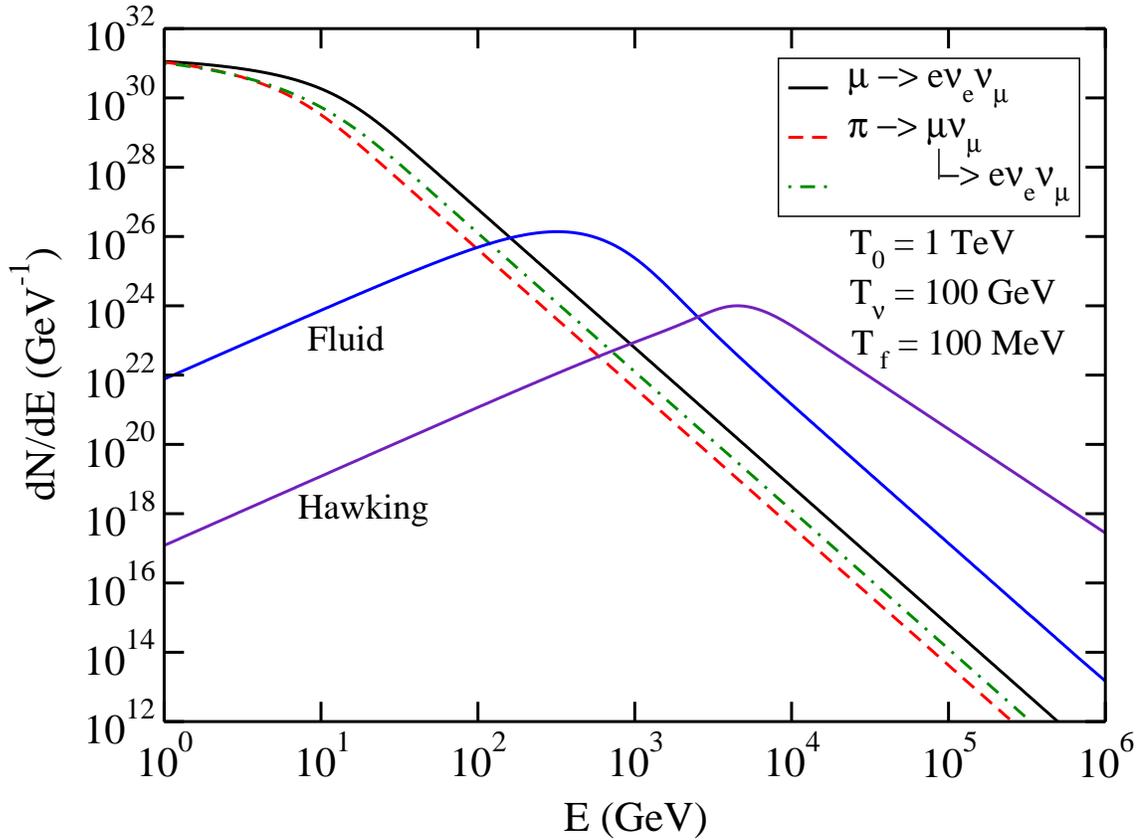}}
\caption{The time integrated neutrino spectra emerging from  
a microscopic black hole.  Here the calculation begins when the black hole 
temperature is $T_0=1$ TeV.  Either the direct Hawking radiation of neutrinos 
or the direct neutrino emission from a neutrino-sphere at a 
temperature of 100 GeV should be used.  All curves are for one species of 
neutrino or anti-neutrino.}
\label{bhnu-f3}
\end{figure}

\begin{figure}[htb]
\centerline{\epsfig{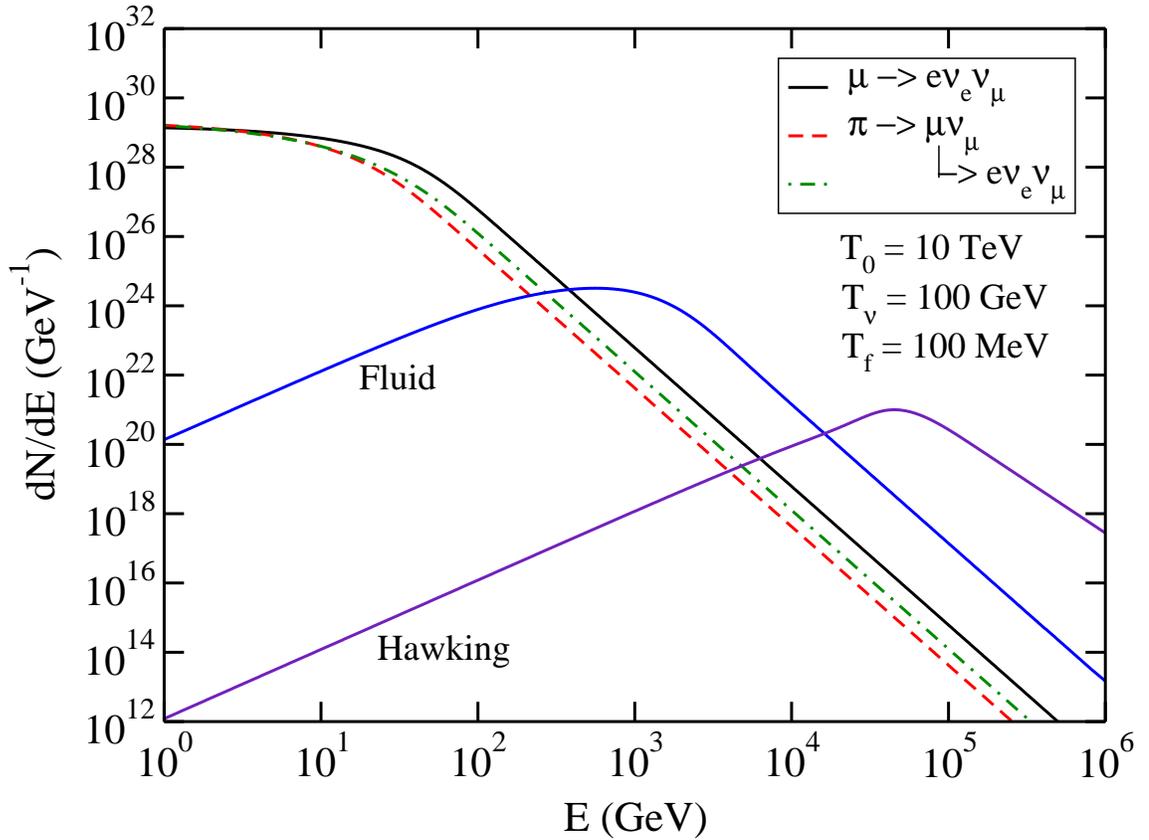}}
\caption{Same as figure \ref{bhnu-f3} but with $T_0=10$ TeV.}
\label{bhnu-f4}
\end{figure}

The time-integrated spectra, starting at the moment when the Hawking temperature 
is 1 and 10 TeV, are shown in figures \ref{bhnu-f3} and \ref{bhnu-f4}, respectively.  The relative 
magnitudes and average energies reflect the trends seen in figures \ref{bhnu-f1} and \ref{bhnu-f2}.  At 
high energy the Hawking spectrum is proportional to $E^{-3}$ while all the 
others are proportional to $E^{-4}$, as was already pointed out in the previous 
sections.  Obviously the greatest number of neutrinos by far are emitted at 
energies less than 100 GeV.  The basic reason is that only about 5\% of the 
total luminosity of the black hole is emitted directly as neutrinos.  About 32\% 
goes into neutrinos coming from pion and muon decay, about 24\% goes into 
photons, with most of the remainder going into electrons and positrons.

\section{Observability of the Neutrino Flux}

We now turn to the possibility of observing neutrinos from a microscopic black 
hole directly.  Obviously this depends on a number of factors, such as the 
distance to the black hole, the size of the neutrino detector, the efficiency of 
detecting neutrinos as a function of neutrino type and energy, how long the 
detector looks at the black hole before it is gone, and so on.

For the sake of discussion, let us assume that one is interested in neutrinos 
with energy greater than 10 GeV and that the observational time is the last 7.7 
minutes of the black hole's existence when its Hawking temperature is 1 TeV and 
above.  As may be seen from figure \ref{bhnu-f3}, most of the neutrinos will come from the 
decay of directly emitted muons.  Integration of Eq. (2.71) from $E_{\rm min} = 
10$ GeV to infinity, and multiplying by 4 to account for both electron and muon 
type neutrinos and anti-neutrinos, results in the total number of
\begin{equation}
N_{\nu} =\frac{4}{750} \frac{m_{\rm P}^2 T_f}{E_{\rm min}^3}
\approx 8\times 10^{31} \, .
\end{equation}
This does not take into account neutrinos directly emitted from the fluid.  For   
$E_{\rm min} = 1$ TeV, for example, Eq. (2.58) should be used in place of Eq. (2.71) 
(see figure \ref{bhnu-f3}), and taking into account tau-type neutrinos too then yields a total 
number of about $3 \times 10^{28}$.  For an exploding black hole located a distance $d$ 
from Earth the number of neutrinos per unit area is
\begin{eqnarray}
N_{\nu}(E > 10\;{\rm GeV}) &=& 
6700 \left(\frac{1\;{\rm pc}}{d}\right)^2 \; {\rm km}^{-2} \, ,\\
N_{\nu}(E > 1\;{\rm TeV}) &=& 
2.5 \left(\frac{1\;{\rm pc}}{d}\right)^2 \; {\rm km}^{-2} \, .
\end{eqnarray}
Although the latter luminosity is smaller by three orders of magnitude, it has 
two advantages.  First, 1/3 of that luminosity comes from tau-type neutrinos.  
Unlike electron and muon-type neutrinos, the tau-type is not produced by the 
decays of pions produced by interactions of high energy cosmic rays with matter 
or with the microwave background radiation.  Hence it would seem to be a much 
more characteristic signal of exploding black holes than any other cosmic source 
(assuming no oscillations between the tau-type and the other two species).  
Second, the tau-type neutrinos come from near the neutrino-sphere, thus probing 
physics at a temperature of order 100 GeV much more directly than the other 
types of neutrinos.

What is the local rate density $\dot{\rho}_{\rm local}$ of exploding black 
holes?  This is, of course, unknown since no one has ever knowingly 
observed a black hole explosion.  The first observational limit was determined 
by Page and Hawking \cite{PH}.  They found that the local rate density
is less than 1 to 10 per cubic parsec per year on the 
basis of diffuse gamma rays with energies on the order of 100 MeV.  This limit 
has not been lowered very much during the intervening twenty-five years.  For 
example, Wright \cite{W} used EGRET data to search for an anisotropic
high-lattitude component of diffuse gamma rays in the energy range from 30 MeV 
to 100 GeV as a signal for steady emission of microscopic black holes.  He 
concluded that $\dot{\rho}_{\rm local}$ is less than about 0.4 per cubic parsec 
per year.  If the actual rate density is anything close to these upper limits 
the frequency of a high energy neutrino detector seeing a black hole explosion 
ought to be around one per year.

\chapter{Cosmic Shells}

\hspace{0.2in} In this chapter we investigate numerical solutions to the combined field equations of gravity and a scalar field with a potential which has two non-degenerate minima.  The absolute minimum of the potential is the true vacuum and the other minimum is the false vacuum.  This potential is shown in figure\ \ref{potential}.  The true vacuum of the potential is at $f_2$ and the false vacuum is at $f_1$.  The two minima are separated by a barrier.

\section{Shell Solutions in Static Coordinates}

\begin{figure}[tbh]
\centering \leavevmode 
\rotatebox{90}{
\epsfxsize=10.0cm \epsfbox{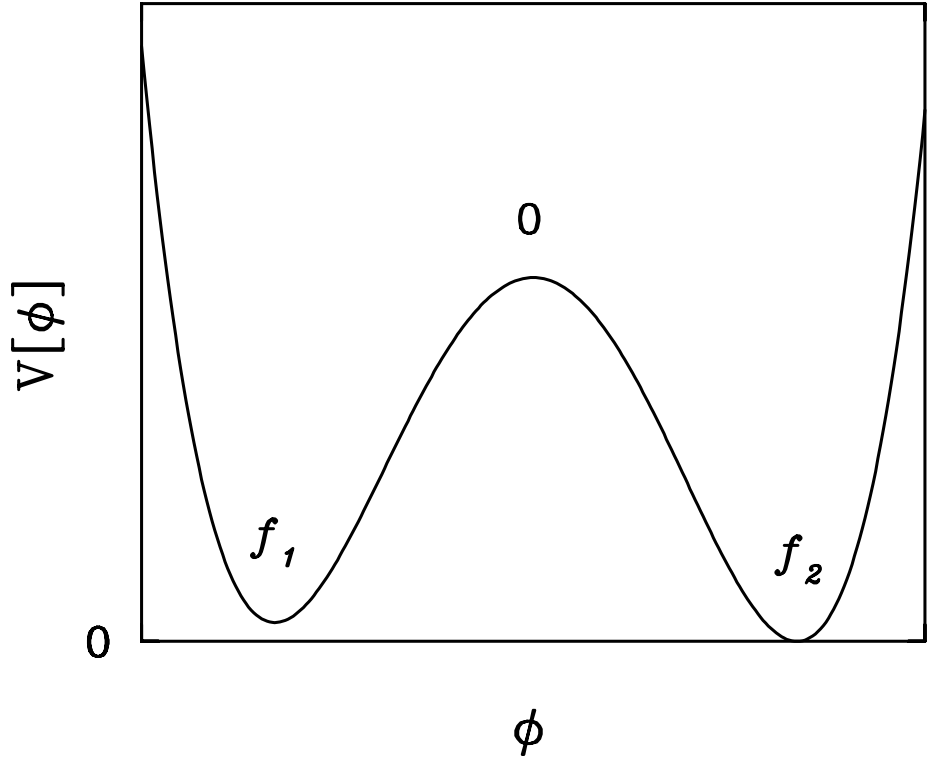} }
\caption{The potential $V[\phi]$.}
\label{potential}
\end{figure}

\hspace{0.2in} Consider a scalar field coupled to gravity with the Lagrangian 
\beeq
{\cal L} = 
{1\over 16\pi G} ~ R
   + {1\over 2}  \phi_{;\mu} \phi^{;\mu}   - V[\phi] 
\eneq
where the potential $V[\phi]$ has
two minima at $f_1$ and $f_2$ separated by a barrier. See figure\ 
\ref{potential}.

We look for spherically symmetric configurations in which 
the metric of space-time is written as
\beeq
ds^2 = 
-\frac{H}{p^2}dt^2 + \frac{dr^2}{H} 
  +r^2 (d\theta^2  + \sin^2 \theta d\vphi^2) ~.
\label{ourmetric1}
\eneq
The functions $\phi$, $H$, and $p$ depend only on $r$ and $t$.
A tetrad basis is chosen, in a region $H>0$, as
\beeq
e_0 = \frac{\sqrt{H}}{p} \, dt \next
e_1 = \frac{1}{\sqrt{H}} \, dr\next
e_2 = r d\theta \next e_3 = r\sin\theta d\vphi ~.
\label{tetrad1}
\eneq
The components of the energy-momentum tensor in the tetrad basis, 
$T_{ab} = {e_a}^\mu {e_b}^\nu T_{\mu\nu}$, are
\beqn
&&\hskip -.8cm
T_{00} = {1\over 2} \bigg( \frac{p^2}{H} \, \dot\phi^2 
  + H \phi'^2 \bigg) + V[\phi] \cr
&&\hskip -.8cm
T_{11} = {1\over 2} \bigg( \frac{p^2}{H} \, \dot\phi^2 
  + H \phi'^2 \bigg) - V[\phi] \cr
&&\hskip -.8cm
T_{22} = T_{33}  
={1\over 2} \bigg( \frac{p^2}{H} \, \dot\phi^2 
   - H \phi'^2 \bigg) - V[\phi] \cr
\noalign{\kern 5pt}
&&\hskip -.8cm
T_{01} = -p \dot\phi \phi'  ~.
\label{tensor1}
\eeqn
Here dots and primes indicate $t$ and $r$ derivatives, respectively.

The scalar field satisfies
\beeq
p \frac{\dd}{\dd t} \bigg( \frac{p}{H} \dot\phi \bigg)
- {p\over r^2} {\dd\over \dd r} \bigg( {r^2 H\over p} \phi' \bigg)
+ V'[\phi] = 0 ~.
\label{scalar1}
\eneq
We introduce the integrated mass function $M(t,r)$ by
\beeq
H = 1 - {2GM\over r} ~.
\label{mass1}
\eneq
The Einstein equations are 
\beqn
&&\hskip -.8cm
M = \int_0^r 4\pi r^2 dr \, T_{00} ~, 
\label{Ein1} \\
&&\hskip -.8cm
{p'\over p} = - 4\pi G r \bigg\{ {p^2\over H^2} \, \dot\phi^2 + 
  \phi'^2 \bigg\} ~,  
\label{Ein2} \\
&&\hskip -.8cm
{\dot H\over H} = - 8\pi G r \dot\phi \phi' ~,
\label{Ein4}\\
&&\hskip -.8cm
{p\over 2} \bigg\{ {\dd\over \dd t} \bigg( {p\dot H\over H^2} \bigg) 
+ {\dd\over \dd r} \bigg( {H'\over p} - {2Hp'\over p^2} \bigg) \bigg\}
+ {1-H\over r^2} 
= 4\pi G\bigg( {p^2\over H} \,\dot\phi^2 - H \phi'^2 \bigg) ~.
\label{Ein3}
\eeqn
One of the equations, Eq.\ (\ref{Ein3}), is redundant as it follows from Eqs.\
(\ref{scalar1}), (\ref{Ein1}), (\ref{Ein2}), and (\ref{Ein4}).

We shall seek static solutions for which the set of equations 
reduces to
\beqn
&&\hskip -.8cm
\phi''(r) + \Gamma_\eff (r)  \phi'(r) = {1\over H} ~ V'[\phi] ~,
\label{scalar2}  \\
&&\hskip -.8cm
M(r) =\int_0^r 4\pi r^2 dr \, \left\{ {1\over 2} H \phi'^2 
   +V[\phi] \right\} ~,
\label{mass2}
\eeqn
where
\beqn
&&\hskip -.8cm
\Gamma_\eff \equiv {2\over r} - {p'\over p} + {H'\over H} ~, \cr
&&\hskip -.8cm
{p'\over p} = - 4\pi G r  \phi'^2  ~.
\label{friction1}
\eeqn 
Equation\ (\ref{scalar2}) can be interpreted as an equation for a particle with
a coordinate $\phi$ and time $r$.  Except for a factor $1/H$ this particle
moves in a potential $U[\phi]=-V[\phi]$.  The coefficient $\Gamma_\eff(r)$
represents time ($r$) dependent friction.  

The potential $V[\phi]$ is supposed to have two minima at $f_1$ and $f_2$.  
We are looking for a solution  which starts at $\phi \sim f_1$, moves close
to $f_2$, and comes back to $f_1$ at $r=\infty$.  The particle's
potential $U[\phi]$ has two maxima.  The particle begins near the top
of one hill, rolls down into the valley and up the other hill, turns
around and rolls down and then back up to the top of the original hill.
This is impossible in flat space, as $\Gamma_\eff$ is positive-definite
so that the particle's energy dissipates and it cannot climb back
to its starting point.

In the presence of gravity the situation changes.  The non-vanishing
energy density can make $H$ a decreasing function of $r$ so that
$\Gamma_\eff$ becomes negative.  The energy lost by the particle during the 
initial rolling down can be regained on the return path by negative
friction, or thrust.  Indeed, this happens.

Let us set up the problem more precisely.  We take a quartic potential
with $f_1 <0 < f_2$;
\beqn
&&\hskip -.8cm
V'[\phi] = \lambda \phi(\phi- f_1)(\phi-f_2) ~,   \cr
\noalign{\kern 5pt}
&&\hskip -.8cm
V[\phi] = \frac{\lambda}{4}(\phi-f_2) 
 \bigg\{ \phi^3 - {1\over 3} (f_2 + 4f_1) \phi^2 
  -{1\over 3} f_2 (f_2- 2f_1 )  (\phi+f_2) \bigg\} ~. 
\label{potential1}
\eeqn
Here $V[f_2] = 0$, and $V[0]$ is a local maximum for
the barrier separating the two minima.
Define $f=(|f_1| + f_2)/2$ and $\Delta f = f_2 - |f_1|$.   In case
$\Delta f > 0$, $\phi=f_1$ corresponds to a false vacuum with the 
energy density $\ep = V[f_1] = {2\over 3} \lambda f^3 \Delta f >0$,
whereas $\phi=f_2$ corresponds to a true vacuum with a vanishing  energy
density. As we shall discuss in detail below, the positivity of the energy
density $\ep$ plays an important role for the presence of shell structure, but
the vanishing $V[f_2]$ is not essential as we see below.   In a more
general potential  it could be that $V[f_2] > V[f_1]$.

We look for solutions with $\phi$ starting at the origin $r=0$ very close to
$f_1$. There is only one parameter to adjust: $\phi_0 \equiv \phi(0)$. 
The  behavior of a solution near the origin is given by
\beqn
&&\hskip -1.cm
\phi = \phi_0 + \phi_2 r^2 + \cdots ~, 
\hskip 2cm \phi_2 = {1\over 6} \, V'[\phi_0] ~,\cr
&&\hskip -1.cm
p = 1 + p_4 r^4 + \cdots ~, 
\hskip 2.3cm p_4 = - 4\pi G \phi_2^2 ~,  \cr
&&\hskip -1.cm
M = m_3 r^3 + \cdots, 
\hskip 2.7cm  m_3 = {4\pi\over 3} \, V[\phi_0] ~, \cr
&&\hskip -1.cm
H = 1 - 2 G m_3 r^2 + \cdots ~.
\label{origin1}
\eeqn
Given $\phi_0$ the equations determine the behavior of a configuration
uniquely.  For most values of $\phi_0$ the corresponding configurations
are unacceptable.   As $r$ increases, $\phi(r)$ either approaches
0 (the local maximum of $V[\phi]$) after oscillation, or comes back to
cross $f_1$ and continues to decrease.
Other than the two trivial solutions, corresponding to the false and true
vacua, we have found a new type of solution.

There are four parameters in the model, one of which, the gravitational
constant $G= m_{\rm P}^{-2}$, sets the scale.   The other three are
$\lambda$, $f_1$, and $f_2$ or, equivalently, the three dimensionless
quantities $f/m_{\rm P}$, $\Delta f/f$, and $\lambda$.  We have  
explored only a limited region of 
 the parameter space.  The moduli 
space of solutions depends critically on $f/m_{\rm P}$ and  $\Delta f/f$.
The $\lambda$ dependence can be absorbed by rescaling.   
The solution $\phi(r; \lambda)$ to Eqs.\ (\ref{mass1}),
(\ref{scalar2})-(\ref{friction1}) for a given $\lambda$ is related 
to the solution for $\lambda=1$ by 
$\phi(r; \lambda) = \phi( \sqrt{\lambda} r ; 1)$.
We shall see that nontrivial solutions appear as $f/m_{\rm P}$ becomes
small.

If $\phi_0 < f_1$, $\phi(r)$ monotonically
decreases as the radius increases. 
On the way $H(r)$ goes to zero at a finite radius $r_m$.  
$\phi'(r)$ may or may not diverge there.    To describe most general
solutions the metric is written as \cite{Volkov}
\beeq
ds^2 = - Q(\rho) dt^2 + {d\rho^2\over Q(\rho)} 
+ R(\rho)^2 (d\theta^2 + \sin^2 d\varphi^2) ~~,
\label{BGmetric1}
\eneq
which is related to Eq. (\ref{ourmetric1}) by
\begin{equation}
R=r ~~,~~ Q = {H\over p^2} ~~,~~
 {d\rho\over dr}  = \pm {1\over |p|} ~~.
\label{BGmetric2}
\end{equation}
If $Q(\rho)$ tends to zero linearly in $\rho$, the space-time has a
nondegenerate Killing horizon, while if $Q(\rho)$ tends smoothly to a nonzero
value, there is a bag of gold solution studied
by Volkov {\it et al.} \cite{Volkov}
in the context of Einstein-Yang-Mills equations.

As
\beeq
|p(r)| = \exp \left\{ -4\pi G
\int_0^r dr ~ r \phi'^2 \right\} ~~,
\label{BGmetric3}
\eneq
$p$ can vanish at $r_m$ and $Q(\rho)$ can approach a nonzero value there if
$\phi'$ diverges fast enough.  In this case
$dR/d\rho=0$.  In other words, as $\rho$ increases, $R(\rho)$ reaches 
a maximum $R_{\rm max}$ at $\rho(r_m)$ and starts to decrease.  Eventually
$R(\rho)$ becomes zero at $\rho_{\rm max}$. 
This would lead to a bag-of-gold  solution.
It is not clear whether such solutions exist in the scalar model
under discussion \cite{Lechner}.

Suppose instead that $f_1 < \phi_0 <0$ and $\phi_0$ is not too close to
$f_1$.  In the particle analogue, the particle starts to roll down the 
hill under the action of $U[\phi]= - V[\phi]$.  It approaches $\phi=0$, and 
oscillates around it.  In the meantime $H(r)$ crosses zero.  We are not
interested in this type of solution either.

\begin{figure}[tbh]
\centering \leavevmode 
\rotatebox{-90}{
\epsfxsize=10.cm \epsfbox{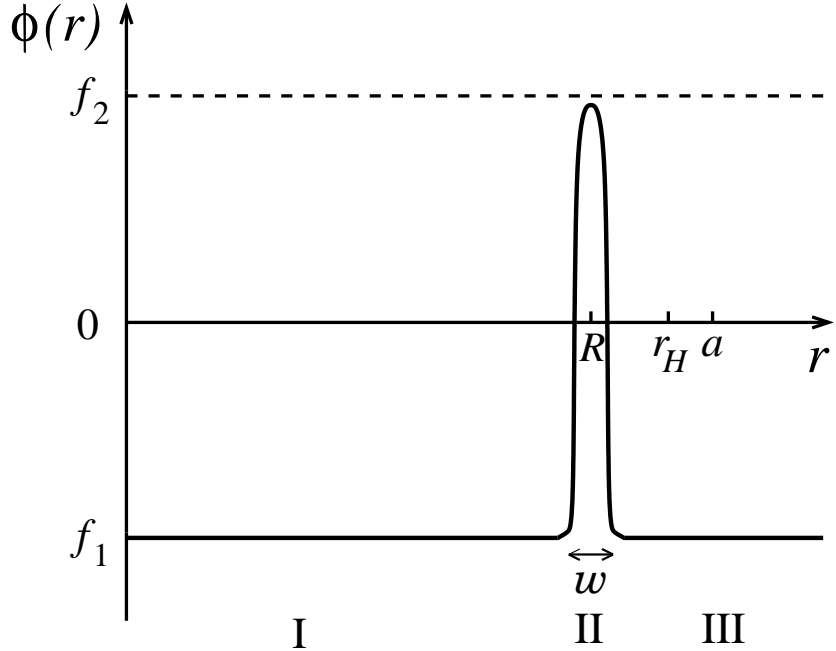}}
\caption{Schematic behavior of $\phi(r)$.}
\label{global-phi}
\end{figure}

\begin{figure}[tbh]
\centering \leavevmode 
\rotatebox{-90}{
\epsfxsize=10.cm \epsfbox{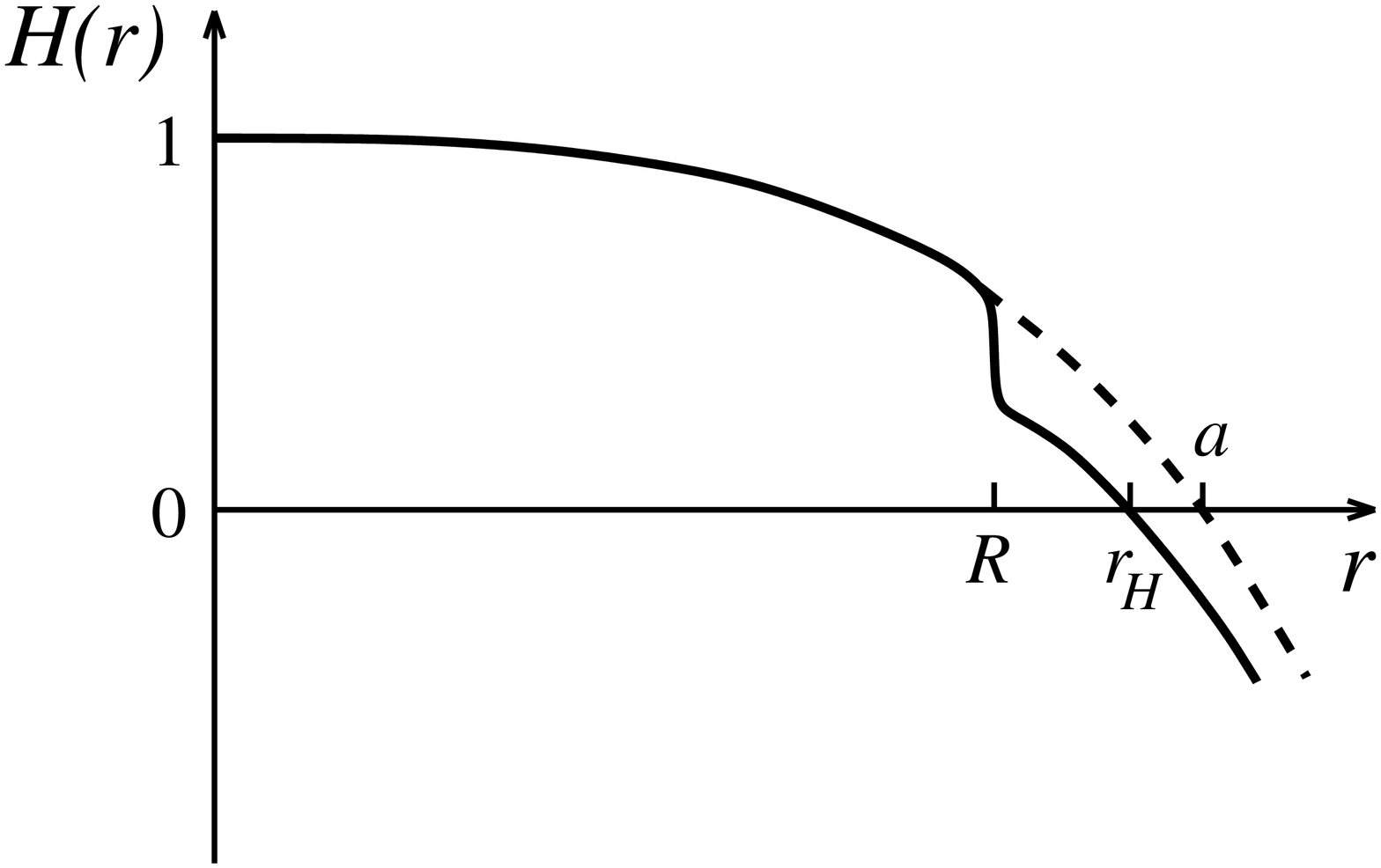}}
\caption{Schematic behavior of  $H(r)$.}
\label{global-H}
\end{figure}

Now suppose that $\phi_0$ is very close to, but still greater than,
$f_1$: $\phi(0) = f_1 + \delta \phi(0)$ with 
$0 < \delta\phi(0)/f \ll 1$.  A schematic of the resulting solution is 
displayed in figures\ \ref{global-phi} and \ref{global-H}.  We divide
space into three regions in the  static coordinates: region I $( 0 \le r <
R_1)$, region II $(R_1 < r < R_2)$, and region III $(R_2 < r)$.  It turns
out that
$\phi(r)$ varies little from 
$f_1$ in regions I and III so that the equation of motion for $\phi$ may be 
linearized in those regions.  In region II the field deviates strongly and the 
full set of nonlinear equations must be solved numerically.  This is the region in which we shall find shell structure.  $H(r)$ deviates from the de
Sitter value significantly.  In region III  the space-time is approximately
de Sitter again.  $H(r)$ crosses zero at $r_H$ where $p(r)$ remains positive.
Consequently $Q(\rho)$ in Eq. (\ref{BGmetric1}) also crosses zero linearly
so that the space-time has a  horizon.

In region I the space-time is approximately de Sitter:
\beeq
T_{00} = \ep \next
H = 1 - {r^2\over a^2} \next a = \sqrt{ {3\over 8\pi G \ep}} \next
p=1 ~.
\label{metric2a}
\eneq
The equation for $\phi(r)$ can be linearized with
$\phi(r) = f_1 + \delta\phi(r)$.  In terms of $z \equiv r^2/a^2$,
\beqn
\Bigg\{ z(1-z) {d^2\over dz^2} 
+ \bigg( {3\over 2} - {5\over 2} z \bigg) {d\over dz}
- {1\over 4} \omega^2 a^2 \Bigg\} \, \delta\phi = 0 ~,
\label{scalar3}
\eeqn
where $\omega^2 = V''[f_1]$.  This is Gauss' hypergeometric equation.
The solution which is regular at $r=0$ is
\beqn
\delta \phi(r) = \delta \phi (0) \cdot
F( \hbox{$\frac{3}{4}$} + i \kappa , \hbox{$\frac{3}{4}$} - i \kappa ,
  \hbox{$\frac{3}{2}$} ; z) ~,
\label{phi1}
\eeqn
where
\beqn
&&\hskip -1.cm
\kappa = \onehalf \sqrt{ \omega^2 a^2 - \hbox{${9\over 4}$} }  \next
\omega^2 a^2 ={9 m_\P^2 \over 8 \pi f \Delta f} 
    \bigg( 1 - {\Delta f\over 2f} \bigg) ~.
\eeqn
We shall soon see that a solution with shell structure appears for
$\omega a \gg 1$ with a particular choice of $\delta \phi(0)$.
The ratio of $\delta \phi'(r)$ to $\delta \phi(r)$ is given by
\beeq
{\delta \phi'(r) \over \delta \phi(r)}
= {4r\over 3 a^2} \Big( \kappa^2 + {9\over 16} \Big)
{F( \hbox{$\frac{7}{4}$} + i \kappa , \hbox{$\frac{7}{4}$} - i \kappa ,
  \hbox{$\frac{5}{2}$} ; z)  \over 
F( \hbox{$\frac{3}{4}$} + i \kappa , \hbox{$\frac{3}{4}$} - i \kappa ,
  \hbox{$\frac{3}{2}$} ; z) } 
\equiv {2r\over a^2} ~ J(z)~.
\label{phi2}
\eneq

The deviation from $f_1$ at the origin, $\delta\phi(0)$, needs to be
very small for an acceptable solution.  The behavior of the
hypergeometric function for $\kappa \gg 1$ and $0<z<1$ is given by 
\cite{Bateman}
\beqn
&&\hskip -1.cm
F(a+ i\kappa, a - i\kappa, c; z) 
\sim {\Gamma(c)\over 2 \sqrt{\pi} } ~ 
\kappa^{{1\over 2} - c} ~
z^{- {c\over 2} + {1\over 4} } (1-z)^{{c\over 2} - {1\over 4} -a } ~
 \exp \left\{ 2\kappa \sin^{-1} \sqrt{z} \right\} .
\label{geometricF}
\eeqn
The ratio $\delta\phi(r)/\delta \phi(0)$ grows exponentially as $r$ increases
like $(4\kappa)^{-1} z^{-1/2} (1-z)^{-1/4} \,\\ 
\exp \left\{ 2\kappa \sin^{-1} \sqrt{z} \right\}$.
At the end of region I, $\delta\phi/|f_1|$ needs
to be very small for the linearization to be valid.  The ratio 
of $F'(z)$ to $F(z)$, $J(z)$ in
Eq. (\ref{phi2}),  is given by
\beeq
J(z) = {\kappa \over \sqrt{ z(1-z) } } \hskip 1cm
 \hbox{for } \kappa \gg 1 ~,~ 0<z<1~.
\label{geometricF2}
\eneq

\begin{figure}[htb]
\centering \leavevmode 
\mbox{
\epsfysize=10.0cm \epsfbox{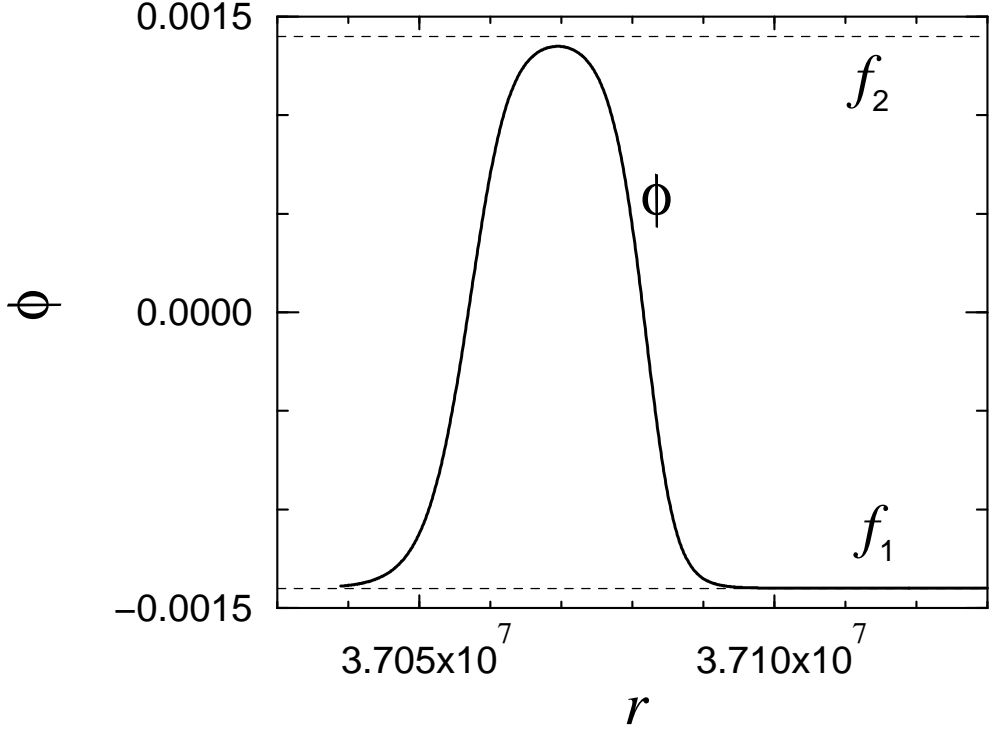}}
\caption{$\phi(r)$ of a solution with $f/m_{\rm P}= 1.40 \times 10^{-3} ,
\Delta f/f = 0.002 , \lambda = 0.01$.  $\phi$ and $r$ are in the units of
$m_{\rm P}$ and $l_{\rm P}$, respectively. The maximum value
of $\phi$ is smaller than $f_2 \sim 0.0014 \cdot m_{\rm P}$.}
\label{figphi}
\end{figure}

In region II, $\phi(r)$ varies substantially and
the nonlinearity of  the equations plays an essential role.
In this region the equations must be solved numerically.  With fine
tuning of the value of $\delta\phi(R_1)$ nontrivial shell solutions
will be found.

The algorithm is the following.  First $\delta\phi(R_1)$ is chosen and
$\delta\phi'(R_1)$ is evaluated by Eqs. (\ref{phi2}) and (\ref{geometricF2}).
To the order in which we work the metric is
$H(R_1) = 1 - (R_1/a)^2$ and $p(R_1)=1$.  With these boundary conditions
Eqs.\ (\ref{scalar2}) and (\ref{mass2}) are numerically solved.

The  behavior of solutions in region II is displayed in figure\ \ref{figphi}.
When the specific values of the input parameters are chosen to be
$\lambda=0.01$, $f/m_\P = 0.002$, and $\Delta f/f = 0.002$, then the output
parameters are $a/l_\P=2.365 \times 10^7$, $\kappa=3.344 \times 10^3$,
$R_1/ l_\P = 1.76104695 \times 10^7$ and $\delta\phi(R_1)/m_\P=10^{-5}$. 
(Here $l_\P$ is the Planck length.)  The field
$\phi(r)$ approaches $f_1$ for $r > R_2$.
In the numerical integration $\delta\phi(R_1)$ is kept fixed while
$R_1$ is varied.  Fine tuning to the ninth digit is necessary.
If $R_1$ is taken to be slightly bigger then $\phi(r)$ starts to deviate
from $f_1$ in the negative direction as $r$ increases.  
If $R_1$ is taken to be slightly smaller then $\phi(r)$ starts 
to deviate from $f_1$ in the positive direction, heading for $f_2$ as
$r$ increases.  With just the right value the space-time becomes nearly de 
Sitter outside the shell.  The value of $\delta\phi$ at the origin ($r=0$) is 
found from Eq. (\ref{phi1}) to be $1.40 \times 10^{-2440}$, which explains why one 
cannot numerically integrate $\phi(r)$ starting from $r=0$.

The behavior of the solution in region III is easily inferred.
  From the numerical integration in region II both $H_2=H(R_2)$
and $p_2 = p(R_2)$ are determined.  In region III the metric can be
written in the form 
\beqn
&&\hskip -1.cm
H(r) = 1 - {2 G \tilde M\over r} - {r^2\over a^2} \cr
&&\hskip -1.cm
p(r) = p_2 ~.
\label{metric2}
\eeqn
Here $\tilde M$ is the mass ascribable to the shell.  Once $T_{00}$
vanishes $H(r)$ must take this form.  The value of $\tilde M$ may be
determined numerically by fitting Eq.\ (\ref{metric2}) just outside
the shell.  
The location of the horizon, $r_H$, is determined by $H(r_H) = 0$.
The field $\phi(r)$ remains very close to $f_1$ in region III. 

At this point due caution is required to continue the solution because 
the static coordinates  defined in Eq. (\ref{ourmetric1}) do not cover the
whole of space-time.  The global structure of the space-time and of the
solutions is  worked out in Sec. 3.4 where different coordinates are
employed.

The solutions we found have shell structure at a radius $R$ whose width $w$ is 
very small.  The radius $R$ is smaller than the horizon radius $r_H$, but is of 
the same order as $a$.  In the next section we shall see that $R$ becomes 
smaller as $f$ becomes smaller, but remains of order $a$ even in the $f \go 0$ 
limit. This implies that the shell structure is cosmic in size.  

An estimate of the order of magnitude of $w$ as well as the conditions necessary
for the existence of the cosmic shell are obtained by a simple
argument.  Return to Eq.\ (\ref{scalar2}),
\beeq
H \phi'' + \left( {2H\over r} + H' \right) \phi'
+ 4\pi G r H \phi'^3 = V'[\phi].
\label{scalar4}
\eneq
In the shell region (region II, $r\sim R$) of typical solutions, $H$
is of order one and drops sharply.   In Eq. (\ref{scalar4}) the $2H/r$ term 
is negligible compared with the $H'$ term.  The remaining terms, $H\phi''$, 
$H'\phi'$, and $V'[\phi]$ are all of the same order of magnitude
so that $f/w^2 \sim \lambda f^3$ or
\beeq
w \sim {1\over \sqrt{\lambda} f} ~.
\label{estimate1}
\eneq
Hence the thickness of the shell is determined by the parameters
in the scalar field potential.  The radius $R$ is smaller than but of the 
same order as $a$.  The solutions exist only if $H'$, which is 
negative, dominates over $4\pi G r H \phi'^2$.  In other words
$w^{-1} > 4\pi G R(f/w)^2$.  Making use of $R\sim a$,  
$a^2 = 9/(16 \pi G \lambda f^3 \Delta f)$ and Eq. (\ref{estimate1}), one finds
\beqn
&&\hskip -1cm 
4\pi \bigg( {f \over m_{\rm P}} \bigg)^2 <
{w\over a} \sim {4\sqrt{\pi}\over 3} {\sqrt{f \Delta f}\over m_{\rm P}} 
\ll 1 \cr
\noalign{\kern 10pt}&&\hskip -1cm 
{\Delta f\over f} > 3 \sqrt{\pi} \bigg( {f \over m_{\rm P}} \bigg)^2 ~.
\label{estimate2}
\eeqn
We shall see in the next section that these relations are satisfied in the 
solutions obtained numerically.

\begin{figure}[htb]
\centering \leavevmode 
\mbox{
\epsfysize=10.0cm \epsfbox{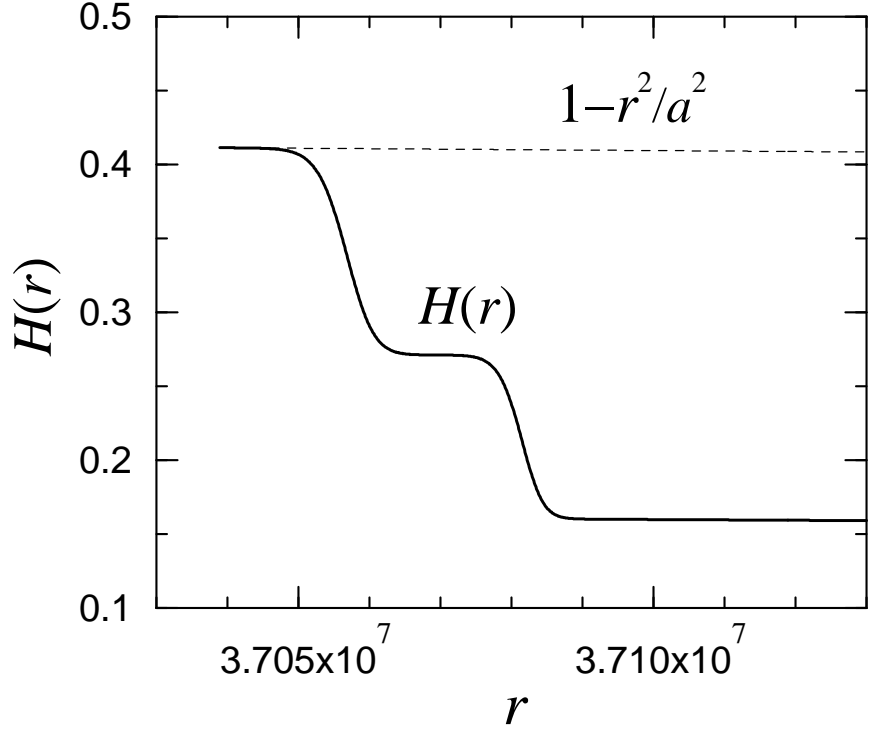}}
\caption{$H(r)$ of a solution with $f/m_{\rm P}= 1.40 \times 10^{-3},
\Delta f/f = 0.002 , \lambda = 0.01$.  $H(r)$ decreases in two steps
in the shell region.  $H(r)$ in the de Sitter space $(=1 - r^2/a^2$) is
plotted in a dotted line.}
\label{figH}
\end{figure}

\begin{figure}[htb]
\centering \leavevmode 
\mbox{
\epsfysize=10.0cm \epsfbox{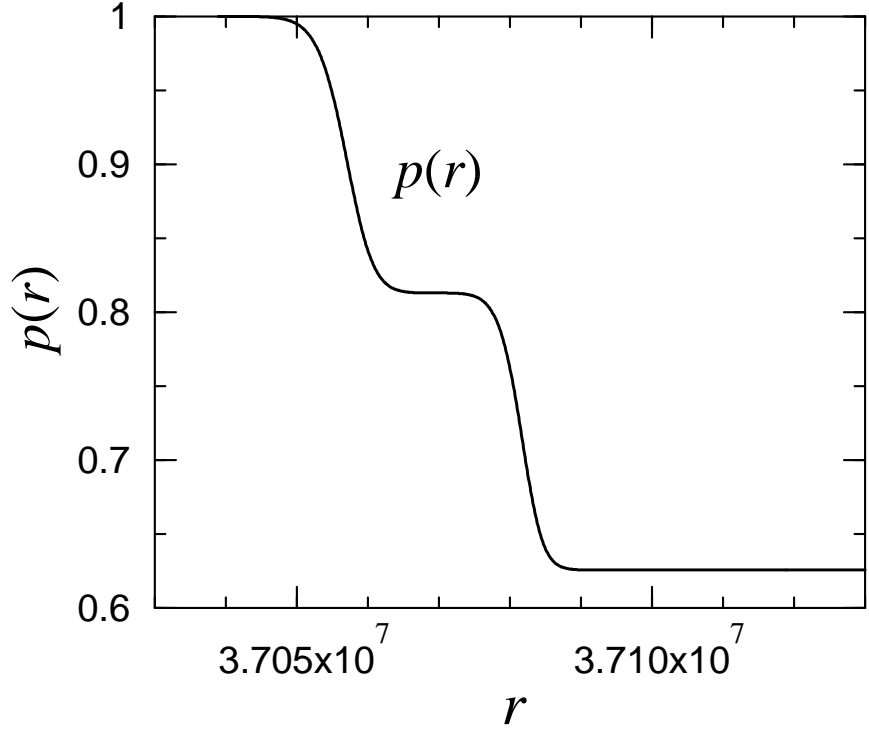}}
\caption{$p(r)$ of a solution with $f/m_{\rm P}= 1.40 \times 10^{-3},
\Delta f/f = 0.002 , \lambda = 0.01$. $p(r)$ decreases in two steps,
from 1 to 0.6256, in the shell region.}
\label{figp}
\end{figure}

\section{Numerical Analysis of the Nonlinear Regime}

\hspace{0.2in} In the preceding section we solved the linearized field equation for $\phi(r)$ in 
regions I and III where the deviation from $f_1$ is small, and we sketched the 
behavior in the nonlinear region II where the shell structure appears.
In this section we present numerical results for region II. As discussed in 
the preciding section the boundary between regions I and II,
located at the matching radius $R_1$, is rather arbitrary,
subject only to the condition that the linearization is accurate 
up to that radius.  Precise tuning is necessary for the pair $R_1$ and
$\delta\phi(R_1)$ to obtain a solution to all the equations.  Technically
it is easier to keep $\delta\phi(R_1)$ fixed and adjust the matching
radius
$R_1$.  
If $R_1$ is chosen too small $\phi$ comes back toward, but cannot reach,  
$f_1$.  What happens is that $\phi$ eventually oscillates around $\phi=0$ as $r$ 
increases.  If $R_1$ is chosen too large $\phi$ comes back to $f_1$ at finite 
$r$ to roll over it and continues to decrease.  
For each  sufficiently small value of $\delta\phi$ chosen as the matching value 
there exists a desired solution with $R_1=R_c$.  The critical value $R_c$ can be 
determined numerically to arbitrary accuracy.

\begin{figure}[tbh]
\centering \leavevmode 
\mbox{
\epsfysize=10.0cm \epsfbox{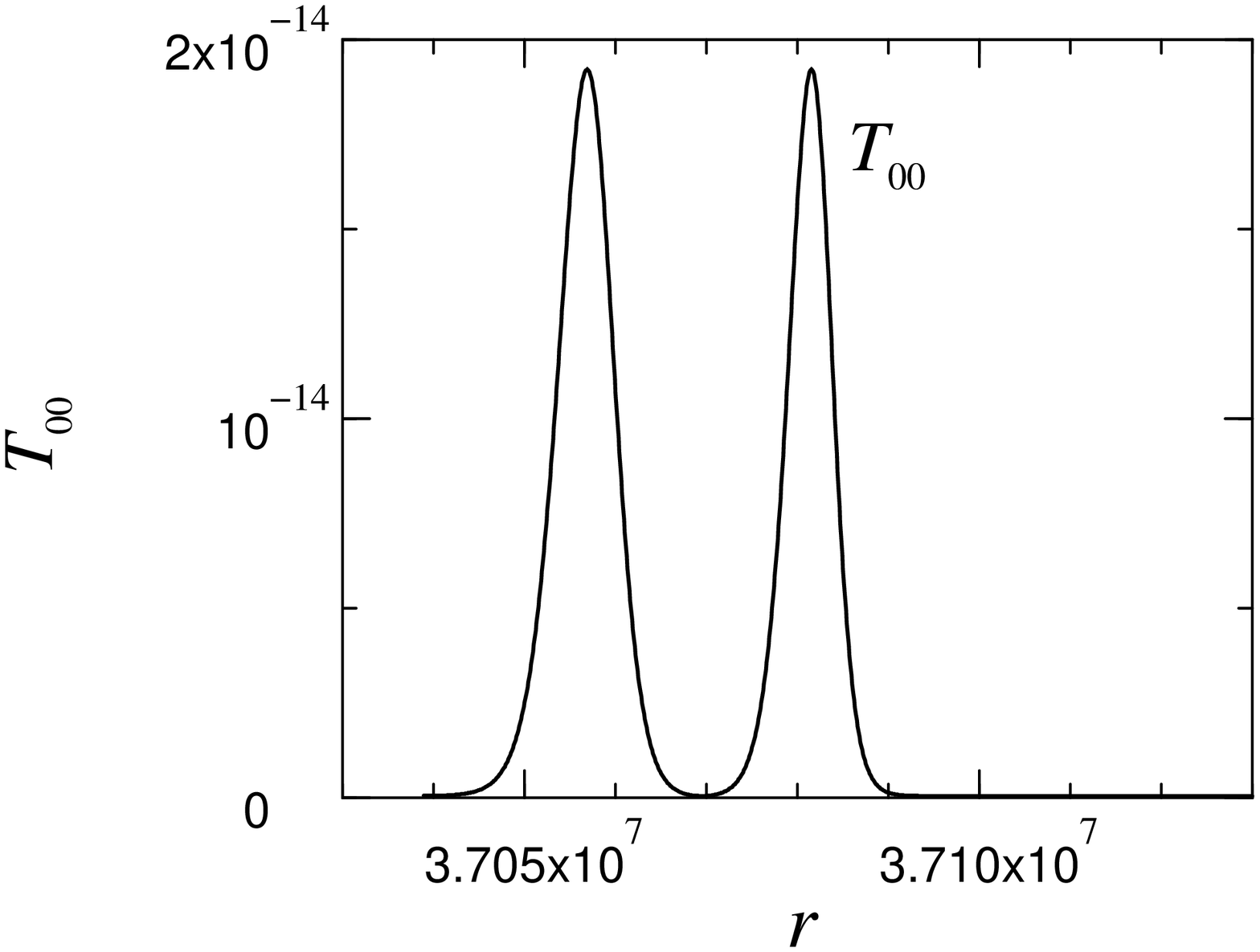}}
\caption{The energy density $T_{00}$ for a shell solution with 
$f/m_{\rm P}= 1.40 \times 10^{-3}$,
$\Delta f/f = 0.002$, $\lambda = 0.01$.  $T_{00}$ and $r$ are in the units
of $m_{\rm P}^4$ and $l_{\rm P}$, respectively.  The energy density is
localized in the two shells over the de Sitter background.}
\label{T00}
\end{figure}

\begin{figure}[htb]
\centering \leavevmode 
\mbox{
\epsfysize=10.0cm \epsfbox{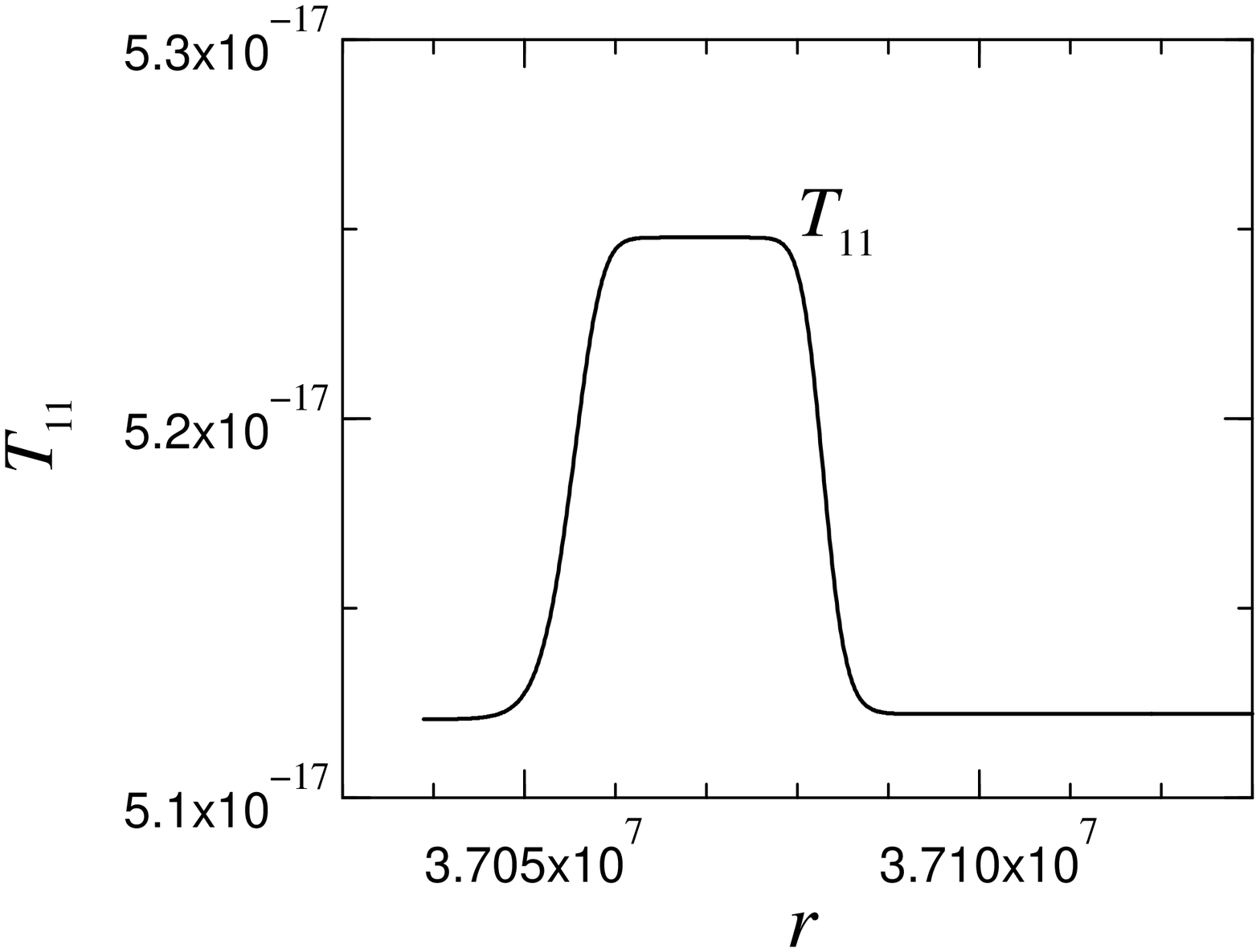}}
\caption{The radial pressure $T_{11}$ for a shell solution with 
$f/m_{\rm P}= 1.40 \times 10^{-3}$,
$\Delta f/f = 0.002$, $\lambda = 0.01$.  $T_{11}$ and $r$ are in the units
of $m_{\rm P}^4$ and $l_{\rm P}$, respectively.  The pressure is higher
and constant  between the two shells.}
\label{T11}
\end{figure}

\begin{figure}[htb]
\centering \leavevmode 
\mbox{
\epsfysize=10.0cm \epsfbox{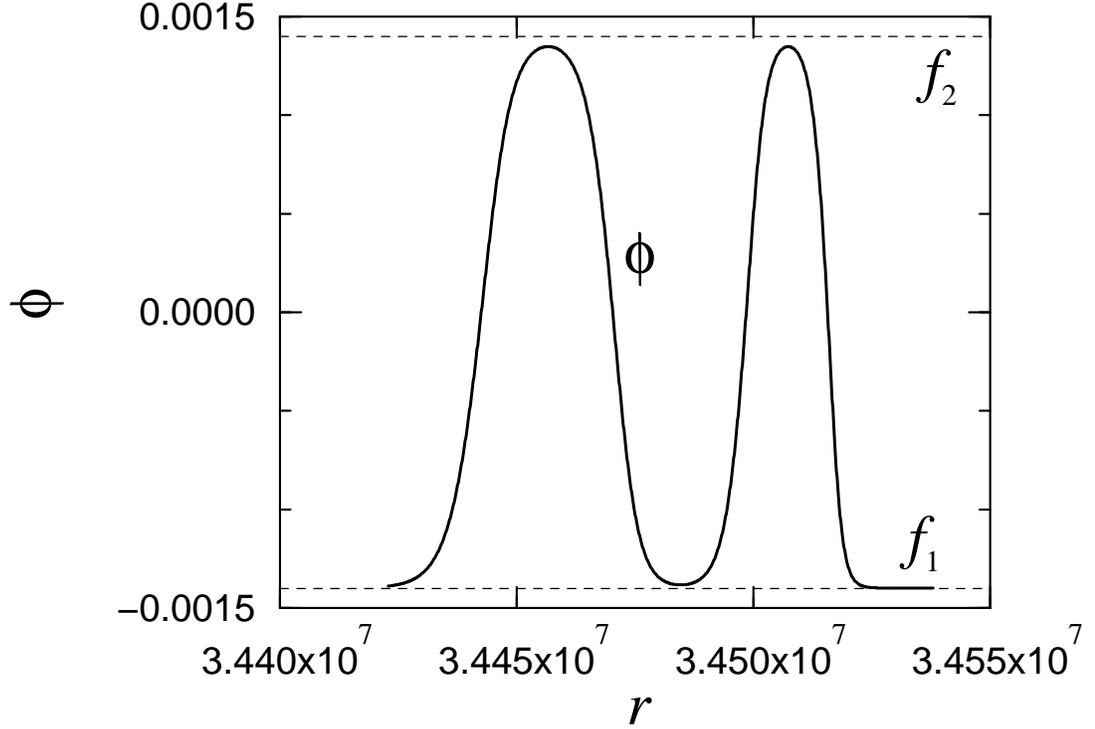}}
\caption{A solution with  double bounces  with  $f/m_{\rm P}= 1.40 \times
10^{-3} , \Delta f/f = 0.002 , \lambda = 0.01$.  A solution with multiple
bounces appears at a smaller shell radius than a solution with a single
bounce.}
\label{double}
\end{figure}

\begin{figure}[htb]
\centering \leavevmode 
\mbox{
\epsfxsize=12.0cm \epsfbox{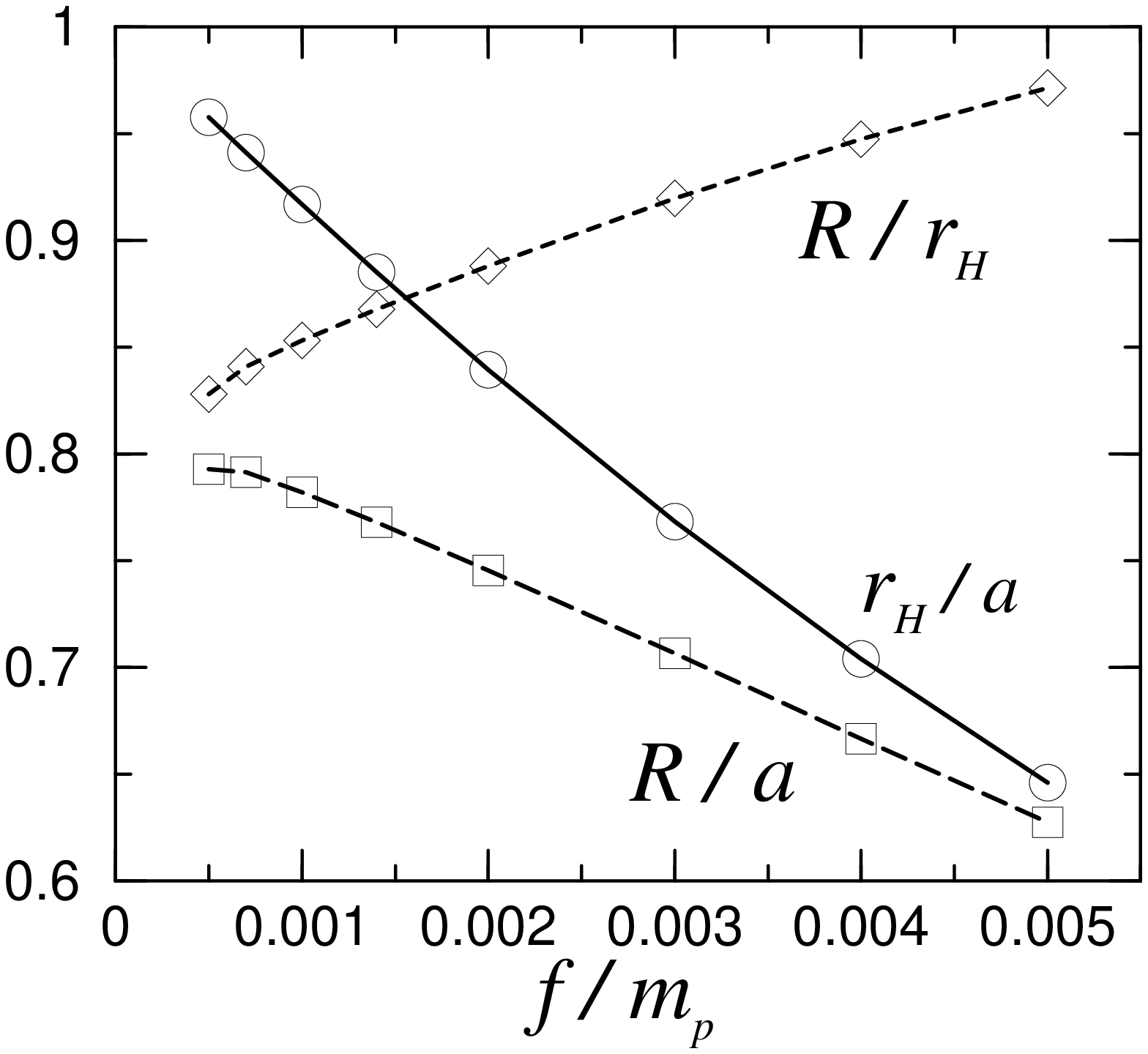}}
\caption{The $f$-dependence of the shell solutions.
$\lambda=0.01$ and $\Delta f/f=0.002$ are fixed.  $R$, $r_H$, and $a$ are
the radius of the shell, the horizon length where $H(r)$ vanishes, and
the horizon length in the de Sitter space, respectively.  The ratios of
various pairs are plotted.  Circles, squares, and diamonds correspond to
data points obtained.
}
\label{fdependence}
\end{figure}

One example of such a solution is displayed in figure\ \ref{figphi} for the
parameters $f/m_\P =1.40 \times 10^{-3}$, $\Delta f/f =0.002$, 
$\lambda=0.01$ and $\delta\phi(R_1)=1.0\times10^{-5}$.  The matching
radius 
$R_1$ is fine-tuned to ten digits: $R_1/l_\P \sim 3.7038855228 \times 10^7$.  
The shell is very thin compared to the radius of the shell, lying in the region
$3.704 \times 10^7 \leq r/l_\P \leq 3.710 \times 10^7$.

In the shell region both $H(r)$ and $p(r)$ decrease in a two-step fashion. 
See figures \ref{figH} and \ref{figp}. Inside and outside the shell $H(r)$
is given by Eqs. (\ref{metric2a}) and (\ref{metric2}), whereas $p(r)$ assumes
the constant values 1 and 0.6256, respectively.

The change is induced by the non-vanishing 
energy-momentum tensor components $T_{00}=- T_{22}=-T_{33}$ and $T_{11}$.
They are 
displayed in figures\ \ref{T00} and \ref{T11}.  The energy density 
$T_{00}$ has two sharp peaks associated with the rapid variation of $\phi$.  
The radial pressure $T_{11}$, on the other hand, steps up quickly, remains 
constant within the shell, and then steps down again.  The value of
$T_{11}$ is very small ($\sim 5 \times 10^{-17} m_{\rm P}^4$)
compared with the maximum value of 
$T_{00} (\sim 2 \times 10^{-14} m_{\rm P}^4$).  The contributions of the 
kinetic energy, $\onehalf H \phi'^2$, and potential energy, $V[\phi]$,
almost  cancel each other.

In this example $\phi(r)$ makes a bounce transition 
$f_1 \go f_2 \go f_1$ once. As the numerical value of $f$ becomes smaller
than $0.002 m_{\rm P}$ a new type of solution with  double bounces
 $f_1 \go  f_2 \go f_1 \go f_2 \go f_1$ emerge.  One example  is  displayed
in figure\ \ref{double}.

The spectrum of the shell solutions depends on the parameter $f$.
In figure\ \ref{fdependence} $R/a$, $r_H/a$, and $R/r_H$ are plotted as
functions of $f$.  (Finding the solution numerically becomes extremely difficult 
as $f$ becomes smaller.)  There are several features to be noted.  First,
with  fixed values of $\lambda$ and $\Delta f /f$ solutions exist only for
$f < f_{\rm max}$.  For example, with $\lambda=0.01$ and $\Delta f /f = 0.002$ 
we find that $f_{\rm max} \sim 0.006$.  Second, as $f$ goes to 0, $r_H/a \go 1$ 
and $R/a \go 0.8$.  The relative size $R/a$ of the shell remains of order
one  even in the $f \go 0$ limit.  Third, as $f$ decreases shell solutions
with multiple bounces become possible. We cannot determine how many times $\phi$ 
can bounce inside the horizon because the numerical evaluation
becomes extremely difficult when $f/m_\P < 0.001$.

In the potential we are analyzing, $\phi=f_1$ and $\phi=f_2$
correspond to the false and true vacua, respectively.  For the 
existence of  shell solutions the fact $V[f_1] > V[f_2]$ is not
crucial, however.   In the example displayed in figure\ \ref{figphi}, for
instance,
$\phi(r)$ swings from $\phi(0)\sim f_1$ to $\phi_{\rm max} < f_2$,
and returns to $f_1$.  Numerically  $V[\phi_{\rm max}] > V[\phi(0)]$.
As we are solving classical differential equations, only the form of the
potential
$V[\phi]$ between $f_1$ and $\phi_{\rm max}$ is relevant.  The form
of the potential for $\phi > \phi_{\rm max}$ does not matter.  The
second minimum can be higher than the first one; 
$V[f_1] < V[f_2] < V[\phi_{\rm max}]$.  
Before concluding this section we would like to add that we have found
no solution in which $\phi$ makes a transition from $f_1$ to $f_2$
as $r$ varies from 0 to $\infty$.

\section{Global Structure of the Space-Time and Solutions}

\hspace{0.2in} In the preceding section we found novel solutions to a theory with a scalar field 
coupled to gravity in static coordinates.  The static coordinates in 
Eq. (\ref{ourmetric1}), however, do not cover all of space-time.  In this
section we  construct coordinates which allow for an extension of the
solution to the full  space-time manifold.  These coordinates are smooth
across the horizon. First we illustrate the construction with de Sitter
space, and then consider the more general case which is applied to the
shell solution.

The $R^1 \times S^3$ metric of the de Sitter space is given by
\beeq
ds^2 = a^2 \left[ -  d\tau^2 + \cosh^2 \tau 
\left( d\chi^2 + \sin^2 \chi \, d\Omega^2 \right) \right] ~.
\label{dS2}
\eneq
One may regard the de Sitter space as a hypersurface in the 
five-dimensional Minkowski spacetime constrained by the condition
\beeq
y_1^2  + y_2^2 + y_3^2 + y_4^2 = a^2 + y_0^2 ~.
\label{dS3}
\eneq
The metric (\ref{dS2}) and hypersurface (\ref{dS3}) cover the entire de Sitter
space, whereas the static metric covers only half of the space.  
Furthermore, the static metric has a coordinate singularity at $r=a$.

The relationship among these coordinate systems are easily found.
The $S^3$ metric (\ref{dS2}) can be transformed to a metric conformal to
the static Einstein universe,
\beqn
\hskip -1cm
&&
ds^2 = {a^2 \over \cos^2 \eta} 
 \left[ -  d\eta^2 
    +  d\chi^2 + \sin^2 \chi \, d\Omega^2 \right] \cr
\noalign{\kern 5pt}
\hskip -1cm
&&
\eta = {\pi\over 2} - 2 \tan^{-1} (e^{-\tau}) ~,~~ 
- {\pi\over 2} \le \eta \le {\pi\over 2}.
\label{dS4}
\eeqn
Suppressing $S^2$, or $\theta$ and $\phi$ variables, one can map the whole de
Sitter space to a square region in the $\chi$-$\eta$ coordinates.  Null 
geodesics
are given by straight lines at 45 degree angles.

The static and hypersurface coordinates are related by
\beqn
\hskip -1cm &&
y_0 = \cases{
\sqrt{a^2 - r^2} \, \sinh \myfrac{t}{a} &for $r<a$  \cr
\sqrt{r^2 - a^2} \, \cosh \myfrac{t}{a}  &for $r>a$   \cr}  \cr
\hskip -1cm &&
y_1 = r \cos \theta \cr
\hskip -1cm &&
y_2 = r \sin \theta \cos\phi \cr
\hskip -1cm &&
y_3 = r \sin \theta \sin\phi \cr
\hskip -1cm &&
y_4 = \cases{
\sqrt{a^2 - r^2} \, \cosh \myfrac{t}{a} &for $r<a$ \cr
\sqrt{r^2 - a^2} \, \sinh \myfrac{t}{a} &for $r>a$ \cr}
\label{connect1}
\eeqn

\begin{figure}[tbh]
\centering \leavevmode 
\rotatebox{-90}{
\epsfxsize=13.cm \epsfbox{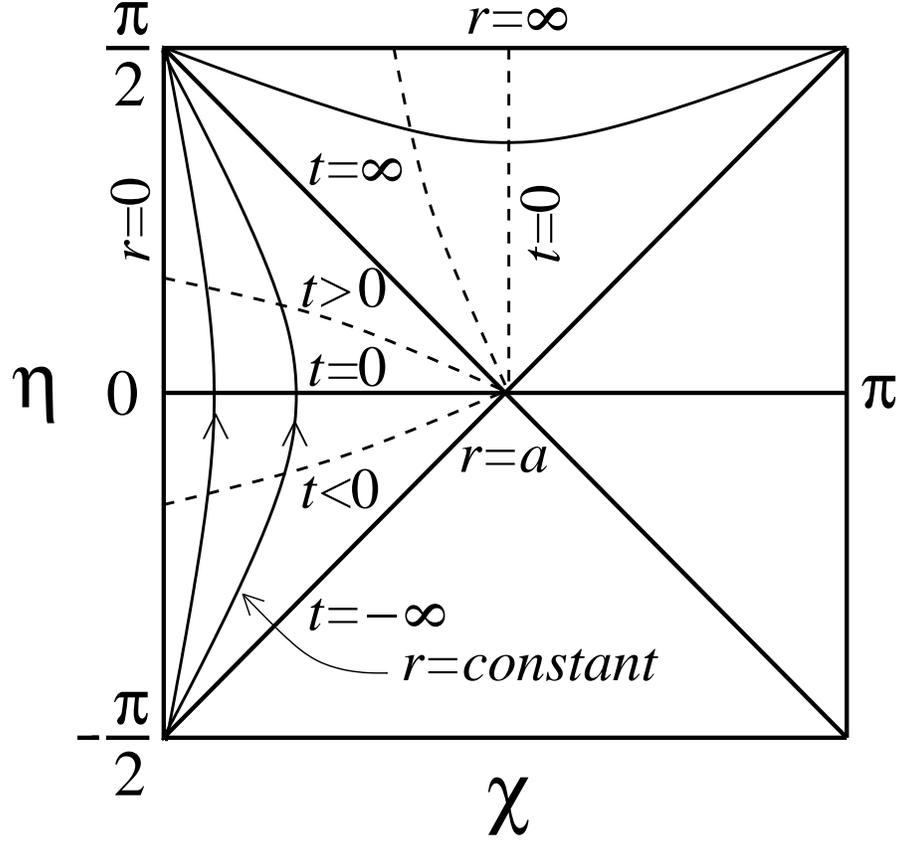}}
\caption{Penrose diagram of the de Sitter space.}
\label{Penrose1}
\end{figure}

Similarly, Eqs. (\ref{dS3}) and (\ref{dS4}) are related by
\beqn
\hskip -1cm &&
y_0 = a \tan \eta   \cr
\hskip -1cm &&
y_1 = a \sec\eta \, \sin\chi \cos \theta \cr
\hskip -1cm &&
y_2 = a \sec\eta \, \sin\chi \sin \theta \cos\phi \cr
\hskip -1cm &&
y_3 = a \sec\eta \, \sin\chi \sin \theta \sin\phi \cr
\hskip -1cm &&
y_4 = a \sec\eta \, \cos\chi \, .  
\label{connect2}
\eeqn
The region inside the cosmological horizon ($r<a$, $-\infty < t < \infty$) in 
the
static metric corresponds to the left quadrant in the conformal metric
($0 \le \chi\le \onehalf \pi$, $|\eta| < \pi/2 - \chi$) with
the relations
\beqn
{r\over a} &=& {\sin\chi\over \cos\eta} \cr
\noalign{\kern 10pt}
{t\over a} &=& {1\over 2} \ln
 {\cos\chi +\sin\eta\over \cos\chi -\sin\eta}  ~.
\label{connect3}
\eeqn
Similarly, the region outside the cosmological horizon ($r>a$, $-\infty < t <
\infty$) in the static metric corresponds to the upper quadrant in the conformal
metric ($0 \le \chi\le  \pi$, $|\eta| > \chi - \onehalf \pi$) with
the relations
\beqn
{r\over a} &=& {\sin\chi\over \cos\eta} \cr
\noalign{\kern 10pt}
{t \over a} &=& 
{1\over 2} \ln { \sin\eta + \cos\chi\over \sin\eta - \cos\chi}~.
\label{connect4}
\eeqn
See figure\ \ref{Penrose1}.

For the general static metric given by Eq. (\ref{ourmetric1}), 
with $H=H(r)$ and 
$p=p(r)$, the construction of the conformal coordinates analogous to Eq. (\ref{dS4})
proceeds as follows.  We suppose that $H(r)$ has a single zero at $r_H$
whereas $p(r)>0$.  The new radial coordinate is defined by
\beeq
r_*(r) = \cases{
\mybig\int_0^r dr' \, \myfrac{p(r')}{H(r')} &for $r< r_H$\cr
\mybig\int_\infty^r dr' \, \myfrac{p(r')}{H(r')} &for $r> r_H$. \cr}
\label{tortoise1}
\eneq
It has a logarithmic singularity at $r_H$, diverging there as
$r_* \sim - \onehalf b\ln |r-r_H|$ where $b= -2 p(r_H)/H'(r_H) >0$.  New
coordinates are introduced by
\beqn
\tan u &=& + e^{+t/b} e^{- r_*/b} \cr
\noalign{\kern 10pt}
\tan v &=& \mp e^{-t/b} e^{- r_*/b} ~.
\label{metric3}
\eeqn

The upper sign is for $r<r_H$ and the lower sign is for $r>r_H$.
The $u$-$v$ coordinates are related to the $U$-$V$ coordinates in the
Kruskal-Szekeres (KS) coordinate system \cite{K-S} for the Schwarschild 
solution, and to the Gibbons-Hawking (GH) coordinate system \cite{G-H} for de 
Sitter space.  The connection between the KS and GH coordinate systems has 
already been discussed in \cite{Guth}.
Essentially $\tan u = U+V$ and $\tan v = V-U$.  
The metric becomes 
\beqn
ds^2 &=& - {4 b^2 F(u,v)\over \cos^2(u+v)} dudv
   + r^2 d {\Omega}^2 \cr
\noalign{\kern 10pt}
F(u,v) &=&{H\over 4 p^2} \Big( 1 - \tan u \tan v \Big)
                          \Big( 1 - \cot u \cot v \Big) \cr
\noalign{\kern 10pt}
&=& {H\over 2 p^2}  \Big( 1 \pm \cosh {2r_*\over b} \Big)
\hskip 1cm {\rm for} \left\{ 
     \matrix{r \le r_H\cr r \ge r_H\cr}  \right. ~.
\label{metric4}
\eeqn
The static metric covers the region interior to the bounding lines $u=0$,
$u+v=\onehalf \pi$, and $u - v = \onehalf \pi$. 
The horizon in the static metric, $r=r_H$,
corresponds to the single point $u=v=0$.  The function $F$ is non-vanishing and 
finite there as $e^{2r_*/b} \sim 1/|r-r_H|$.  In the
$u$-$v$ coordinates the metric is regular on $u=0$ so that the extension to the
region $u<0$ can be made, whereas the static metric covers only half of the 
space.

Applied to the de Sitter space we find
\beeq
r_H=b=a ~,~~e^{-r_*/b} = \left| {a -r\over a+ r} \right|^{1/2} ~,~~
F = 1~,
\label{connect5}
\eneq
and the metric (\ref{dS4}) is recovered by
\beeq
\eta = u + v ~,~~ \chi = v - u + {\pi\over 2}  ~.
\label{connect6}
\eneq

The shell solution found in the preceeding sections can be extended to 
the entire space-time.  In the region III ($R_2 < r $) defined  in the
static metric $H(r)$ is given by Eq. (\ref{metric2}), and 
the location of the horizon is determined by $H(r_H)=0$.  In region III,
$\delta \phi(r)$ is very small so that its equation of motion  can
be linearized. We divide region III into two;  region IIIa ($R_2 < r <
2r_H - R_2$) and  region IIIb ($r > 2 r_H - R_2$).   
In region IIIa, $H(r)$ can be approximated by
\beeq
H(r) \sim A \bigg(  1 - {r^2\over r_H^2} \bigg) ~,
\label{metric6}
\eneq
where
\beeq
A = {1\over 2} \bigg( {3r_H^2\over a^2} - 1 \bigg) ~.
\eneq
The coefficient $A$ has been chosen such that both Eqs. (\ref{metric2})
and (\ref{metric6}) have the same slope $H'(r_H)$ at the horizon.  In the 
examples described in Sec. 3.3, errors caused by Eq. (\ref{metric6}) are less than 
15\% in region IIIa.  Now we write the linearized version of
Eq. (\ref{scalar2}) in  terms of $y = 1 - (r^2/r_H^2)$,
\beqn
&&\hskip -1.cm
\Bigg\{ y(1-y) {d^2\over dy^2} 
+ \bigg( 1 - {5\over 2} y \bigg) {d\over dy}
-  \bigg( \tilde \kappa^2 + {9\over 16} \bigg) \Bigg\} \, \delta\phi = 0  \cr
\noalign{\kern 5pt}
&&\hskip -1.cm
A \bigg( \tilde \kappa^2 + {9\over 16} \bigg)
 = \kappa^2 + {9\over 16} = {1\over 4} \omega^2 a^2. 
\label{scalar5}
\eeqn
Since $\delta\phi(r)$ must be regular at $r=r_H$ ($y=0$), the solution is
\beeq
\delta\phi(r) = \delta\phi(r_H) \cdot
F( \hbox{$\frac{3}{4}$} + i\tilde\kappa, \hbox{$\frac{3}{4}$} - i\tilde\kappa,
 1; y)  ~.
\label{phi3}
\eneq
The normalization $\delta\phi(r_H)$ must be such that $\delta\phi(r)$ matches at
$r=R_2$ with the value determined by numerical integration in region II.
Essentially $\delta\phi(r)$ decreases exponentially when $R_2 < r < r_H$.

Near $r=R_2$ the hypergeometric function behaves as 
\beeq
F( \hbox{$\frac{3}{4}$} + i\tilde\kappa, 
   \hbox{$\frac{3}{4}$} - i\tilde\kappa, 1; y)
\sim {1\over 2 \sqrt{\pi \tilde\kappa} } \,
   y^{-1/4} (1-y)^{-1/2} \, 
  \exp \Big\{ 2 \tilde\kappa \sin^{-1} \sqrt{y} \Big\} 
\label{geometricF3}
\eneq
so that
\beeq
{\delta \phi'(r) \over \delta \phi(r)} 
= - {2r\over r_H^2} \cdot {\tilde\kappa\over \sqrt{y(1-y)} }  ~.
\label{phi4}
\eneq
The value of the right-hand side of Eq. (\ref{phi4}) at $r=R_2$ can be compared with the
value obtained by direct numerical integration in region II.  In one example
with $f/m_{\rm P}= 2\times 10^{-3}, \Delta f/f= 2\times 10^{-3}, 
\lambda= 0.01$, $R_2 = 1.765 \times 10^7, a = 2.365 \times 10^7, 
r_H = 1.985 \times 10^7$, $\kappa = 3344$, and $\tilde\kappa = 4481$.
The numerical value for $\delta \phi'(R_2) / \delta \phi(R_2)$ is
$-0.000848$, whereas the value from Eq. (\ref{phi4}) is $-0.000985$.  
With the uncertainty in the value of $\tilde\kappa$ caused by the
approximation (\ref{metric6}) taken into account, one may conclude that the
agreement is rather good.  To gauge the difficulty of determining the
solution numerically for all values of $r$, we note that 
$\delta \phi(r_H) \sim 10^{-1850} \cdot \delta\phi(R_2)$.

\begin{figure}[tbh]
\centering \leavevmode 
\rotatebox{-90}{
\epsfxsize=13.cm \epsfbox{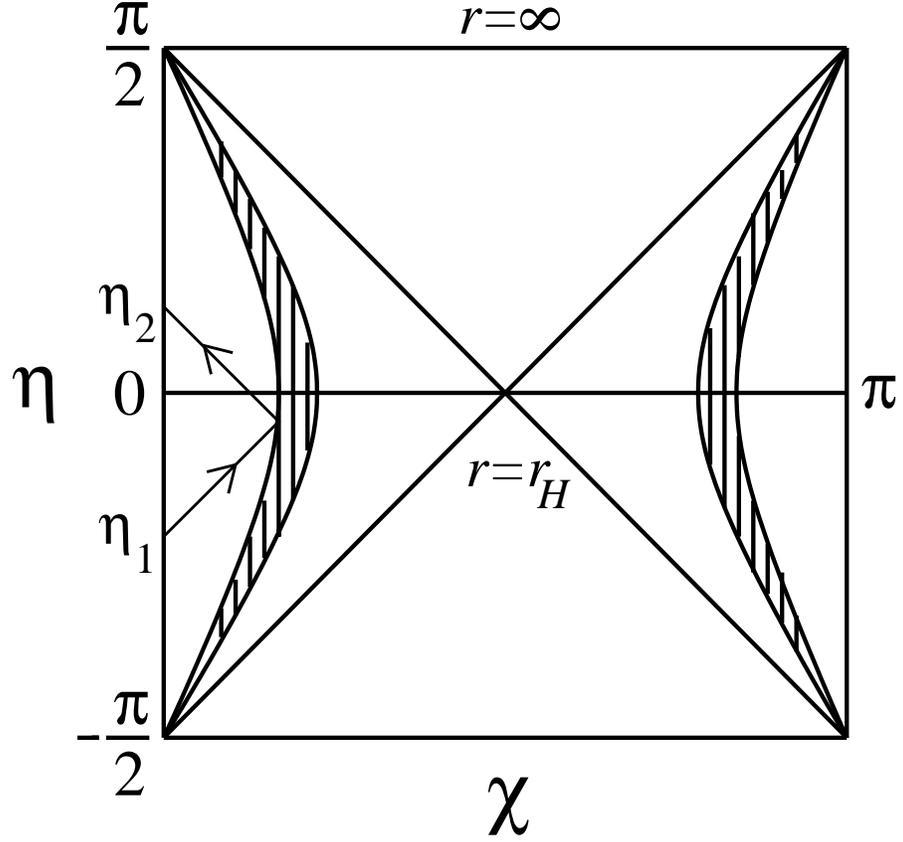}}
\caption{Global structure of the shell solution in the conformal coordinates.
The mirror image appears in the other hemisphere.}
\label{Penrose2}
\end{figure}

The most important observation is that the solution $\phi(r)$ is regular at
the horizon, $r=r_H$, though the static metric $(t,r,\theta,\phi)$ is not:
$r=r_H$ in the static metric is a coordinate singularity.  As seen above,
$r=r_H$ corresponds to $u=v=0$ in the conformal metric (\ref{metric4})
which is non-singular.   Near the horizon
$\tan u \tan v \sim r - r_H$ so that
\beeq
\phi = f_1 + \delta \phi(r_H) 
\bigg\{ 1 - {2\over r_H} \Big( \tilde \kappa^2 + {9\over 16} \Big)
  \, uv + \cdots  \bigg\} ~.
\label{phi5}
\eneq
Furthermore, as $\tan u \tan v = e^{-2 r_*(r)/b}$, the static solution
$\phi(u,v)$ depends only on $\tan u \tan v$ in the entire space-time.  This 
implies that
\beeq
\phi(-u, -v) = \phi(u, v) ~.
\label{phi6}
\eneq
In the $\eta$-$\chi$ coordinates ($S^3$ metric) defined by Eq. (\ref{connect6})
\beeq
\phi(\eta, \chi) = \phi(-\eta, \pi - \chi) = \phi(\eta, \pi-\chi) ~.
\label{phi7}
\eneq
This relation shows that there is a mirror image of the shell structure
in the other hemisphere in the $S^3$ metric.  See figure\ \ref{Penrose2}.  The shaded
regions in the figure represent the shells.  The size of the shells 
is invariant, as is obvious in the static coordinates.  In the 
$S^3$ or conformal metric an observer at the center ($\chi = 0 $ or $\pi$)
sends a light signal toward the shell.  Another observer at the shell,
upon receiving the signal from the center, sent a signal back to
the observer at the center.  The total elapsed time, in terms of the
proper time of the observer at the center $a\tau$ where $\tau$
is related to $\eta$ by Eq. (\ref{dS4}) at $\chi=0$, is 
$a \ln (a+R)/(a-R)$, independent of when the signal is emitted  
initially, as straightforward manipulations show.

Finally, in the region outside the horizon in the static metric, or
in the upper or lower quadrant in the Penrose diagram,  $\delta\phi(r)$
decays as $r^{-3/2}$.    The proof goes as follows.
For $r \gg r_H$,
$H \sim 1 - (r^2/a^2)$, so that $\delta\phi(r)$ satisfies Eq.\ 
(\ref{scalar3}).  Hence it  is a linear combination of 
$u_1 = F(\frac{3}{4} + i \kappa , \frac{3}{4} - i \kappa ,\frac{3}{2} ; z)$
and $u_2 = z^{-1/2} 
F(\frac{1}{4} + i \kappa , \frac{1}{4} - i \kappa ,\frac{1}{2} ; z)$.
These can be written as
\beqn
&&\hskip -1cm
\left[ \matrix{u_1 \cr u_2\cr} \right] =
\left[ \matrix{\onehalf \Gamma( \hbox{${3\over 4}$} - i\kappa )^{-2} \cr
    \Gamma( \hbox{${1\over 4}$} - i\kappa )^{-2} \cr} \right]
{\sqrt{\pi} \,  \Gamma (-2i\kappa) \over (-z)^{(3/4) + i\kappa}}
~ F \Big( \hbox{$\frac{3}{4}$} + i\kappa, 
   \hbox{$\frac{1}{4}$} + i\kappa, 1 + 2i\kappa ; {1\over z} \Big)  \cr
\noalign{\kern 12pt}
&&\hskip .1cm
+ \left[ \matrix{\onehalf \Gamma( \hbox{${3\over 4}$} + i\kappa )^{-2} \cr
    \Gamma( \hbox{${1\over 4}$} + i\kappa )^{-2} \cr} \right]
{\sqrt{\pi} \,  \Gamma (+ 2i\kappa) \over (-z)^{(3/4) - i\kappa}}
~ F \Big( \hbox{$\frac{3}{4}$} - i\kappa, 
   \hbox{$\frac{1}{4}$} - i\kappa, 1 - 2i\kappa ; {1\over z} \Big) ~.  
\label{geometricG4}
\eeqn
From this one can write
\beeq
\phi \sim f_1 + A \left( {a\over r} \right)^{3/2} \,
\sin \Big( 2\kappa \ln {r\over a} + \delta \Big)
\label{phi8}
\eneq
when $r \gg r_H$.

\section{False Vacuum Decay}

\hspace{0.2in} If the region inside the shell is in a false vacuum state one should consider 
its quantum decay to the true vacuum state.
The lifetime of the false vacuum may be determined semiclassically using the
methods of Coleman {\it et al.} without \cite{without} or with \cite{Coleman}
gravity taken into account.  The rate per unit volume for making a transition
from the false vacuum to the true vacuum is expressed as
\begin{equation}
\frac{\Gamma}{V} = Ae^{-B/\hbar}[1+{\cal O}(\hbar)]
\end{equation}
where Planck's constant has been used here to emphasize the semiclassical 
nature of the tunneling rate.  For the potential being used in this paper
we find the O(4), Euclidean space, bounce action (neglecting gravity) to be
\begin{equation}
B_0 = \frac{36\pi^2}{\lambda}\left(\frac{f}{\Delta f}\right)^3
\end{equation}
where the radius of the critical size bubble which nucleates the transition is
\begin{equation}
\rho_c=\frac{3}{\Delta f} \sqrt{\frac{2}{\lambda}} \, .
\end{equation}
For any sensible estimate of the coefficient $A$ the lifetime of the false
vacuum will exceed the present age of the universe when the condition
\begin{equation}
\lambda \left(\frac{\Delta f}{f}\right)^3 < 1
\end{equation}
is satisfied.  This calculation is based on the thin wall approximation, which 
is valid when the critical radius is large compared to the coherence length of 
the potential, namely, $1/\sqrt{|V^{\prime\prime}|}$.  This condition translates 
into
\begin{equation}
\Delta f \ll 6f \, .
\end{equation}
With gravity included the bounce action is
\begin{equation}
B = \frac{B_0}{\left[1+(\rho_c/2R)^2 \right]^2} \, .
\end{equation}
Gravitational effects are negligible when $\rho_c < R$.  When this condition
is not fulfilled the shell radius is too small to accommodate even a single 
nucleation bubble and therefore nucleation is further suppressed.

\chapter{Summary}

The increasing energy of the gamma rays and neutrinos emitted by exploding microscopic black holes and
the disappearance of such point sources in a certain period of time 
are unique characteristics because astrophysical sources normally cool at late times.  
This would 
directly reflect the increasing Hawking temperature as the black hole explodes 
and disappears.  We should emphasize in particular the usefulness of 
distinguishing between the electron and muon-type neutrinos and the tau-type 
ones.  The latter are much less likely to be produced by high energy cosmic 
rays.  If the rate density of exploding microscopic black holes in our vicinity 
is anywhere close to the current limit based on gamma rays, it should be 
possible to observe them with present and planned large astrophysical neutrino 
detectors.  Observation of high energy neutrinos, especially in conjunction with 
high energy gamma rays, may provide a window on physics well beyond the TeV 
scale.  
The results of this research will bear directly on the
 observational program on astroparticle physics that involves gamma ray detectors such as Milagro and neutrino detectors such as AMANDA and IceCube, as well as detectors for ultra high energy
cosmic rays, such as Auger, to observe and detect possible signals
of microscopic black holes.

Still, there is much work to be done in determining whether the 
matter surrounding a black hole can reach and maintain thermal equilibrium. 
The previous calculations should be improved by including the effect 
of gravity, which was neglected in the calculation, which may lead to 
corrections of up to $10\%$.  The 
equation of state should be improved, and the viscosities computed using the 
relaxation times for self-consistency of the transition from viscous fluid flow to free streaming.  Also, there should be a more fundamental investigation of 
the relaxation times starting from the microscopic interactions.  Our next step is to calculate the flux of other particles such as antiprotons and positrons radiated by microscopic black 
holes in order to help the observation of these black holes.   Positrons and antiprotons are expected to freeze out at a temperature around 100 MeV.  Another worthwhile project is to carry out cascade simulations of the 
spherically expanding matter around the microscopic black hole at a level of 
sophistication comparable to that of high energy heavy ion collisions. This 
project is much more complicated than the cascade simulation in heavy ion 
collision, though, because we need to deal with a much wider range of energies 
and particles involved in exploding microscopic black holes.   

The study of microscopic black holes might well lead to great
advancements in fundamental physics.  Because the highest
temperatures in the universe exist in the vicinity of microscopic black holes,
matter at extremely high temperatures can be studied, and physics
beyond the four dimensional Standard Model and above the electroweak
scale can be tested.  In addition, because it is believed that
baryon number is violated at high temperatures, the study of
microscopic black holes could possibly answer the question of why
our universe became matter-dominated.  Because microscopic black
holes explode, they are an ideal model for studying the Big Bang
and the birth of our universe.  Finally, the study of microscopic
black holes will help us to determine whether they are the source
of the highest energy cosmic rays.  The origin of these cosmic
rays is still one of the biggest mysteries today. Observation and
experimental detection of microscopic black holes with satellite and
ground based neutrino observatories may lead to a novel paradigm
for understanding new sources of the highest energy cosmic and
gamma rays in the Universe.  Observation and experimental detection of microscopic black holes will be one 
of the great challenges in the new millennium.


In the second part of this thesis we have reported the discovery of shell-like solutions to the 
combined field equations of gravity and a scalar field with a double-minima 
potential.  These solutions exist in a space that is asymptotically de Sitter.
The range of parameters which allow such solutions are very broad.  If anything 
like these structures exist in nature they most likely would have been created 
in the early universe and are therefore cosmological.  We know of no other way 
to produce them.

To make matters interesting, let us suppose that the cosmological constant
suggested by recent observations of distant Type Ia supernovae \cite{super}
arises from the universe actually being in a false vacuum state.  A best fit to
all cosmological data \cite{science} reveals that the present energy density of
the universe has the critical value of $\epsilon_c = 3H_0^2/8\pi G$,
with one-third of it consisting of ordinary matter and two-thirds of it
contributed by the cosmological constant.  Suppose that the cosmological
constant arises from a potential of the form we have been analyzing.
Then with a present value of the
Hubble constant of $H_0=65$ km/s$\cdot$Mpc we find that
\begin{equation}
a = \sqrt{\frac{3}{2}}\frac{1}{H_0} = 1.7 \times 10^{26} \,\, {\rm m}
\end{equation}
and so
\begin{equation}
\left( \lambda f^3 \Delta f \right)^{1/4}=2.4 \times 10^{-3} \,\, {\rm eV} \, .
\end{equation}
This is a constraint on the parameters of the potential, $\lambda$, $f$ and 
$\Delta f$.  Although they cannot be determined individually from this data,
we can place limits on them such that a shell structure might arise.  Recalling
(\ref{estimate2}) we find
\begin{eqnarray}
f &<& 30 \lambda^{-1/6} \, {\rm MeV} ~, \nonumber \\
w &>& 6.3 \lambda^{-1/3} \, {\rm fm} ~.
\end{eqnarray}
This is pure speculation of course.  A cosmological constant, if
it exists, may have its origins elsewhere.  But if it does arise from a false
vacuum, a variety of questions immediately present themselves.  Is $\phi$ a new
field, not present in the standard model of particle physics, whose only purpose
is this?  Where does the energy scale of 2.4 meV come from?  Why should
$V[\phi]$ have a global minimum of 0, especially when quantum mechanical
fluctuations are taken into account?  To these questions we have no answers.

\appendix

\appendix
\chapter{Numerical Calculations}

In the numerical calculations of Eqs. (2.11), (2.13), (2.14), (2.17) and (2.18), we use the unitless variables $t=T/T_H$ and $x=r/r_s$.  In terms of these unitless variables, we can rewrite the equations as   
\begin{eqnarray}
s_t &=&  \frac{4\pi^2}{90} t^3 \,
\left\{ \begin{array}{ll}
101.5, \hspace{0.2in} t_{EW} \leq t ,\\
56.5 + 45 \,{\rm e}^{-(t_{EW}-t)/t}, \hspace{0.2in} t_{QCD}\leq t < t_{EW},\\
2 + 3.5 \,{\rm e}^{\frac{-m_e/T_h}{t}}+27.25\,{\rm e}^{-(t_{QCD}-t)/t},  \hspace{0.2in} t < t_{QCD},
\end{array}\right.
\end{eqnarray}
\begin{equation}
\eta_t = \frac{82.5}{\frac{4}{3} \frac{\pi^2}{30}101.5}s_t,
\end{equation}
\begin{equation}
\xi_t = 10^{-4}\eta_t,
\end{equation}
\begin{equation}
\gamma uts_t
-\frac{4}{3} (4\pi)\eta_t \gamma u \left( \frac{du}{dx} - \frac{u}{x} \right) 
-(4\pi)\zeta_t \gamma u \left( \frac{du}{dx} + \frac{2u}{x} \right)
         = \frac{4 \pi L/T_H^2}{x^2},
\end{equation}
and
\begin{equation}
\frac{d}{dx}(x^2 u s_t)  = 4\pi\frac{x^2}{t} \left[
\frac{8}{9} \eta_t \left( \frac{du}{dx} - \frac{u}{x} \right)^2
+ \zeta \left( \frac{du}{dx} + \frac{2u}{x} \right)^2 \right]
\end{equation}
where $s_t=s/T_H^3$, $\eta_t=\eta/T_H^3$ and $\xi_t=\xi/T_H^3$.  Here $L=64\pi^2\alpha_hT_H^2$ where $\alpha_h=4.43\times 10^{-4}$ includes neutrinos and gravitons.  In the case of calculating gamma ray spectra, gravitons and neutrinos should be ignored since gravitons never contribute to the fluid and neutrinos freeze-out at a temperature around $100$ GeV.  In the case of calculating neutrino spectra, only gravitons need to be ignored for direct neutrinos which freeze-out at $100$ GeV.  For the indirect neutrinos both gravitons and neutrinos need to be ignored.   In our calculations these small numerical differences in $\alpha_h$ have been neglected.

In order to find the asymptotic behavior of $u$ and $t$ in the limit where $x \rightarrow 0$, we look for a behavior of 
the form $u=Ax^\alpha$ and $t=Bx^\beta$ with $\gamma \approx 1$.  For black holes with  
temperatures greater than $T_{EW}=100$ GeV, we can write 
the equation of state as 

\begin{eqnarray}
s_t=\frac{4}{3}at^3, & \eta_t=b_St^3 & \mbox{and } ~ \xi_t=b_Bt^3, 
\end{eqnarray}
where $a=\frac{\pi^2}{30}101.5$, $b_S= 82.5$ and $b_B= b_S\times 10^{-4}$. The best solution in the small $x$ limit is

\begin{eqnarray}
\alpha &=& \frac {2}{5},  \nonumber \\   
\beta &=& -\frac {3}{5} , \nonumber \\ 
A &=& \frac {5a}{\pi(8b_S+144b_B)}
        \Biggl\{ \frac{4\pi L}{T_H^2} \nonumber \\
         &&\times \left(\frac{4}{3}a+4\pi(\frac{4}{5}b_S
            -\frac{12}{5}b_B)
             \frac {5a}{\pi(8b_S+144b_B)}\right)^{-1} 
     \left(\frac {5a}{\pi(8b_S+144b_B)}\right)^{-1} \Biggl\}^{1/5}
,\nonumber \\
\mbox{and} \nonumber \\
B &=& \Biggl\{ \frac{4\pi L}{T_H^2} \left(\frac{4}{3}a+4\pi(\frac{4}{5}b_S
             -\frac{12}{5}b_B)
             \frac {5a}{\pi(8b_S+144b_B)}\right)^{-1}\nonumber \\
      &&\times \left(\frac {5a}{\pi(8b_S+144b_B)}\right)^{-1}\Biggl\}^{1/5}
.
\end{eqnarray} 
  Typical numerical values for $u$ and $t$ in this limit are $u = 0.105715x^{2/5}$ and $t = 1.31521x^{-3/5}$.

In order to find the asymptotic behavior of $u$ and $t$ in the limit where $x \rightarrow \infty$, we again look for solutions of the form 
$u=Ax^{\alpha}$ and $t=Bx^{\beta}$ with $\gamma \approx u$. In this limit the equation of state can be written in the same form as in Eqn. (A.6), where $a=\frac{\pi^2}{30}2$, $b_S= 82.5\frac{2}{101.5}$ and $b_B= b_S\times 10^{-4}$. The asymptotic solution in the large $x$ limit is  
\begin{eqnarray}
\alpha &=& \frac {1}{3},  \nonumber \\   
\beta &=& -\frac {2}{3} , \nonumber \\ 
A &=& \frac {9a}{\pi(32b_S+441b_B)}
\Biggl\{\frac{4\pi L}{T_H^2} \nonumber \\ 
&& \times \left(\frac{4}{3}a+4\pi( \frac{8}{9}b_S- \frac{7}{3}b_B)
\frac {9a}{\pi(32b_S+441b_B)}\right)^{-1}
\left(\frac {9a}{\pi(32b_S+441b_B)}\right)^{-2}\Biggl\}^{1/6}
,\nonumber \\
\mbox{and} \nonumber \\
B &=& \Biggl\{ \frac{4\pi L}{T_H^2} 
\left(\frac{4}{3}a+4\pi( \frac{8}{9}b_S- \frac{7}{3}b_B)
\frac {9a}{\pi(32b_S+441b_B)}\right)^{-1} \nonumber \\
&& \times \left(\frac {9a}{\pi(32b_S+441b_B)}\right)^{-2}\Biggl\}^{1/6}
.
\end{eqnarray} 
The typical numerical values for $u$ and $t$ in this limit are $u= 0.184460x^{1/3}$ and $t=5.09763x^{-2/3}$.  

\begin{figure}[tbh]
\centerline{\epsfig{figure=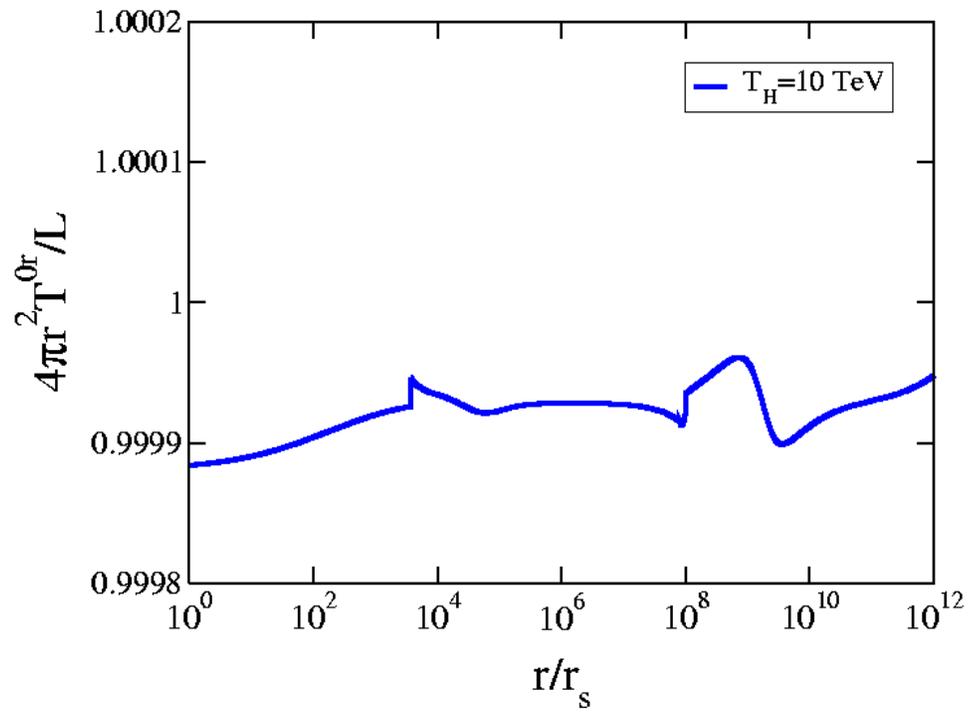,width=11.0cm,angle=270}}
\caption{$4\pi r^2 T^{0r}/L$ as a function of $r/r_s$ for a black hole temperature of $10$ TeV.  The curve begins at $r_s$ and terminates when the local temperature reaches $T=m_e$.}
\end{figure}

To solve Eqs. (A.1)-(A.5) we use a 4th order vector Runge-Katta 
method.  In our program we apply a shooting type method to find the numerical solution over a wide range of $x$ such that $u$ has the correct asymptotic behavior when $x \rightarrow \infty$.  In the program this is done by comparing the numerical solution for $u$ to its correct asymptotic behavior at large $x$.  The initial value of $t$ at $x=0.1$ is fixed and the initial value of $u$ at $x=0.1$ is changed by the program until the correct solution is obtained.  In order to check the validity of the numerical solution, we calculate $4\pi r^2 T^{0r}/L$ using finite difference method.  The result, which is very close to 1 over a wide range of x, is plotted in figure A.1 for a black hole temperature of 10 TeV.  For more information see the numerical recipes mbhwind.c and subroutinewind.h attached to this thesis.

\end{document}